\documentclass[aps,pra,twocolumn,superscriptaddress,showpacs,floatfix]{revtex4-1}

\usepackage{times,amssymb,amsmath}
\usepackage{txfonts}
\usepackage[usenames,dvipsnames]{color}
\usepackage{array}
\usepackage{simplewick}
\usepackage[pdftex]{graphicx}
\renewcommand{\vec}[1]{\mathbf{#1}}

\mathchardef\mhyphen="2A

\renewcommand{\Re}{\text{Re}\,}

\newcommand{\sv}{\varv}
\newcommand{\sV}{\mathcal{V}}

\newcommand{\ket}{\rangle}
\newcommand{\bra}{\langle}
\newcommand{\be}{\begin{equation}}
\newcommand{\ee}{\end{equation}}
\newcommand{\bea}{\begin{eqnarray}}
\newcommand{\eea}{\end{eqnarray}}

\newcommand{\mC}{\mathcal{C}}

\newcommand{\mU}{\mathcal{U}}

\newcommand{\hpsi}{\hat{\psi}}
\newcommand{\hpsid}{\hat{\psi}^{\dagger}}

\newcommand{\br}{\mathbf{r}}

\newcommand{\x}{\textrm{x}}

\newcommand{\s}{\sigma}

\newcommand{\nn} {\nonumber}

\def\a{\alpha}
\def\b{\beta}

\def\s{\sigma}

\newcommand{\iu}{\mathrm{i}}
\newcommand{\plus}{ {\scriptscriptstyle +}}
\newcommand{\minus}{{\scriptscriptstyle -}}
\makeatletter
\let\amstexbig\big
\def\newbig#1{%
  \ifx#1|%
    \expandafter\@firstoftwo
  \else
    \expandafter\@secondoftwo
  \fi
  {\big@bar}%
  {\amstexbig{#1}}%
}
\AtBeginDocument{\let\big\newbig}
\def\big@bar{\bBigg@{1.1}|}
\makeatother
\begin{document}


\title{Single and double electron emission: combination of projection operator and
  nonequilibrium Green's function approaches} \author{Y. Pavlyukh}
\email[]{yaroslav.pavlyukh@physik.uni-halle.de} \author{M. Sch{\"u}ler}
\author{J. Berakdar} \affiliation{Institut f\"{u}r Physik, Martin-Luther-Universit\"{a}t
  Halle-Wittenberg, 06120 Halle, Germany} \date{\today}
\begin{abstract}
This work provides a unified theoretical treatment of the single and correlated
double-electron emission from a general electronic system.  Using Feshbach projection
method, the states of interest are selected by the projection operator; the Feshbach-Schur
map determines the effective Hamiltonian and the optical potential for the emitted
electrons. On the other hand, the nonequilibrium Green's functions method is demonstrated
to be a complementary approach and an explicit correspondence between both methods is
established.  For a self-contained exposition some results on single electron emission are
re-derived using both formalisms. New insights and results are obtained for the correlated
electron-pair emission: This includes the effective two-electron Hamiltonian, the explicit
form of the Feshbach self-energy in terms of the many-body self-energies, and the
diagrammatic expansion of the two-particle current. As an illustration of the diagrammatic
technique the process of the two-particle emission assisted by the excitation of plasmons
is explicitly worked out.
\end{abstract}
\maketitle
%
%
\section{Introduction}
Scattering experiments deliver the most detailed information on the structure of matter.
For instance, the fully resolved spectra of an electron emitted from an electronic system
upon photon or particle impact encode the spin and momentum-resolved spectral properties
of the sample
\cite{cardona_photoemission_1978,hufner_photoelectron_2003,schattke_solid-state_2002,
  schmidt_electron_1997,weigold_electron_1999}. For direct information on the two-particle
properties the detection of a correlated electron pair is necessary which is usually
performed in a one-photon double-electron emission \cite{schmidt_electron_1997} or in a
swift particle-impact double-electron emission experiment
\cite{berakdar_electron-impact_2003}.  Calculations of the electron emission spectra from
atomic and molecular systems~\cite{cardona_photoemission_1978,manson_photoelectron_1982,
  amusia_atomic_1990,schmidt_electron_1997,eland_dynamics_2009} as well as from condensed
matter~\cite{cardona_photoemission_1978,hufner_photoelectron_2003,
  schattke_solid-state_2002} are done routinely. The underlying theories and techniques
differ, however.  The issue addressed here concerns the formulation of a unified and
numerically accessible theoretical framework of single and double photoelectron emission
(SPE and DPE) from finite and extended electronic systems.  A method of choice for this
purpose is the nonequilibrium Green's functions (NEGF)
approach~\cite{kadanoff_quantum_1962,devreese_linear_1976,danielewicz_quantum_1984,
  stefanucci_nonequilibrium_2013}.  In full generality the response function describing
electron emission is more involved than the optical response which is related to
\emph{time-ordered} particle-hole ($p\mhyphen h$) Green's function (GF) for which well
established approximations exist. Even for a \emph{single} electron emission the response
function can only be defined on the Keldysh contour and after performing the calculations,
the times are projected on the real \emph{observable} times. The second complication is
that for a fixed energy and momentum of the detected electron the sample maybe left in an
excited state.  A typical example is the plasmon satellites in core-level
photoemission~\cite{wohlfarth_ionization_2002}. There, the target is left with one excited
plasmon ~\cite{inglesfield_plasmon_1983}. The conservation of energy and momentum allows
to focus on, e.~g., the \emph{no-loss current}. The response function is then determined
by the product of two vertex functions and three single-particle Green's
functions~\cite{almbladh_theory_1985}. If an approximation is made for one of the
constituents, it has to be taken over consistently to the others. The notion of a
\emph{conserving} approximation is rooted in this requirement.

First theories of electron emission were empirical: E.g. for surfaces, following Berglund and
Spicer~\cite{berglund_photoemission_1964} the photoemission is regarded as a three stages
process: excitation, transport to the surface (during this stage the particle may loose
energy), and the transformation into a scattering (detector) state.  In 1970
Gerald~D.~Mahan wrote ``we have not yet been able to derive a simple, time-ordered,
correlation function which would serve as the starting point for a closed-loop type of
calculation. That is, we have not yet found a "Kubo formula for
photoemission."''~\cite{mahan_theory_1970}. Shortly thereafter Schaich and
Ashcroft~\cite{schaich_model_1971} and Langreth~\cite{langreth_scattering_1971} employed a
time-ordered formalism for the response function, and Caroli \emph{et
  al.}~\cite{caroli_inelastic_1973} introduced the nowadays standard NEGF formulation. The
well-known Fermi Golden rule expression for the photocurrent
\[
J_\vec
p=2\pi\int_{-\infty}^{\mu}\!\!d\varepsilon\,\delta(\varepsilon_{\vec p}-\varepsilon-\omega)
\langle\chi_{\vec p}^{(-)}|\hat \Delta \hat A(\varepsilon)\hat\Delta^\dagger|\chi_{\vec p}^{(-)}\rangle
\]
derives  rigorously  from the response-function formalism. In 1985 Carl-Olof Almbladh
obtained the following modifications of the \emph{no-loss} current:
\[
\begin{split}
J_\vec
p&=2\pi\int_{-\infty}^{\mu}\!\!d\varepsilon\,\delta(\varepsilon_{\vec p}-\varepsilon-\omega)
\langle\chi_{\vec p}^{(-)}|\hat\Lambda(\varepsilon+\omega,\varepsilon)
\hat A(\varepsilon)\\&\quad\times\hat\Lambda^\dagger(\varepsilon+\omega,\varepsilon)|\chi_{\vec p}^{(-)}\rangle.
\end{split}
\]
In these formulas an interaction with an electromagnetic field of the frequency $\omega$ is assumed.
 $\chi_{\vec p}^{(-)}$ denotes the final scattering state with the momentum $\vec p$ and
energy $\varepsilon_{\vec{p}}$, and  $\hat A(\epsilon)$ is the spectral function. $\hat
\Lambda(\epsilon+\omega,\epsilon)$ is the so-called vertex function which,  for noninteracting
systems,  reduces to the operator of the light-matter interaction $\hat \Delta$. In interacting
systems it describes the screening of the optical field by the sample electrons  and
the accompanying polarization effects~\cite{almbladh_importance_1986}.

The physics beyond \emph{no-loss} has many facets.  There are two prominent examples: the
plasmon satellites~\cite{inglesfield_plasmon_1983,
  campbell_interference_2002,guzzo_valence_2011} and the Auger
effect~\cite{cini_density_1976,sawatzky_quasiatomic_1977,tarantelli_many_1991,tarantelli_aggregation_1994}. In
both cases the system is left in an excited state that relaxes subsequently either due
many-body effects or results in the emission of a secondary electron. It should be noted,
however, that the borderline in such a classification is blurred: one can consider the
Auger effect as a two-step process, in which the decay is treated independently from the
primary ionization or as the \emph{no-loss} double
photoemission~\cite{verdozzi_auger_2001}. The former point of view yields a description of
the Auger effect in terms of an equilibrium two-hole Green's
function~\cite{sawatzky_quasiatomic_1977,cini_theory_1978,tarantelli_greens_1985}.

The goal here is to generalize the nonequilibrium approach as to treat single and double
electron emission. We will mostly discuss processes related to the absorption of
\emph{one} photon. Particle impact is discussed only in the optical limit as specified in
the Appendix~\ref{ap:EI}.  In particular, this work provides a detailed discussion of DPE,
a process that was experimentally realized for various
systems~\cite{schmidt_electron_1997,herrmann_two_1998}.  For a self-contained presentation
we start by defining observables and introducing basic formulas solely based on the
time-dependent perturbation theory and the assumption of adiabatic switching of the
light-matter interaction (Sec.~\ref{sec:basic}). Already on this level one can reformulate
these expressions in the Fermi golden rule form and demonstrate how the sudden
approximation can be used to reduce the many-body to two-body description
(Sec.~\ref{sec:sudden}). Such reduction, however, neglects the energy loss of an emitted
electron on its way to detector. These \emph{extrinsic} losses are treated by means of the
projection operator technique (Sec.~\ref{sec:prj}). For single photoemission (SPE) this
approach was established in works of Almbladh~\cite{almbladh_theory_1985}, Bardyszewski
and Hedin~\cite{bardyszewski_new_1985}, Fujikawa and Hedin~\cite{fujikawa_theory_1989},
Hedin, Michiels and Inglesfield~\cite{hedin_transition_1998}, and for DPE by Brand and
Cederbaum~\cite{brand_extended_1996}. The notion of the optical potential is central to
this approach. While the case of elastic scattering was considered in a classical work of
Bell und Squires~\cite{bell_formal_1959}, the inelastic case, which is especially relevant
for photoemission, is more involved and has a long history with a recent progress due to
Cederbaum~\cite{cederbaum_optical_2000,cederbaum_optical_2001}. In Sec.~\ref{sec:fgr} we
closely follow the derivation of Almbladh and extend the theory to the two-electron
case. There are important differences as compared to the single-electron emission. Under
some assumptions DPE is only possible for interacting
systems~\cite{berakdar_emission_1998}. We demonstrate that the vertex function is the
source of this electronic correlation effect. Finally, we corroborate our findings by
performing a diagrammatic expansion of the derived DPE response function in terms of
Green's function on the Keldysh contour (Sec.~\ref{sec:diag}). We consistently use atomic
units.

\section{The two-electron current\label{sec:theory}}
For DPE from atomic and molecular systems
\cite{schwarzkopf_energy-_1993,briggs_differential_2000} a variety of very successful
techniques, based on a full numerical solution or using approximate correlated scattering
states of the few-body Schr\"odinger equation, were put forward. The wave-function-based
methods and, consecutively, the scattering approach are less suitable for extended
degenerate fermionic systems. Such DPE experiments were first performed for Cu(001) and
Ni(001) crystals~\cite{herrmann_two_1998} and meanwhile for a variety of other samples.
Here comes the \emph{response} formalism into play: the expectation values of products of
the creation and annihilation operators are computed over the ground state of a
(many-body) system, and perturbative expansions are evaluated with the help of Wick's
theorem. If the studied process can be regarded as a multi-step event, then the rate
equations are often a very efficient tool. They can be derived either from the density
matrix or from the NEGF formalisms using some additional assumptions.  For instance the
generalized Kadanoff-Baym Ansatz has been used to derive the quantum master equations
starting from NEGF approach to describe the transport in molecular
systems~\cite{esposito_transport_2009}.

Here we present a self-contained derivation of the two-particle current starting from the
time-dependent perturbation theory. The resulting formula (Eq.~\eqref{eq:golden}) is,
however, less useful for practical applications because it requires (generally unknown)
many-body states.  One has either a choice to completely neglect the target-ejected
particles interaction which still might be relevant for higher energies
(Sec.~\ref{sec:sudden}), or, as will be demonstrated in the next section (\ref{sec:prj})
to properly reduce the formulations as to work with effective residual interactions
(i.e. optical potentials).
\subsection{Basic definitions\label{sec:basic}}
\paragraph{Hamiltonian:} A system of interacting fermions is considered  that has the Hamiltonian
%
\begin{eqnarray}
\hat{H}&=&\int d x
\,\hpsid(x)h(x)\hpsi(x) \nn\\ &+&\frac12\int d x d x^\prime
\hpsid(x)\hpsid(x^\prime)v(x,x^\prime)\hpsi(x^\prime)\hpsi(x),
\label{ham}
\end{eqnarray}
%
where the field operator $\hpsi$ ($\hpsi^{\dag}$) with argument $x\equiv(\br,\s)$
annihilates (creates) a fermion in position $\br$ with spin $\s$. Needed below is the
anti-symmetrized interaction
\begin{multline}
V(x_1, x_2,x_3,x_4)=v(\vec r_1,\vec r_2)\big(\delta(x_2-
x_3)\delta(x_1-x_4)\\ -\delta(x_1-x_3)\delta(x_2-x_4)\big).
\label{eq:sv}
\end{multline}
One may  wish also to change the basis for the representation of creation and
annihilation operators via
\begin{equation}
\hpsi(x)=\sum_i\langle x| i\rangle c_i,
\label{eq:cdef}
\end{equation}
where the sum runs over a complete set of one-particle states and we consistely skip
$\hat{\cdots}$ on $c_i$ and $c_i^\dagger$. To study photoemission we need to further
classify the states according to their geometric character. A state will be called
\emph{bound} ($\phi_i\in \mathcal{B}$) if for any $\epsilon>0$ there is a compact set $B
\subset \mathbb{R}^3$ such that for all times $t$ the state remains in $B$: $\lVert
\chi_{B^{c}} e^{i t \hat{H}} \,\phi_i  \rVert <\epsilon$, where $B^c$ is the complement of B,
$\chi_{B^{c}}$ denotes the corresponding characteristic function. Analogically for the
\emph{scattering} states ($\phi_{\vec{k}}\in \mathcal{C}$) we adopt the following
definition: they are the vectors for which
$\lim_{T\rightarrow\infty}\frac1{2T}\int_{-T}^T\lVert\chi_B e^{it \hat{H}} \phi_{\vec
  {k}}\rVert\,dt=0$ for all compact sets $B \subset \mathbb{R}^3$, i.e. they leave any bounded
region. It is clear that $\mathcal{B}\perp\mathcal{C}$ and according to the RAGE
theorem~\cite{demuth_determining_2005} all the states from the discrete (point) spectrum
are bound, whereas the continuum states (absolutely continuous and singularly continuous)
are the scattering states. Thus, parallels between the geometric and the spectral
classification allows us to use continuum and scattering, and point and bound terms
interchangeably, although for the purpose of the present work the geometric classification
is preferred. Finally we note that if our theory is to be applied to solids the use of
localized Wannier functions~\cite{marzari_maximally_2012} is preferred, at least for
systems where their existence can be proved~\cite{brouder_exponential_2007}.

 We will use the letters $(abcd)$ for general orbitals, $(ijnm)$ for bound orbitals and
 bold-face letters for continuum states. In these notations:
%
\bea
  \label{eqhamiltonian2ndq2}
  \hat{H} &=& \sum_{a b} t_{a b} c^\dagger_a c_b + \frac12
  \sum_{abcd}\sv_{abcd} c^\dagger_a c^\dagger_b c_d c_c \\
  &=&\sum_{a b} t_{a b} c^\dagger_a c_b + \frac14
  \sum_{abcd}\sV_{abcd} c^\dagger_a c^\dagger_b c_d c_c \ .
\eea
%
\paragraph{Initial state preparation:}
The above Hamiltonian determines the quantum state of the target (wave-function $|\Psi_0
\rangle$ with corresponding energy $E_0$) in the remote past ($t=-\infty$). When the
system is perturbed by the interaction with external fields it evolves to a new state. As
a typical mechanism we consider here the light-matter interaction
%
\be \hat{V}(t)=(\hat \Delta e^{-i\omega
  t}+\hat \Delta^{\dagger}e^{i\omega t})e^{\eta t},\quad
\hat\Delta=\sum_{ab}\Delta_{ab}c_a^\dagger c_b.
\label{eq:V}
\ee
%
In this expression $\hat{V}(t)$ is adiabatically turned on allowing to introduce a typical
interaction time $\sim (2\eta)^{-1}$. The form \eqref{eq:V} permits generalizations: In
Appendix~\ref{ap:EI} we consider the process of impact ionization caused a charged
projectile particle (e.g. an electron) impinging on the target system. At high energy the
projectile can be regarded as distinguishable from electrons of the system. This allows to
average the projectile-target interaction over the projectile's states and write the
perturbation in essentially the same form as in Eq.~\eqref{eq:V}, i.e. as a
single-particle operator.

From the first-order time-dependent perturbation theory we obtain the \emph{approximate}
eigenstate $|\tilde{\Psi}^{(+)}\rangle$ of the full Hamiltonian $\hat{H}+\hat{V}(t)$ at
time $t=0$:
\begin{equation}
|\tilde{\Psi}^{(+)}\rangle=|\Psi_0\rangle+\lim_{\eta\rightarrow 0}\frac{1}
 {E_0+\omega-\hat{H}+i\eta}\hat \Delta|\Psi_0\rangle.\label{eq:pt}
\end{equation}

Readers will immediately notice parallels of Eq.~\eqref{eq:pt} with the scattering theory
where the M{\o}ller operators $\hat{\Omega}^{(\pm)}$ convert an eigenstate of $\hat H$
(the Hamiltionian of the target system) at $t=\mp\infty$, into an eigenstate of $\hat
H+\hat V(0)$ (the full Hamiltonian)
$|\Psi^{(\pm)}_\alpha\rangle=\hat{\Omega}^{(\pm)}|\Psi_\alpha\rangle$ at time $t=0$
(cf. Eqs.~(14.66) of Joachain~\cite{joachain_quantum_1975}). The scattering theory is
required when electromagnetic fields are quantized. For classical fields Eq.~\eqref{eq:pt}
follows from the first order expansion (in $\hat\Delta$) of the M{\o}ller operator
$\hat{\Omega}^{(+)}$. To emphasize the similarity we denote the state given by
Eq.~\eqref{eq:pt} as the scattering state. In what follows we omit the tilde which we used
to denote its approximate character.

\paragraph{Observables:}
Assuming we know the quantum state of the target at $t=0$ some observables can be
computed. Since we are interested in photoemission these are the expectation values of the
current operators. The safe way to introduce them is to use the continuity equation which
is gauge-invariant. The one-electron current $J_\vec{k}$ is defined as the number of
electrons $N_\vec{k}$ with a given momentum $\vec k$ outside the target divided by the
effective interaction time $(2\eta)^{-1}$. There is a detailed
discussion~\cite{almbladh_theory_1985} on why electrons in the sample give a negligible
contribution to the current. Same arguments are valid for the two electron case. Thus, we
analogically define the two-electron current as
\begin{equation}
J_{\vec{k_1},\vec{k_2}}=\lim_{\eta\rightarrow0}2\eta \langle
\hat{N}_{\vec{k_1}}\hat{N}_{\vec{k_2}}-\delta_{\vec k_1,\vec k_2}\hat{N}_{\vec{k_1}}\rangle.
\label{eq:j2}
\end{equation}
In the expression above (and all subsequent derivations) we do not explicitly spell out
the spin quantum numbers. The dependence on the spin can be recovered by substituting the
continuum quantum numbers like $\vec k$ by $\vec k \s$ (likewise for bound indices).  The
second term excludes the one-electron current in the case when two momenta are equal.
Eq.~\eqref{eq:j2} gives access to the differential cross-section through the following
relation:
\be
\frac{d^2\sigma}{d\vec{k_1}d\vec{k_2}}=\frac{\omega}{I}J_{\vec{k_1},\vec{k_2}},
\ee
where $I/\omega$ is the photon flux density~\cite{berakdar_concepts_2003}. For the
velocity gauge $\hat \Delta=\frac{1}{c}\vec{A_0}\cdot\hat{\vec{p}}$, $I=\frac{\omega^2A_0^2}{2\pi
  c}$, where $\vec{A_0}$ is the amplitude of the vector potential. Similar expressions can
be given for the length gauge.

 The average in Eq.~\eqref{eq:j2} is performed over the perturbed state~\eqref{eq:pt}:
\begin{multline}
J_{\vec{k_1},\vec{k_2}}=\lim_{\eta\rightarrow0}2\eta
\big\langle \Psi_0\big|\hat \Delta^\dagger\frac{1} {E_0+\omega-\hat{H}-i\eta} c_\vec{k_1}^\dagger c_\vec{k_2}^\dagger
c_\vec{k_2} c_\vec{k_1}\\\times\frac{1}{E_0+\omega-\hat{H}+i\eta}\hat\Delta\big|\Psi_0\big\rangle,
\label{eq:jresp}
\end{multline}
where we used the usual anti-commutation relations for the fermionic operators. The
current is quadratic in $\hat \Delta$ or linear in the number of absorbed photons. The
first order in $\hat \Delta$ gives the linear conductivity current and is of no interest
here.~\cite{caroli_inelastic_1973}

To derive the Fermi golden rule for DPE we insert a complete set of the $(N-2)$-particle
states and use the scattering theory to evaluate matrix elements of the type:
\[
M^*_{\vec{k_1},\vec{k_2},\beta}=\big\langle \Psi_0\big|\hat \Delta^\dagger\frac{1} {E_0+\omega-\hat{H}-i\eta}
c_\vec{k_1}^\dagger c_\vec{k_2}^\dagger \big|\Psi_\beta^{2+}\big\rangle.
\]
We will generally use lower indices to distinguish quantum states and upper indices to
indicate the charge of the system or the nature of the state ($\pm$), i.e., incoming or
outgoing. For a scattering process with the following energy balance
\[
E_i=E_0+\omega\rightarrow E_f=\varepsilon_\vec{k_1}+\varepsilon_\vec{k_2}+E^{2+}_\beta
\]
the M{\o}ller operator $\hat{\Omega}^{(-)}$ translates a wave-function in the remote
future into a \emph{incoming} (they are sometimes called inverted LEED
states~\cite{hedin_transition_1998}) scattering state at $t=0$:
\[
|\Psi^{(-)}_\beta\big\rangle=\hat{\Omega}^{(-)}c_\vec{k_1}^\dagger c_\vec{k_2}^\dagger
\big|\Psi_\beta^{2+}\big\rangle =\lim_{\eta\rightarrow
  0}\frac{-i\eta}{E_f-\hat{H}-i\eta}c_\vec{k_1}^\dagger c_\vec{k_2}^\dagger
\big|\Psi_\beta^{2+}\big\rangle.
\]
Following Almbladh~\cite{almbladh_theory_1985} we obtain:
\begin{equation}
M^*_{\vec{k_1},\vec{k_2},\beta}=\frac{1}{E_i-E_f-i\eta}\big\langle \Psi_0\big|\hat \Delta^\dagger\big|
\Psi^{(-)}_\beta\big\rangle,\label{eq:T_DPE}
\end{equation}
resulting in the Fermi golden rule for DPE for an adiabatic switching of $\hat{V}(t)$:
\begin{multline}
J_{\vec{k_1},\vec{k_2}}=\lim_{\eta\rightarrow0}2\eta \sum_\beta \big|M_{\vec{k_1},\vec{k_2},\beta}\big|^2\\
=2\pi \sum_\beta \delta(E_i-E_f)\big|\big\langle\Psi^{(-)}_\beta|\hat \Delta|\Psi_0\big\rangle\big|^2.
\label{eq:golden}
\end{multline}

This is essentially an exact equation if strong field effects are neglected, i.e. if the
first-order perturbation theory in field strength is adequate. Now we discuss some common
approximations. In the \emph{sudden approximation} the M{\o}ller operator is set to the
identity operator and it follows $|\Psi^{(-)}_\beta\rangle\approx c_\vec{k_1}^\dagger
c_\vec{k_2}^\dagger|\Psi_\beta^{2+}\rangle$ leading, e.g., to Eq.~(1) of Napitu and
Berakdar~\cite{napitu_two-particle_2010}.  The sudden approximation is broadly used to
interpret the single photoemission. However, it is easy to construct an example when it
completely fails: Consider photoemission from a system surrounded by a impenetrable
potential barrier. Irrespective of the photon energy there will be zero current in the
detector. Thus, it is \emph{extrinsic losses}~\cite{hedin_transition_1998} that are
missing in the sudden approximation.

\subsection{Sudden approximation\label{sec:sudden}}
In the sudden approximation for SPE it is possible to reduce the many-body description to
a single-particle picture which also allows to approximately treat the M{\o}ller operator
and accommodate extrinsic losses. The central object in such an approach are the
\emph{Dyson orbitals}~\cite{pavlyukh_communication:_2011}. The \emph{hole} Dyson orbital
is defined as an overlap of $(N-1)$ many-particle state with the $N$-particle initial
state:
\begin{multline}
\phi_\alpha(x_1)=\sqrt{N}\int\!d(x_2 \ldots x_N) [\Psi^+_\alpha(x_2,\ldots,x_{N})]^*\\
\times \Psi_0(x_1,\ldots,x_{N})
=\langle \Psi^+_\alpha| \hpsi(x_1) |\Psi_0\rangle,
\label{eq:1h}
\end{multline}
A rather extensive review of such overlap operators as well as the proof on the last
"dressed in the fancy outfit of the occupation number formalism'' identity can be found in
Ref.~\cite{bang_one-_1985}. Practical approaches for their computation are overviewed in
Refs.~\cite{cederbaum_correlation_1986,deleuze_new_1999}. By introducing a similar
\emph{two-hole Dyson orbital}:
\begin{multline}
\phi^{(2)}_\beta(x_1,x_2)
=\sqrt{\frac{N(N-1)}{2!}}\int\!\!d (x_3\ldots x_N) [\Psi^{2+}_\beta(x_3,\ldots,x_{N})]^*\\
\times \Psi_0(x_1,\ldots,x_{N})
=\frac{1}{\sqrt{2}} \big\langle \Psi^{2+}_\beta| \hpsi(x_1) \hpsi(x_2) |\Psi_0\big\rangle,
\label{eq:2h}
\end{multline}
and neglecting the M{\o}ller operator we obtain for the two-particle
current~\eqref{eq:golden}:
\begin{equation}
J_{\vec{k_1},\vec{k_2}}=2\pi \sum_\beta \delta(E_i-E_f)
\big|\big\langle\vec{k_1}\vec{k_2}|\hat \Delta|\phi^{(2)}_\beta\big\rangle\big|^2,
\label{eq:golden0}
\end{equation}
where $|\vec{k_1}\vec{k_2}\rangle$ is asymptotic two-particle state, i.e. anti-symmetrized
product of two plane-waves. The two-hole orbital is anti-symmetric with respect to the
interchange of particle coordinates and in general has norm $\le1$.  To derive
\eqref{eq:golden0} it is instructive to consider first a corresponding matrix element for
SPE:
\[
M_{\vec{k},\alpha}\approx\frac{1}{E_i-E_f+i\eta}
\sum_{ab}\Delta_{ab}\big\langle \Psi^{+}_\alpha\big|
c_\vec{k}c^\dagger_ac_b\big|\Psi_0\big\rangle.\label{eq:T0_SPE}
\]
Now we have
$c_\vec{k}c^\dagger_ac_b\big|\Psi_0\big\rangle=\delta_{\vec{k},a}c_b\big|\Psi_0\big\rangle
+c^\dagger_ac_bc_\vec{k}\big|\Psi_0\big\rangle$ and it is time to make another very
important assumption:
\begin{equation}
c_\vec{k}|\Psi_0\big\rangle\approx0.\label{eq:ap1}
\end{equation}
It is not valid in general, however, one can use the same arguments as Almbladh (see
discussion around his Eq.~(11)) to demonstrate that it gives a vanishing contribution. For
homogeneous electron gas this is even a generally valid statement. Besides allowing to
compute the matrix elements the assumption \eqref{eq:ap1} also justifies why terms
resulting from the second-order perturbation theory give vanishing contributions to the
current.

In this way (see
Appendix~\ref{sec:idem}) $M_{\vec{k},\alpha} =\frac{1}{E_i-E_f+i\eta}\langle\vec{k}|\hat
\Delta |\phi_\alpha\rangle$ and
\[
J_{\vec{k}}=2\pi\sum_{\alpha} \delta(E_i-E_f)\left|\langle\vec{k}|\hat \Delta|\phi_\alpha\rangle\right|^2.
\]
For DPE we analogically analyze the matrix element entering Eq.~\eqref{eq:T_DPE} and
neglect terms with two holes at momenta $\vec{k_1}$ and $\vec{k_2}$
(i.e. $c_\vec{k_2}c_\vec{k_1}\big|\Psi_0\big\rangle\approx0$) as compared to the terms
with only one hole (Appendix~\ref{sec:idem}). Notice that for SPE we neglected one hole
term as compared to zero hole contribution (cf. Eq.~\eqref{eq:ap1}).

It is obvious that the sudden approximation is only valid for large momenta
$k_{1,2}$ and it is indifferent to the state in which the system is left in (the final
double ionized state can be an excited state). Thus, it is desirable to generate improved
approximations to Eq.~\eqref{eq:golden} by rewriting it in the two-particle form, but with
an improved final state (such as Eq.~(4) of Fominykh \emph{et
  al.}~\cite{fominykh_theory_2000} or Eq.~(2) of Fominykh \emph{et
  al.}~\cite{fominykh_spectroscopy_2002}).

\section{Extrinsic effects\label{sec:prj}}
A many-body target interacts with light such that certain number of electrons is emitted.
Here, the fundamental question is whether it is legitimate to describe the process in such
a way that only quantum numbers of ejected particles are considered and remaining degrees
of freedom are traced out, i.e. put into some effective interactions. The projection
operator formalism is a general method to treat this kind of problems. In this section we
introduce the basic concepts of this theory and demonstrate the reader that a deep
connection with the nonequilibrium Green's function formalism exist. We conclude this
rather mathematical section by considering two examples. Based on these examples the Fermi
golden rule is derived in the subsequent section.
 \subsection{Nonequilibrium Green's functions}
\begin{figure}[b!]
	\centering
       \includegraphics[width=\columnwidth]{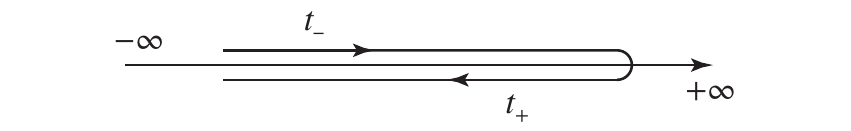}
       \caption[]{The Keldysh time-loop contour $\mC$. The forward branch is denoted with
         a ``$-$'' label while the backward branch is denoted by a ``$+$'' label.}
      \label{fig:contour}
\end{figure}

In the Keldysh formalism~\cite{stefanucci_nonequilibrium_2013} the field operators evolve
on the time-loop contour $\mC$ shown in Fig.\,\ref{fig:contour}.  Operators on the {\em
  minus}-branch are ordered chronologically while operators on the {\em plus}-branch are
ordered anti-chronologically.
Letting $z_{1}$ and $z_{2}$ be two contour-times, the Green's function $G(x_1z_1,x_2 z_2)$
can be divided into different components $G^{\a\b}(x_1t_1,x_2 t_2)$ depending on the
branch $\a,\b=+/-$ to which $z_{1}$ and $z_{2}$ belong. As before, $x_i$, denote a
composite coordinate comprising space and spin variables.  For $\a=\b=-$ we have the
\emph{time-ordered} Green's function
%
\begin{equation}
G^{\minus\minus} (x_1t_1,x_2 t_2) = -i \bra T
\left[ \hpsi_H (x_1 t_1) \hpsid_H (x_2 t_2)\right] \ket.
\label{eq:gmm}
\end{equation}
%
In this expression the average $\bra\ldots \ket$ is taken over a given density matrix
$\hat{\rho}$ and $T$ is the time-ordering operator.  The subscript ``$H$'' attached to a
general operator $\hat{O}$ signifies that that operator is in the Heisenberg picture
%
\begin{equation}
\hat{O}_{H}(t)=\hat{\mU}(t_0,t)\hat{O}\,\hat{\mU}(t,t_{0}),
\label{hp}
\end{equation}
%
where $\hat{\mU}(t_1,t_2)$ is the time-evolution operator and $t_{0}$ is an arbitrary
initial time. Reversing the time arrow the $G^{\minus\minus}$ is converted into the
\emph{anti-time-ordered} Green's function
%
\begin{equation}
G^{\plus\plus} (x_1 t_1,x_2 t_2) = -i
\bra \bar{T} \left[ \hpsi_H (x_1 t_1) \hpsid_H (x_2 t_2)\right] \ket,
\end{equation}
%
where
$\bar{T}$ orders the operators anti-chronologically. Finally, choosing $z_{1}$ and $z_{2}$
on different branches we have
%
\begin{subequations}
\label{g><}
\bea G^{\minus\plus} (x_1 t_1,x_2 t_2) &=&  i \bra
\hpsid_H (x_2t_2) \hpsi_H (x_1 t_1) \ket,
\label{g<}\\
G^{\plus\minus} (x_1 t_1,x_2t_2) &=&-i \bra \hpsi_H (x_1 t_1) \hpsid_H (x_2 t_2) \ket.
\label{g>}
\eea
\end{subequations}
%
The last two components are equivalently written as $G^{\minus\plus}=G^{<}$ (\emph{lesser}
Green's function) and $G^{\plus\minus}=G^{>}$ (\emph{greater} Green's function), and
describe the propagation of an added hole ($G^{<}$) or particle ($G^{>}$) in the medium.

It is often convenient in addition to time ordered and anti-ordered functions to introduce
the retarded and advanced components:
\begin{subequations}
  \bea
    \label{eqretardedgfdef1}
    G^\mathrm{R}(x_1, x_2; t) &=& \theta(t) \left[
      G^>(x_1, x_2; t)-G^<(x_1,x_2; t)\right],\\
    G^\mathrm{A}(x_1, x_2; t) &=& \theta(-t) \left[
      G^<(x_1,x_2; t)-G^>(x_1, x_2; t)\right].
  \eea
\end{subequations}
In order to find their representation in frequency space we multiply the retarded GF by
$e^{-\eta t}$ with $\eta \rightarrow 0+$ in order to enforce the convergence and compute
the Fourier integral:
\begin{multline}
  \label{eqretardedgffourier1}
  G^\mathrm{R}(x_1,x_2;\omega) = \langle \hat\psi(x_1)
  \frac{1}{\omega + E_0 - \hat{H} + \iu \eta}\hat{\psi}^\dagger(x_2) \rangle\\
 + \langle \hat\psi^\dagger(x_2)
  \frac{1}{E_0 - \omega - \hat{H} - \iu \eta}\hat{\psi}(x_1)  \rangle \ .
\end{multline}
Let further introduce (for general $z\in \mathbb{C}$) the
particle-type and hole-type GF by
\begin{subequations}
  \label{eq:phdef}
\bea
    \label{eqparticlegfdef1}
    G^\mathrm{(p)}(x_1,x_2;z) &=& \langle \hat\psi(x_1)
  \frac{1}{z - \hat{H}}\hat{\psi}^\dagger(x_2)  \rangle \ ,\\
     \label{eqholegfdef1}
    G^\mathrm{(h)}(x_1,x_2;z) &=& \langle \hat\psi^\dagger(x_2)
  \frac{1}{z - \hat{H} }\hat{\psi}(x_1) \rangle \ .
  \eea
\end{subequations}
From Eqs.~\eqref{eq:phdef} follows
\begin{align*}
  G^\mathrm{R/A}(x_1,x_2;\omega) =
  G^\mathrm{(p)}(x_1,x_2;E_0 + \omega \pm \iu \eta)\\-
  G^\mathrm{(h)}(x_1,x_2; E_0 - \omega \mp \iu \eta) \ .
\end{align*}
Finally, let us present the equation of motion (EOM) for the retarded GF in the form:
\begin{multline}
\label{eqretardedEOMph}
  (\omega + \iu \eta) G^\mathrm{R}(x_1,x_2; \omega)  =
  \delta(x_1-x_2) \\+\langle[\hat\psi(x_1),\hat{H}]
    \frac{1}{E_0+\omega - \hat{H} + \iu \eta} \hat\psi^\dagger(x_2)\rangle
  \\-\langle \hat\psi^\dagger(x_2)
    \frac{1}{E_0+\omega - \hat{H} + \iu \eta} [\hat\psi(x_2),\hat{H}]\rangle.
\end{multline}

The two-particle Green's functions are much more diverse. However, we will only need those
containing creation operators with the same time argument and the same holds for
annihilation operators. To specify the relative order of creation (or annihilation)
operators infinitesimally small times are added. Because such Green's functions depend on
two times only, the same nomenclature as in the single-particle case can be used. Thus, we
define
\begin{eqnarray*}
    G^\mathrm{(pp)}(x_1,x_2; \bar{x}_1,\bar{x}_2;
    z) = \langle \hat\psi(x_1)\hat\psi(x_2)
    \frac{1}{z-\hat{H}} \hat\psi^\dagger(\bar{x}_2)\hat\psi^\dagger(\bar{x}_1) \rangle \ ,\\
    G^\mathrm{(hh)}(x_1,x_2; \bar{x}_1,\bar{x}_2; z) =
    \langle\hat\psi^\dagger(\bar{x}_2)\hat\psi^\dagger(\bar{x}_1)
    \frac{1}{z-\hat{H}} \hat\psi(x_1)\hat\psi(x_2) \rangle \ .
  \end{eqnarray*}
They are the constituents of the retarded and advanced two-particle Green's functions:
\begin{align*}
  i G^\mathrm{R/A}(x_1,x_2; \bar{x}_1,\bar{x}_2;\omega) =
  G^\mathrm{(pp)}(x_1,x_2; \bar{ x}_1,\bar{x}_2;E_0 + \omega \pm \iu \eta)\\
-G^\mathrm{(hh)}(x_1,x_2; \bar{x}_1,\bar{x}_2;E_0 - \omega \mp \iu \eta) \ .
\end{align*}
For the retarded function the following equation of motion can be derived:
\begin{multline}
  \label{eqeomretarded2pgfsimpler}
  (\omega + \iu \eta)G^\mathrm{R}(x_1,x_2; \bar{x}_1,\bar{
    x}_2; \omega) \\= \delta(x_1-\bar{x}_1) G^<(x_2,\bar{x}_2,0)
  -\delta(x_1-\bar{x}_2) G^>(x_2,\bar{x}_1,0)\\
  +\delta(x_2-\bar{x}_2) G^>(x_1,\bar{x}_1,0)
  -\delta(x_2-\bar{x}_1) G^<(x_1,\bar{x}_2,0)\\
- i\langle \left[\hat\psi(x_1) \hat\psi(x_2),\hat{H}\right]
    \frac{1}{E_0+\omega - \hat{H} +\iu \eta} \hat\psi^\dagger(\bar{x}_2)\hat\psi^\dagger(\bar{x}_1)\rangle \\
    - i\langle\hat\psi^\dagger(\bar{x}_2) \hat\psi^\dagger(\bar{x}_1)\frac{1}{E_0-\omega - \hat{H} -\iu \eta}
     \left[\hat\psi(x_1) \hat\psi(x_2),\hat{H}\right]\rangle \ .
\end{multline}
\subsection{Two projection operators\label{subsec:projectionspedpe}}
In the previous section we have seen that relevant types of Green's functions can be
written in the form of a \emph{resolvent} $\langle (z-\hat H)^{-1}\rangle$, $z\in \mathbb
C$.  To be more specific about the state over which the averaging is performed we select
from all possible states of the target and emitted particles the relevant ones for the
effect of interest by employing projection operators. In the following we consistenly skip
$\hat{\cdots}$ when writing these operators and use $1$ to denote the identity
operator. Hence $P+Q=1$ are two complementary projection operators with the idempotence
($P^2=P$, $Q^2=Q$) as their defining property and the basis formula for computing
resolvents
\be
\begin{split}
P\frac{1}{z-\hat H}&=\frac{P}{z-\hat H_P-\hat \Sigma_P(z)}\\
&\quad\times\left[1+PHQ\frac{1}{z-\hat H_Q}\right],
\end{split}
\label{eq:res}
\ee
where $\hat H_P=P\hat HP$, $\hat H_Q=Q\hat HQ$, and the
self-energy operator is defined as:
\begin{equation}
\hat \Sigma_P(E)=P\hat HQ\frac{1}{E-\hat H_Q}Q\hat HP.
\label{eq:Sgm}
\end{equation}
The map $F_p:\, \hat H\rightarrow \hat H_p+\hat \Sigma_P(E)$ is called the Feshbach-Schur
map, it relates the eigenvalue problem on the full Hilbert space and to that on its
subspace. We summarize relevant matrix identities in Appendix~\ref{sec:matrix_id}. Due to
the presence of the bath Hamiltonian $\hat H_Q$ in Eq.~\eqref{eq:Sgm} this definition
cannot be used for practical computation of the self-energy. Fortunately, a connection
with the many-body perturbation theory (MBPT)
exists~\cite{domcke_theory_1991,capuzzi_projection_1996}. If, for example, starting from
the $N$-particle Schr{\"o}dinger equation $\hat{H}|\Psi_0\rangle=E_0|\Psi_0\rangle$ we use
a projector
\[
P=\hpsi^\dagger(\vec r)\big|\Psi_\alpha^+\big\rangle
\frac{1}{\bar n_\alpha(\vec r)}\big\langle\Psi_\alpha^+\big|\hpsi(\vec r),
\]
where $\bar n(\vec r)$ is the hole-density of ionized state $\alpha$, i.e. $\bar n(\vec
r)\equiv\big\langle\Psi_\alpha^+\big|\hpsi(\vec r)\hpsi^\dagger(\vec
r)\big|\Psi_\alpha^+\big\rangle$, the eigenvalue problem on the
$P$-subspace~\eqref{eq:Heff} ($\big\langle\Psi_\alpha^+\big|\psi(\vec r) (\hat H_P+\hat
\Sigma_P(E)-E\hat I_P)P\big|\Psi_0\big\rangle=0$) is the Lipmann-Schwinger equation for
the hole Dyson orbital~\eqref{eq:1h}. Notice that $\hat H_P$ contains the electrostatic and
exchange part of self-energy, whereas $\hat \Sigma_P(E)\rightarrow 0$ for
$E\rightarrow\pm\infty$. Similarly, in 1959 Bell and Squires~\cite{bell_formal_1959}
considered a one-body potential for the scattering of a particle incident on a complex
(many-body) target. They demonstrated that this \emph{optical potential} is exactly given
by the sum of all proper linked diagrams, i.e. many-body self-energy in the
\emph{time-ordered} formulation. In fact, their Eq.~(7) directly corresponds to
Eq.~\eqref{eq:Heff} when $P$ is a projection yielding a \emph{particle} Dyson orbital.

In order to study single and double photoemission we introduce two special projection
operators. The main goal of this section is to establish an equivalence between the
abstractly defined self-energy (Eq.~\eqref{eq:Sgm}) and the self-energy of the many-body
perturbation theory.  We consider the expression appearing in the first line of
Eq.~\eqref{eq:res} i.~e. resolvents of the type
\[
P\frac{1}{z-\hat H}P=P\frac{1}{z-\hat H_P-\hat \Sigma_P(z)}P.
\]
We will demonstrate that the formalism of nonequilibrium Green's functions is easily
paralleled with the Feshbach projection algebra (FPA). The basic relation for the
subsequent derivations are the operator identities
\begin{subequations}
\label{eq:mi}
\bea
  \label{eqmatrixiden}
  (\hat{A}-\hat{B})^{-1} &=& \hat{A}^{-1}+\hat{A}^{-1}\hat{B} (\hat
  A-\hat B)^{-1} \ , \\
  (\hat A-\hat B)^{-1} &=& \hat A^{-1}+(\hat A-\hat B)^{-1} \hat B\hat
  A^{-1} \ .
\eea
\end{subequations}
We will show below that with
\begin{subequations}
  \label{eqchoiceAB1}
  \bea
  \hat A &=& z - P \hat{H} P \equiv z-\hat{H}_P, \\
  \hat B &=& Q\hat{H} P+P \hat{H} Q + Q \hat{H} Q \ ,
  \eea
\end{subequations}
the operator identity~\eqref{eq:mi} has a structure of the Dyson equation for certain
Green's functions.

For SPE we consider the projection operator
\begin{equation}
  \label{eqproj1}
  P_\alpha = \sum_{\vec k} c^\dagger_\vec{k} |\Psi^{+}_\alpha \rangle \langle \Psi^{+}_\alpha | c_\vec{k} ,
\end{equation}
where the sum runs over scattering states. It is common to select these single-particle
states $|\varphi_\vec{k}\rangle$ to be eigenfunctions of some reference Hamiltonian with
proper boundary conditions. We request that $|\Psi^{+}_\alpha \rangle$ is a completely
bound remainder of the ionization event and does not emit a second electron at a later
stage (Auger electrons are a typical example for these kind of processes). There are many
equivalent ways to impose this restriction, for instance we will assume
\begin{equation}
  \label{eqckzero}
  c_\vec{k} |\Psi^{+}_\alpha \rangle = 0,
\end{equation}
i.e., implying $|\Psi^{+}_\alpha \rangle$ is a vacuum state for photoelectrons. From the
assumption follows the indempotency ($P_\alpha^2=P_\alpha$, see Appendix~\ref{sec:idem}
for proof) and, thus, $P_\alpha$ represents a true projection operator. The application of
$P_\alpha$ restricts the possible processes which might occur upon excitation to the
definite emission of \emph{one} photoelectron, whereas the ionized system is left in a
(possibly excited) bound state $|\Psi^{+}_\alpha \rangle$. From the assumption
Eq.~\eqref{eqckzero} follows another restriction:
\begin{eqnarray}
\lim_{r\rightarrow\infty}\hat \psi(x,t)|\Psi^{+}_\alpha \rangle
&=&\lim_{r\rightarrow\infty}\sum_i\langle x|i\rangle c_i(t)|\Psi^{+}_\alpha\rangle \nonumber\\
&+&\lim_{r\rightarrow\infty}\sum_{\vec k}\langle x |\vec k\rangle c_\vec{k}(t)|\Psi^{+}_\alpha\rangle =0,
\label{eqckzero2}
\end{eqnarray}
where the first term is equal to zero because each bound state ($i$) is necessarily given
by a square integrable function (converse is not true). In the following we will use
another consequence of assumptions Eqs.~(\ref{eqckzero},~\ref{eqckzero2}):
\bea
\label{eq:restriction}
G^<_{\vec{k}a}(\omega)=0,&\quad& G^<_{a\vec{k}}(\omega)=0,\\
\lim_{r_1\rightarrow \infty}G^<(x_1 t_1,x_2 t_2)
&=&\lim_{r_1\rightarrow \infty}G^<(x_2 t_2,x_1 t_1)=0.
\eea

 The projection operator for DPE we define as
\begin{equation}
  \label{eqprojDPE}
  P_\beta = \frac12 \sum_{\vec p \vec p^\prime} c^\dagger_{\vec p} c^\dagger_{\vec
    p^\prime} |\Psi^{2+}_\beta \rangle \langle \Psi^{2+}_\beta | c_{\vec p^\prime}
  c_{\vec p} \ .
\end{equation}
Here, $|\Psi^{2+}_\beta \rangle $ is the doubly-ionized reference state, to which two
photoelectrons with continuum quantum numbers $\vec p$ and $\vec p^\prime$ are added. We
can easily show the indempotency of the projection operator~\eqref{eqprojDPE} if we
require, similar to Eq.~\eqref{eqckzero},
\begin{equation}
  \label{eqckzeroDPE}
  c_\vec{p} |\Psi^{2+}_\beta \rangle = 0 \ .
\end{equation}
\subsection{Example of SPE}
\paragraph{Equation of motion (EOM):}
As a starting point let us use the following operator identity which can be derived from
Eq.~\eqref{eqmatrixiden} or verified by direct computation
%
\begin{align*}
  (z - E^+_\alpha) P_\alpha \frac{1}{z-\hat{H}} P_\alpha =  P_\alpha
+ P_\alpha(\hat{H}-E^+_\alpha)\frac{1}{z-\hat{H}}P _\alpha\ .
\end{align*}
With the definition of the SPE projection operator
$P_\alpha$ in Eq.~\eqref{eqproj1}, we find
\begin{align*}
  P_\alpha \frac{1}{z-\hat{H}} P_\alpha &= \sum_{\vec p \vec q}
  c^\dagger_{\vec p} |\Psi^+_\alpha \rangle \langle \Psi^+_\alpha |
  c_{\vec p}\frac{1}{z-\hat{H}}  c^\dagger_{\vec q}  |\Psi^+_\alpha \rangle
  \langle \Psi^+_\alpha | c_{\vec q} \\
 &= \sum_{\vec p \vec q}
  c^\dagger_{\vec p} |\Psi^+_\alpha \rangle G^\mathrm{(p)}_{\vec p
    \vec q}(z) \langle \Psi^+_\alpha | c_{\vec q}\ ,
\end{align*}
where we applied the definition of the particle-type GF Eq.~\eqref{eqparticlegfdef1}. Note
that the GF is defined for a particular subspace spanned by the operator $P_\alpha$ and
should therefore always be understood as the GF associated to $|\Psi^+_\alpha \rangle
$. For brevity, however, we omit labelling GF by $\alpha$.

Using these notations the operator identity reads
%
\begin{multline*}
  (z - E^+_\alpha) \sum_{\vec p \vec q}
  c^\dagger_{\vec p} |\Psi^+_\alpha \rangle G^\mathrm{(p)}_{\vec p \vec q}(z) \langle \Psi^+_\alpha | c_{\vec q} =
  \sum_{\vec k} c^\dagger_\vec{k} |\Psi^{+}_\alpha \rangle \langle \Psi^{+}_\alpha | c_\vec{k} \\
  + \sum_{\vec p \vec q} c^\dagger_{\vec p} |\Psi^+_\alpha \rangle
  \langle \Psi^+_\alpha |
  c_{\vec p}(H-E^+_\alpha)\frac{1}{z-\hat{H}}  c^\dagger_{\vec q}  |\Psi^+_\alpha \rangle
  \langle \Psi^+_\alpha | c_{\vec q}  \ .
\end{multline*}
%
With the help of our assumption Eq.~\eqref{eqckzero} we can now remove the sum by applying
$\langle \Psi^+_\alpha | c_{\vec p^\prime}$ from the left and $c_{\vec q^\prime}^\dagger|
\Psi^+_\alpha \rangle $ from the right as Eq.~\eqref{eqckzero} implies $\langle
\Psi^+_\alpha | c_{\vec p^\prime} c^\dagger_{\vec p} | \Psi^+_\alpha \rangle = \delta_{\vec p \vec
  p^\prime}$. Furthermore, we note that $\langle \Psi^+_\alpha | c_{\vec p}(\hat{H}-E^+_\alpha)
= \langle \Psi^+_\alpha | [c_{\vec p},\hat{H}]$ because of $\hat{H} |\Psi^+_\alpha \rangle
=E^+_\alpha |\Psi^+_\alpha \rangle $. Hence, we obtain
%
\begin{equation}
  \label{eqFPAEOM1}
  (z-E^+_\alpha) G^\mathrm{(p)}_{\vec p \vec q}(z) = \delta_{\vec p
    \vec q} + \langle \Psi^+_\alpha |
\left[c_{\vec p},\hat{H}\right]\frac{1}{z-\hat{H}} c^\dagger_{\vec q} |
\Psi^+_\alpha \rangle \ .
\end{equation}
%
As stated above, we can think of $|\Psi^+_\alpha \rangle $ as a vacuum state for free
particles (cf. Eq.~\eqref{eqckzero}). The hole-type GF is identically zero. Therefore,
\[
G^\mathrm{(p)}_{\vec p \vec q}(E^+_\alpha +\omega + \iu \eta) =G^\mathrm{R}_{\vec p \vec q}(\omega),
\]
Substituting $z=E^+_\alpha + \omega + \iu \eta$ in Eq.~\eqref{eqFPAEOM1} we realize its
equivalence to Eq.~\eqref{eqretardedEOMph}. In other words, by applying the FPA we can
derive EOM for the retarded Green's function.

\paragraph{Effective Hamiltonian:}
In Eq.~\eqref{eqmatrixiden} $ \hat A^{-1}$ plays the role of the reference Green's
function. Correspondingly, $P\hat{H}P$ is the effective Hamiltonian.  Using the standard
anti-commutation algebra and the assumption~\eqref{eqckzero}, we find
%
\begin{multline}
  \label{eqeffsingleparticleham}
  \langle \Psi^+_\alpha | c_{\vec p} \hat{H}
  c^\dagger_{\vec q} | \Psi^+_\alpha \rangle =
  E^+_\alpha \delta_{\vec p \vec q} +\langle \Psi^+_\alpha |
  \left[ c_{\vec p}, \hat{H} \right]c^\dagger_{\vec q} | \Psi^+_\alpha \rangle \\
  = E^+_\alpha \delta_{\vec p \vec q}+t_{\vec p \vec q} + \sum_{nm}\left(\sv_{\vec p n m \vec q}
    -\sv_{n \vec p  m \vec q}  \right)\langle \Psi^+_\alpha | c^\dagger_n c_m | \Psi^+_\alpha \rangle\\
  = E^+_\alpha \delta_{\vec p \vec q} +\tilde t_{\vec p \vec q},
\end{multline}
i.e. it consists of the total energy of the ionized system and the Hartree-Fock
Hamiltonian for continuum states. The latter is computed with the density matrix of the
target:
%
\be
\tilde t_{\vec p \vec q}
=t_{\vec p \vec q}+\sum_{(nm)\in \mathcal{B}^2} \sV_{\vec p n \vec q m}
\langle c^\dagger_n c_m \rangle.\label{eq:hf}
\ee
%
Let $\hat{h}$ be an operator acting on the subspace of continuum states with matrix elements
given by Eq.~\eqref{eqeffsingleparticleham}. Its resolvent
%
\begin{equation}
  g^{\mathrm{(p)}}_{\vec p \vec q}(z) =  \langle \Psi^+_\alpha |
  c_{\vec p} \frac{1}{z - \hat{h}} c^\dagger_{\vec q} |
  \Psi^+_\alpha \rangle
  \label{eq:gp0}
\end{equation}
%
relates to the reference retarded GF as $g^{\mathrm{R}}_{\vec p \vec q}(\omega) =
g^{\mathrm{(p)}}_{\vec p \vec q}(E^+_\alpha + \omega + \iu \eta) $.

\paragraph{Self-energy and the Dyson equation:}
The second correlator in the EOM~\eqref{eqFPAEOM1} amounts to
%
\begin{multline*}
  \langle \Psi^+_\alpha |  [c_\vec{p},\hat{H}]
    \frac{1}{z-\hat{H}} c^\dagger_{\vec q} |\Psi^+_\alpha \rangle = \sum_{a} t_{\vec p a}
    \langle \Psi^+_\alpha | c_a  \frac{1}{z-\hat{H}} c^\dagger_{\vec
      q} |\Psi^+_\alpha \rangle \\ \quad + \sum_{n} \sum_{ab} \sv_{\vec p n a b}
    \langle \Psi^+_\alpha | c^\dagger_n c_a c_b
    \frac{1}{z-\hat{H}} c^\dagger_{\vec q}| \Psi^+_\alpha \rangle \ .
\end{multline*}
%
With Eq.~\eqref{eqchoiceAB1} inserted into the identity Eq.~\eqref{eqmatrixiden} we apply
$P_\alpha$ from left and right, use the same trick to multiply with suitable states from
left and right, and find
%
\begin{align}
\label{eq:de_spe_pf}
  G^\mathrm{(p)}_{\vec p \vec q}(z) &= g^\mathrm{(p)}_{\vec p \vec q}(z)
  - \sum_{\vec k \vec k^\prime} g^\mathrm{(p)}_{\vec p \vec k}(z)\tilde t_{\vec k \vec k^\prime}G^\mathrm{(p)}_{\vec k^\prime \vec q}(z)\nonumber\\
   +& \sum_{\vec k }\sum_{a} g^\mathrm{(p)}_{\vec p \vec k}(z) t_{\vec k a} G^\mathrm{(p)}_{a \vec q}(z)\\
    +&\sum_{\vec k}\sum_{n} \sum_{ab} g^\mathrm{(p)}_{\vec p \vec k}(z) \sv_{\vec k n a b}
 \langle \Psi^+_\alpha | c^\dagger_n c_a c_b \frac{1}{z-\hat{H}} c^\dagger_{\vec q}| \Psi^+_\alpha \rangle \ .\nonumber
\end{align}
%
With $z=E^+_\alpha + \omega + \iu \eta$ Eq.~\eqref{eq:de_spe_pf} has a structure of a
Dyson equation for the retarded Green's function in the subspace of continuum states:
%
\be
G^\mathrm{R}_{\vec p \vec q}(\omega)=g^{\mathrm{R}}_{\vec p \vec q}(\omega)+ \sum_{\vec k
  a} g^{\mathrm{R}}_{\vec p \vec k}(\omega)\Sigma^\mathrm{R}_{\vec k a}(\omega)
G^\mathrm{R}_{a \vec q}(\omega).
\label{eq:de_spe_gf}
\ee
%
The second sum runs over the full set of orbitals (bound and continuum). This is the most
general form and without additional analysis it cannot be reduced to the Dyson equation
with the self-energy from the projection formalism (cf.~Eq.~\eqref{eq:Sgm}). Let us
compare Eq.~\eqref{eq:de_spe_pf} and Eq.~\eqref{eq:de_spe_gf}. At first we notice that
Eq.~\eqref{eq:hf} defines the reference Hamiltonian only on the subspace of scattering
states. We might extend the definition and request, for instance, that all the basis
functions (bound and scattering) are the eigenstates of the reference Hamiltonian. This
implies $\tilde t_{\vec p\vec q}=\varepsilon_{\vec{p}}\delta_{\vec{p}\vec{q}}$ and $\tilde
t_{n \vec q}=0$. Thus, mean-field terms of the Hartree-Fock Hamiltonian are then cancelled
by the frequency independent part of the last correlator in Eq.~\eqref{eq:de_spe_pf}. In
the case when the reference Hamiltonian is not diagonal in the chosen basis the embedding
self-energy terms additionally appear. In the simplest case (no interaction), they can be
written as $\Sigma^{\mathrm{em}}_{\vec p\vec q}(z)=\sum_{mn}t_{\vec pn}g_{nm}^{\text{(p)}}(z)t_{n\vec
  q}$.  Let us now assume that the single-particle basis is such that no embedding
self-energy appear. What would be the diagrammatic structure of the
self-energy~\eqref{eq:Sgm}? From the Dyson equation in the bound-continuum sector
%
\be
G^\mathrm{R}_{l \vec q}(\omega)=\sum_{m\vec{k}} g^\mathrm{R}_{l
  m}(\omega)\Sigma^\mathrm{R}_{m \vec k}(\omega) G^\mathrm{R}_{\vec k \vec
  q}(\omega)+\sum_{mn} g^\mathrm{R}_{l m}(\omega)\Sigma^\mathrm{R}_{m n}(\omega)
G^\mathrm{R}_{n \vec q}(\omega), \label{eq:bc}
\ee
%
we determine the Green's function in
this sector ($G_{bc}$) and substitute in Eq.~\eqref{eq:de_spe_gf}:
%
\bea G^\mathrm{R}_{\vec
  p \vec q}(\omega)&=&g^{\mathrm{R}}_{\vec p \vec q}(\omega)+ \sum_{\vec k
  \vec{k}^\prime}g^{\mathrm{R}}_{\vec p \vec
  k}(\omega)\nonumber\\ &\quad\times&\left[\Sigma_{cc}+\Sigma_{cb}\frac{g_b}{1-g_b\Sigma_{bb}}\Sigma_{bc}\right]_{\vec
  k \vec{k}^\prime} G^\mathrm{R}_{\vec{k}^\prime\vec{q}},
\label{eq:de_cc_gf}
\eea
%
where for brevity the subscripts $b$ and $c$ denote the bound and the continuum sectors.
Expression in square brakets (Eq.~\eqref{eq:de_cc_gf}) can now be compared with the
self-energy from the projection formalism~\eqref{eq:Sgm}. Notice, that the reference
Green's function was assumed to be diagonal, i.e. $g_b\equiv g_{bb}$ and $g_{bc}=0$.

\paragraph{Dominant scattering mechanisms:}
\begin{figure}[t!]
\centering
       \includegraphics[width=0.99\columnwidth]{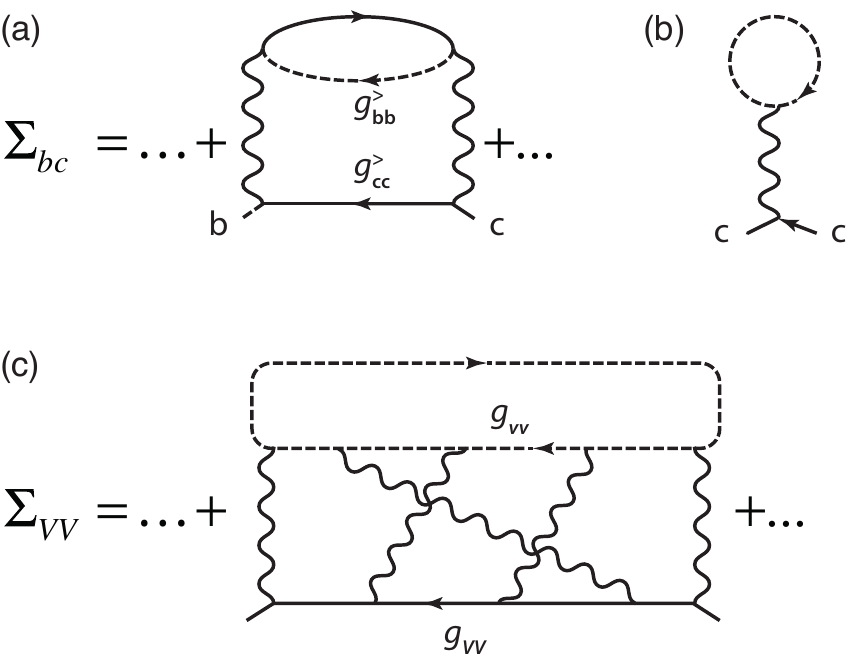}
       \caption[]{(a) Example of self-energy diagram that mixes bound and continuum states
         and is the building block of the second term in brackets in
         Eq.~\eqref{eq:de_cc_gf}; (b) Mean-field Hartree contribution to the effective
         Hamiltonian Eq.~\eqref{eqeffsingleparticleham}; (c) A typical contribution to the
         electron self-energy in continuum-continuum sector in the case when the
         photoelectron is completely screened in the sample.
      \label{fig:diagrams}}
\end{figure}
Let us recapitulate what led us to Eq.~\eqref{eq:de_cc_gf}. We have chosen a projection
operator in the form~\eqref{eqproj1}. This specifies the state of a system after the
photoionization as containing one photoelectron in the scattering state plus the bound
ionized target. Next, we obtained an effective Hamiltonian~\eqref{eqeffsingleparticleham}
acting on the $P$ subspace and used it to define the reference Green's
function~\eqref{eq:gp0}. We want to understand what is the diagrammatic content of the
Feshbach self-energy~\eqref{eq:Sgm}. It is not possible to use this equation directly
because it involves the effective Hamiltonian on the complementary $Q$-subspace. However,
it is possible to use another matrix identity~\eqref{eqmatrixiden} and to formulate the
Dyson equation for the full Green's function in the $P$ subspace~\eqref{eq:de_spe_pf}
avoiding the use of the $Q\hat{H}Q$ resolvent. This equation can be put in a direct
correspondence with the Dyson equation for the retarded GF from the many-body perturbation
theory. The difference between them is the domain where the self-energies are defined: the
Feshbach self-energy operates on the continuum sector only, whereas many-body perturbation
theory does not impose such a restriction. By writing another Dyson equation~\eqref{eq:bc}
in the bound-continuum sector we can finally obtain the Dyson equation with an effective
self-energy in the continuum-continuum sector. This self-energy is an exact counterpart of
the Feshbach self-energy~\eqref{eq:Sgm}. To the best of our knowledge it is the first
explicit example of such correspondence. Critical for our derivation was the choice of the
single-particle basis. We have demonstrated that it is the projection operator that
determines the effective Hamiltonian, and if the basis is such that the Hamiltonian is
diagonal the embedding self-energy vanishes and one arrives at Eq.~\eqref{eq:de_cc_gf}. No
further assumptions have been made and Eq.~\eqref{eq:de_cc_gf} is so far exact.

Let us analyze the meaning of different terms of the photoelectron self-energy
(Fig.\,\ref{fig:diagrams}). As discussed in details by Bardyszewski and
Hedin~\cite{bardyszewski_new_1985}, Almbladh~\cite{almbladh_theory_1985} and Fujikawa and
Hedin~\cite{fujikawa_theory_1989} scattering states vanish in the sample (damped)
represent the real photoelectron states more precisely. One can derive explicitly the
residual interaction that they experience. The reasoning is easier to perform in real
space where the Coulomb interaction depends on two coordinates only
(cf. Eq.~\eqref{eq:sv}) as opposite to the Coulomb matrix elements which are four index
quantities. Since the scattering states are damped in the sample, there are only two
nonvanishing Green's functions $G_{vv}$ and $G_{VV}$ operating exclusively in the inner
($v$), outer ($V$) spaces, respectively. The Green's function starting in the sample and
ending outside of it ($G_{Vv}$) and the reverse ($G_{vV}$) vanish. We can rewrite
Eq.~\eqref{eq:de_cc_gf} in these new notations, however, it is not even necessary as it
amounts to mere replacement $b\rightarrow v$ and $c\rightarrow V$. What has changed is the
interaction lines in the diagrammatic expansion of the self-energy. They can connect $v$
and $V$ domains and generate therefore nonzero contributions. It is easy to see, however,
that the second self-energy term vanishes: a diagrammatic expansion of $\Sigma_{vV}$
necessarily contains at least one $g_{vV}$ line which is zero according to our
assumption. Thus, only $\Sigma_{VV}$ needs to be analyzed. By explicitly forbidding the
particle exchange with the sample we arrived exactly at the case of \emph{elastic}
electron scattering considered in the seminal paper of Bell and
Squires~\cite{bell_formal_1959}. We will see below that the structure of $\Sigma_{VV}$ is
quite general and appears in the diagrammatic consideration of other processes,
remarkably, in the parquet diagram treatment of the Fermi edge
singularities~\cite{roulet_singularities_1969}. There, however, a similar diagrammatic
expansion arises due to the specific choice of the interaction between the deep hole
(labeled by $m$) and the conduction electrons:
$\hat{H}_1=\sum_{\vec{k}\vec{k}^\prime}V_{\vec{k}\vec{k}^\prime}c_{\vec{k}}^\dagger
c_{\vec{k}^\prime}c_mc_m^\dagger$.  In contrast to their work, what induces a special
structure of diagrams for $\Sigma_{VV}$ is not a specific form of the interaction matrix
elements, but rather the absence of the off-diagonal blocks in $g$. It is easy to
construct the electron self-energy fulfilling these restrictions: it consists of
\emph{one} open photoelectron line (depicted as solid line on Fig.\,\ref{fig:diagrams} and
a number of closed bound electron loops (depicted as dashed lines). Because of the
restriction~\eqref{eq:restriction} there are no photoelectron loops.

The topic of the present section is quite extensive and such an aspect as the Lehmann
representation of the Green's functions mentioned here was completely left out of our
discussion. This is, however, very relevant for the treatment of finite systems, with
important recent progress, e.g.,~\cite{feuerbacher_direct_2005}.
\subsection{Example of DPE}
\begin{figure}[t!]
\centering
       \includegraphics[width=0.99\columnwidth]{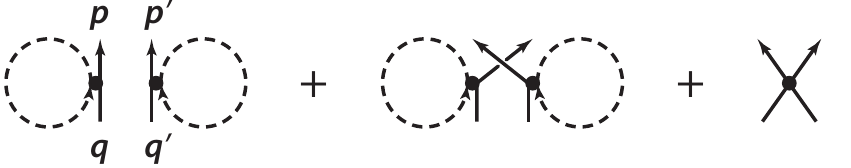}
       \caption[]{Interaction between the photoelectrons incorporated in the effective
         Hamiltonian Eq.~\eqref{eq:h2new}. Dashed lines denote bare bound state
         propagators. Dots denote the anti-symmetrized Coulomb interaction~\eqref{eq:sv}.
      \label{fig:diagrams2}}
\end{figure}

\paragraph{Equation of motion:}
The derivation for the two-particle case goes along the same lines. We insert the
definition of the projection operator (Eq.~\ref{eqprojDPE}) in the identity
\begin{align*}
  (z - E^{2+}_\beta) P_\beta \frac{1}{z-\hat{H}} P_\beta
= P_\beta + P_\beta(\hat{H}-E^{2+}_\beta)\frac{1}{z-\hat{H}}P _\beta\ ,
\end{align*}
replace $\langle \Psi^{2+}_\beta | c_{\vec p^\prime} c_{\vec p} (\hat{H}-E^{2+}_\beta)
=\langle \Psi^{2+}_\beta |[ c_{\vec p^\prime} c_{\vec p},\hat{H}]$, and as for SPE compute
the matrix elements of the whole expression. The final results read as
\begin{multline}
  \label{eqFPAEOMDPE1}
  (z - E^{2+}_\beta)G^\mathrm{(pp)}_{\vec p\vec p^\prime  \vec
   q\vec q^\prime }(z) = \delta_{\vec p \vec q}\delta_{\vec p^\prime
   \vec q^\prime}-\delta_{\vec p \vec q^\prime}\delta_{\vec p^\prime \vec q} \\
  + \langle\Psi^{2+}_\beta |\left[ c_{\vec p}
 c_{\vec p^\prime} , H \right]\frac{1}{z-\hat{H}}  c^\dagger_{\vec q^\prime}
 c^\dagger_{\vec q}  |\Psi^{2+}_\beta \rangle \ .
\end{multline}
The prefactor $1/4$ originating from the product of two projection operators is cancelled
because of the symmetries of the particle-particle GF and of the second term on the
right-hand side of Eq.~\eqref{eqFPAEOMDPE1}:
\begin{equation}
  \label{eqppgfsymmetries}
  G^\mathrm{(pp)}_{\vec p^\prime \vec p \vec
   q^\prime \vec q}(z) = G^\mathrm{(pp)}_{\vec p\vec p^\prime  \vec
   q\vec q^\prime }(z) = - G^\mathrm{(pp)}_{\vec p^\prime \vec p \vec
   q \vec q^\prime}(z) = - G^\mathrm{(pp)}_{\vec p \vec p^\prime \vec
   q^\prime \vec q}(z) \ .
\end{equation}
Inserting $z = E^{2+}_\beta + \omega + \iu \eta$ shows the equivalence of
Eq.~\eqref{eqFPAEOMDPE1} to the equation of motion~\eqref{eqeomretarded2pgfsimpler}.
\paragraph{Effective two-particle Hamiltonian:}
Analogically to the SPE case we consider the Feshbach-projected Hamiltonian in the
subspace defined by $P_\beta$ and describing two electrons including their interaction and
their mean-field interaction with the ionized system:
\begin{widetext}
%
\be
  \langle \Psi^{2+}_\beta | c_{\vec p^\prime} c_{\vec p} \hat{H}
  c^\dagger_{\vec q} c^\dagger_{\vec q^\prime} | \Psi^{2+}_\beta
  \rangle = E^{2+}_\beta \left( \delta_{\vec p \vec q}
    \delta_{\vec p^\prime \vec q^\prime}-\delta_{\vec p \vec q^\prime}
    \delta_{\vec p^\prime \vec q} \right) + \langle \Psi^{2+}_\beta |
  \left[c_{\vec p^\prime} c_{\vec p} ,H\right] \,
  c^\dagger_{\vec q} c^\dagger_{\vec q^\prime} | \Psi^{2+}_\beta
  \rangle  \ ,\label{eq:h2}
\ee
%
where the last term can be expressed as follows
\begin{align}
  \langle \Psi^{2+}_\beta |
  \left[c_{\vec p^\prime} c_{\vec p} ,H\right] \,
  c^\dagger_{\vec q} c^\dagger_{\vec q^\prime} | \Psi^{2+}_\beta
  \rangle &= t_{\vec p \vec q} \delta_{\vec p^\prime \vec q^\prime} +
  t_{\vec p^\prime \vec q^\prime} \delta_{\vec p \vec q} - t_{\vec p
    \vec q^\prime} \delta_{\vec p^\prime \vec q} - t_{\vec p^\prime
    \vec q} \delta_{\vec p \vec q^\prime} + \sv_{\vec p \vec p^\prime \vec q \vec q^\prime} -
  \sv_{\vec p \vec p^\prime \vec q^\prime \vec q}\nonumber\\ &+
  \sum_{n} \sum_{ab} \left[
  \sv_{\vec p n a b}\langle
  \Psi^{2+}_\beta | c^\dagger_n c_{\vec p^\prime} c_a c_b c^\dagger_{\vec q}
  c^\dagger_{\vec q^\prime} | \Psi^{2+}_\beta
  \rangle  -
  \sv_{\vec p^\prime n a b} \langle
  \Psi^{2+}_\beta | c^\dagger_n c_{\vec p} c_a c_b c^\dagger_{\vec q}
  c^\dagger_{\vec q^\prime} | \Psi^{2+}_\beta\rangle  \right] \ .
\label{eq:k1s}
\end{align}
\end{widetext}
The first correlator in the square brackets evaluates in terms of the density matrix with
respect to $|\Psi^{2+}_\beta \rangle$ with bound state indices to:
\[
 \sum_{nm} \big[ \sV_{\vec p n \vec q m}\delta_{\vec p^\prime \vec q^\prime} - \sV_{\vec p
     n \vec q^\prime m}\delta_{\vec p^\prime \vec q} \big] \langle c^\dagger_n c_m \rangle\ .
\]
Here we have written it in terms of the matrix elements of the anti-symmetrized Coulomb
interaction~\eqref{eq:sv} $\sV_{abcd}\equiv\sv_{abcd}-\sv_{abdc}$. Similarly, the
second correlator is obtained from this expression by the index exchange $\vec
p\leftrightarrow \vec p^\prime$. The effective two-particle Hamiltonian~\eqref{eq:h2} is
so expressible as a Hartree-Fock Hamiltonian~\eqref{eq:hf} for two independent electrons
plus the interaction (Fig.\,\ref{fig:diagrams2}):
%
\bea
h_{\vec p^\prime\vec p\vec q^\prime \vec q}&=&E^{2+}_\beta
\left( \delta_{\vec p \vec q}\delta_{\vec p^\prime \vec q^\prime}
-\delta_{\vec p \vec q^\prime}\delta_{\vec p^\prime \vec q} \right)
 + \left(\tilde t_{\vec p \vec q} \delta_{\vec p^\prime \vec q^\prime}
 +\tilde t_{\vec p^\prime \vec q^\prime}\delta_{\vec p \vec q}\right) \nonumber\\
&-& \left(\tilde t_{\vec p \vec q^\prime} \delta_{\vec p^\prime \vec q}
 +\tilde t_{\vec p^\prime \vec q} \delta_{\vec p \vec q^\prime} \right)
 + \sV_{\vec p^\prime  \vec p\vec q^\prime \vec q}.
\label{eq:h2new}
\eea
%
\paragraph{Kernel and Dyson equation:}
We return to the matrix identity~\eqref{eqmatrixiden} and insert the
splitting~\eqref{eqchoiceAB1} with $P = P_\beta$ (Eq.~\eqref{eqprojDPE}):
\begin{align*}
  P_\beta \frac{1}{z-\hat{H}} P_\beta = P_\beta \frac{1}{z-\hat{h}}
  P_\beta +  P_\beta \frac{1}{z-\hat{h}}
  P_\beta \hat{H} \frac{1}{z-\hat{H}} P_\beta \\-P_\beta \frac{1}{z-\hat{h}}
  P_\beta \hat{H} P_\beta\frac{1}{z-\hat{H}} P_\beta \ ,
\end{align*}
and define the reference two-particle GF
\begin{equation}
  \label{eqrefgfDPE1}
  g^{\mathrm{(pp)}}_{\vec p \vec p^\prime \vec q \vec q^\prime}(z) =
  \langle \Psi^{2+}_\beta | c_{\vec p} c_{\vec p^\prime}
  \frac{1}{z-\hat{h}} c^\dagger_{\vec q^\prime} c^\dagger_{\vec q}
  | \Psi^{2+}_\beta \rangle \ ,
\end{equation}
Invoking again the symmetries~\eqref{eqppgfsymmetries}, which also hold for the reference
GF, and applying the same states from left and right, we obtain
\begin{multline}
  \label{eqppgfdysoneq1}
  G^\mathrm{(pp)}_{\vec p \vec p^\prime \vec q \vec q^\prime}(z)
  = g^{\mathrm{(pp)}}_{\vec p \vec p^\prime \vec q \vec q^\prime}(z) +
  \sum_{\vec k \vec k^\prime} g^{\mathrm{(pp)}}_{\vec p \vec p^\prime \vec k \vec k^\prime}(z)\\
   \times\Big[ \langle \Psi^{2+}_\beta | \left[ c_{\vec k} c_{\vec k^\prime}, \hat{H} \right] \frac{1}{z-\hat{H}}
    c^\dagger_{\vec q^\prime} c^\dagger_{\vec q} | \Psi^{2+}_\beta \rangle \\
     -\frac12\sum_{\vec n \vec n^\prime}
    \langle \Psi^{2+}_\beta | \left[ c_{\vec k} c_{\vec k^\prime}, \hat{H} \right]
    c^\dagger_{\vec n^\prime} c^\dagger_{\vec n}  | \Psi^{2+}_\beta \rangle
    G^\mathrm{(pp)}_{\vec n \vec n^\prime\vec q \vec q^\prime}(z) \Big]\ .
\end{multline}
It is instructive to divide the kernel entering the equation of motion (second line of
Eq.~\eqref{eqFPAEOMDPE1}) or the Dyson equation (second line of
Eq.~\eqref{eqppgfdysoneq1}) into the terms containing higher correlation functions and those
expressible in terms of two-particle GFs:
\begin{multline*}
\langle\Psi^{2+}_\beta | \left[c_{\vec k} c_{\vec k^\prime},\hat{H} \right]
  \frac{1}{z-\hat{H}} c^\dagger_{\vec q^\prime}  c^\dagger_{\vec q} | \Psi^{2+}_\beta \rangle
=T_{\vec k \vec k^\prime \vec q \vec q^\prime}(z)\\
+\sum_{b} \left(t_{\vec k^\prime b} G^\mathrm{(pp)}_{\vec k b \vec q \vec q^\prime}(z)
  - t_{\vec k b} G^\mathrm{(pp)}_{\vec k^\prime b \vec q \vec q^\prime}(z)\right)
  +\sum_{ab}\sv_{\vec k \vec k^\prime a b} G^\mathrm{(pp)}_{ a b \vec q \vec q^\prime}(z).
\end{multline*}
The latter gives rise to the particle-particle embedding self-energy. We can now formally
introduce the correlated frequency-dependent and the static kernels:
\begin{multline}
  T_{\vec k \vec k^\prime \vec q \vec q^\prime}(z)=\sum_{\vec n \vec n^\prime} \Big[
   \mathcal{K}^{\mathrm{c}}_{\vec k \vec k^\prime \vec n \vec n^\prime}(z)
   +\frac12\mathcal{K}^{\infty}_{\vec k \vec k^\prime \vec n \vec n^\prime}
  \Big]G^\mathrm{(pp)}_{\vec n \vec n^\prime \vec q \vec q^\prime}(z)\\=
  \sum_{n} \sum_{ab} \sv_{\vec k n a b} \langle\Psi^{2+}_\beta |
  c^\dagger_n c_{\vec k^\prime} c_a c_b \frac{1}{z-\hat{H}} c^\dagger_{\vec q^\prime}
  c^\dagger_{\vec q} | \Psi^{2+}_\beta \rangle \\
  -\sum_{n} \sum_{ab} \sv_{\vec k^\prime n a b} \langle\Psi^{2+}_\beta |
  c^\dagger_n c_{\vec k} c_a c_b \frac{1}{z-\hat{H}} c^\dagger_{\vec q^\prime}
  c^\dagger_{\vec q} | \Psi^{2+}_\beta \rangle \ .
\label{eq:k1v}
\end{multline}
The static part is exactly cancelled by the density-dependent part of the effective
Hamiltonian:
\begin{multline}
\mathcal{K}^{\infty}_{\vec k \vec k^\prime\vec q \vec q^\prime }=
\sum_{nm} \langle c^\dagger_n c_m \rangle
\Big[
\sV_{\vec k n \vec q m}\delta_{\vec k^\prime \vec q^\prime}
+\sV_{\vec k^\prime n \vec q^\prime m}\delta_{\vec k\vec q}\\
-\sV_{\vec k n \vec q^\prime m}\delta_{\vec k^\prime \vec q}
-\sV_{\vec k^\prime n \vec q m}\delta_{\vec k\vec q^\prime}
\Big].
\label{eq:kinf}
\end{multline}
The embedding self-energy originates from the kernel as well as from the effective
Hamiltonian~\eqref{eq:h2new}:
\begin{multline}
\sum_{\vec n \vec n^\prime} \mathcal{K}^{\mathrm{em}}_{\vec k \vec k^\prime\vec n \vec n^\prime }
  G^\mathrm{(pp)}_{\vec n \vec n^\prime \vec q \vec q^\prime}(z) \\
=\sum_{m} \left(\tilde{t}_{\vec k^\prime m} G^\mathrm{(pp)}_{\vec k m \vec q \vec q^\prime}(z)
 -\tilde{t}_{\vec k m} G^\mathrm{(pp)}_{\vec k^\prime m \vec q \vec q^\prime}(z) \right)\\
+\sum_{ab} \sv_{\vec k \vec k^\prime a b}
  G^\mathrm{(pp)}_{ a b \vec q \vec q^\prime}(z)-\sum_{\vec p\vec p^\prime} \sv_{\vec k \vec k^\prime \vec p\vec p^\prime}
  G^\mathrm{(pp)}_{ \vec p\vec p^\prime \vec q \vec q^\prime}(z) .
\label{eq:kem}
\end{multline}

With the results~(\ref{eq:h2new},~\ref{eq:k1v},~\ref{eq:kinf},~\ref{eq:kem}) we can cast
the Dyson Eq.~\eqref{eqppgfdysoneq1} in the final form
%
\bea
  G^\mathrm{R}_{\vec p \vec p^\prime \vec q \vec q^\prime}(\omega)
  &=&g^{\mathrm{R}}_{\vec p \vec p^\prime \vec q \vec q^\prime}(\omega) +
  \sum_{\vec k \vec k^\prime} \sum_{\vec n \vec n^\prime}
  g^{\mathrm{R}}_{\vec p \vec p^\prime \vec k \vec k^\prime}(\omega)\nonumber\\
  &\times&
  \left(\mathcal{K}^{\mathrm{em}}_{\vec k \vec k^\prime \vec n \vec n^\prime}
  +\mathcal{K}^{\mathrm{c}}_{\vec k \vec k^\prime \vec n \vec n^\prime} (\omega) \right)
  G^\mathrm{R}_{\vec n \vec n^\prime \vec q \vec q^\prime}(\omega) \ .
  \label{eqppgfdysoneq2}
\eea
%
Eq.~\eqref{eqppgfdysoneq2} has a form of the Dyson equation for the two-particle Green's
function, however, the reference GF $g^{\mathrm{R}}_{\vec p \vec p^\prime \vec q \vec
  q^\prime}(\omega)$ is not given as a product of fully-interacting single-particle GFs,
but rather is the full two-particle GF --- the resolvent of the effective
Hamiltonian~\eqref{eq:h2} which includes the full electron-electron repulsion and the
mean-field contribution from the ionized system.

\section{Fermi golden rule\label{sec:fgr}}
\subsection{Single photoemission\label{sec:spe}}
SPE was treated by several authors. We recapitulate the main points.  The total observed
current is proportional to the expectation value of the electron number operator
$\hat{N}_\vec{k}=c_\vec{k}^\dagger c_\vec{k}$. Out of all possible final states of the
target we discard all unbound states, i.e. $c_\vec{k}|\Psi_\alpha^{+}\rangle=0$ and choose
only those relevant for a specific experiment. Let $\lambda_\alpha$ be a corresponding
distribution function. For instance when the target is left in the ground state we can set
$\lambda_0=1$ and $\lambda_\alpha=0$ for all excited states. Modified particle number
operator for this process reads:
\[
\hat{\widetilde N}_\vec{k}=\sum_\alpha \lambda_\alpha c_\vec{k}^\dagger |\Psi_\alpha^{+}\rangle
\langle \Psi_\alpha^{+}|c_\vec{k}=\sum_\alpha\lambda_\alpha
P_\alpha c_\vec{k}^\dagger c_\vec{k}P_\alpha.
\]
The same expression can be obtained from the Langreth approach starting from the Wigner
distribution function~\cite{langreth_scattering_1971}.  Let now the SPE current be the
expectation value of this operator
\begin{multline}
J_\vec{k}=\lim_{\eta\rightarrow0}2\eta\sum_\alpha\lambda_\alpha\big\langle \Psi_0\big|\hat \Delta^\dagger
\frac{1} {E_0+\omega-\hat{H}-i\eta}P_\alpha c_\vec{k}^\dagger c_\vec{k} P_\alpha\\
\times\frac{1}{E_0+\omega-\hat{H}+i\eta}\hat \Delta\big|\Psi_0\big\rangle.
\label{eq:j1resp2}
\end{multline}
We only consider the case
\be
P_\alpha\frac{1}{E_i-\hat{H}+i\eta}\approx\frac{P_\alpha}{E_i-\hat{H}_P-\hat{\Sigma}^{(+)}_P(E_i)},
\label{eq:approx1}
\ee
where we neglect the off-diagonal term in Eq.~\eqref{eq:res} and define
$\hat{\Sigma}^{(\pm)}_P(\omega)=\hat{\Sigma}_P(\omega\pm i\eta)$. We omit the subscript
$\alpha$ where it does not cause a confusion. A simple calculation leads to the modified
matrix element
\begin{equation}
M_{\vec{k},\alpha}=\langle \Psi_\alpha^{+}|c_\vec{k}\frac{1}{E_i-\hat{H}_P-\hat{\Sigma}^{(+)}_P(E_i)}
P_\alpha\hat \Delta\big|\Psi_0\big\rangle.
\end{equation}
Using the same assumption for the computation of the matrix element of $\hat \Delta$,
$\langle \Psi_\alpha^{+}|c_\vec{p}\hat \Delta|\Psi_0\rangle =\big\langle \vec{p}|\hat
\Delta|\phi_\alpha\rangle$ and the definition of the Green's function on the $P_\alpha$
subspace:
\[
G_{\vec{p}\vec{k},\alpha}^{(\mathrm{p})}(\omega+\varepsilon_\alpha\pm i\eta)=\langle
\Psi_\alpha^{+}|c_\vec{p}\frac{1}{E_i-\hat{H}_P-\hat{\Sigma}^{(\pm)}_P(E_i)}c_\vec{k}^\dagger|\Psi_\alpha^{+}\rangle
\]
we obtain for the current
\begin{multline}
J_\vec{k}=\lim_{\eta\rightarrow0}2\eta\sum_\alpha\lambda_\alpha\sum_{\vec{p}\vec{q}}\langle\phi_\alpha|
\hat\Delta^\dagger|\vec{p}\rangle G_{\vec{p}\vec{k},\alpha}^{(\mathrm{p})}(\omega+\varepsilon_\alpha-i\eta)\\
\times G_{\vec{k}\vec{q},\alpha}^{(\mathrm{p})}(\omega+\varepsilon_\alpha+i\eta)\langle
\vec{q}|\hat \Delta|\phi_\alpha\rangle,
\label{eq:jk1}
\end{multline}
where $\varepsilon_\alpha=E_0-E_\alpha^+$.  As shown in Appendix~\ref{ap:GF} we can
express the particle Green's functions in terms of M{\o}ller operators
\begin{subequations}
\label{eq:GFM}
\begin{eqnarray}
G_{\vec{p}\vec{k},\alpha}^{(\mathrm{p})}(\omega+\varepsilon_\alpha-i\eta)&=&
\frac{1}{\omega+\varepsilon_\alpha-\varepsilon_{\vec k}-i\eta}
\langle \vec{p}|\chi_{\vec{k},\alpha}^{(-)}\rangle,\\
G_{\vec{k}\vec{q},\alpha}^{(\mathrm{p})}(\omega+\varepsilon_\alpha+i\eta)&=&
\frac{1}{\omega+\varepsilon_\alpha-\varepsilon_{\vec k}+i\eta}
\langle\chi_{\vec{k},\alpha}^{(-)}|\vec{q}\rangle.
\end{eqnarray}
\end{subequations}
This finally leads to the current
\[
J_\vec{k}=2\pi\sum_\alpha\lambda_\alpha\langle\chi_{\vec{k},\alpha}^{(-)}|\hat \Delta|
 \phi_\alpha\rangle \delta(\omega+\varepsilon_{\alpha}-\varepsilon_{\vec{k}})\langle\phi_\alpha|\hat \Delta^\dagger|
\chi_{\vec{k},\alpha}^{(-)}\rangle.
\]
A standard definition of the spectral function entails to
\[
\hat A(\zeta)=\sum_\alpha|\phi_\alpha\rangle\delta(\zeta-\varepsilon_\alpha)\langle\phi_\alpha|.
\]
Therefore, we can recast the expression for the current in a more familiar response form
\[
J_\vec{k}=2\pi\int_{-\infty}^{\mu} d\zeta\delta(\omega+\zeta-\varepsilon_{\vec{k}})
\langle\chi_{\vec{k},\alpha}^{(-)}|\hat \Delta \hat{\tilde{A}}(\zeta)\hat \Delta^\dagger|\chi_{\vec{k},\alpha}^{(-)}\rangle,
\]
where the tilde denotes a spectral function with restrictions imposed by the weighting
factors $\lambda_\alpha$ and $\mu$ is the chemical potential, or in the Fermi golden rule form:
\[
J_\vec{k}=2\pi\sum_\alpha\lambda_\alpha\delta(\omega+\varepsilon_{\alpha}-\varepsilon_{\vec{k}})
\big|\langle\chi_{\vec{k},\alpha}^{(-)}|\hat \Delta|\phi_\alpha\rangle\big|^2.
\]
 The major distinction from other approaches is that both, initial and final states are
 dependent on the final state of the target $\alpha$. Formally,
 $|\chi_{\vec{k},\alpha}^{(-)}\rangle$ is the incoming scattering state of an electron in the
 optical potential of the ionized target in the state $|\Psi_\alpha^+\rangle$. Notice that the
 current has been obtained using the approximation~\eqref{eq:approx1}. Exact calculation
 leads to the appearance of the vertex functions that describe a screening of the optical
 field by the electrons of the target~\cite{almbladh_theory_1985}. We will stop on this
 point when treating DPE process.

\subsection{Double photoemission \label{sec:dpeI}}
The total observed current is given in terms of the expectation value of the electron
number operators $\hat{N}_{\vec{k_1}\vec{k_2}}=\hat{N}_{\vec{k_1}}\hat{N}_{\vec{k_2}}-\delta_{\vec k_1,\vec
  k_2}\hat{N}_{\vec{k_1}}$, viz. Eq.~\eqref{eq:j2}. Out of all possible final states of the
target we discard all unbound states, i.e. $c_\vec{k}|\Psi_\beta^{2+}\rangle=0$ and
introduce weights $\lambda_\beta$ selecting the relevant ones. The modified
observable reads:
\begin{multline}
\hat{\widetilde{N}}_{\vec{k_1}\vec{k_2}}
=\sum_\beta\lambda_\beta c_\vec{k_1}^\dagger c_\vec{k_2}^\dagger|\Psi_\beta^{2+}\rangle
\langle \Psi_\beta^{2+}|c_\vec{k_2} c_\vec{k_1}\\
=\sum_\beta P_\beta c_\vec{k_1}^\dagger c_\vec{k_2}^\dagger c_\vec{k_2} c_\vec{k_1}P_\beta.
\end{multline}
This allows us to improve upon Eq.~\eqref{eq:golden0}:
\begin{multline}
J_{\vec{k_1},\vec{k_2}}=\lim_{\eta\rightarrow0}2\eta\sum_\beta\lambda_\beta
\big\langle \Psi_0\big|\hat \Delta^\dagger\frac{1} {E_0+\omega-\hat{H}-i\eta}\\
P_\beta c_\vec{k_1}^\dagger c_\vec{k_2}^\dagger
c_\vec{k_2} c_\vec{k_1}P_\beta\frac{1}{E_0+\omega-\hat{H}+i\eta}\hat \Delta\big|\Psi_0\big\rangle,
\label{eq:jresp2}
\end{multline}
Using assumption~\eqref{eq:approx1} Eq.~\eqref{eq:jresp2} can be written in the Fermi
golden rule form with a modified matrix element
\[
M_{\vec{k_1}\vec{k_2},\beta}=\big\langle\Psi_\beta^{2+}\big|c_\vec{k_2}c_\vec{k_1}
\frac{1}{E_i-\hat{H}_P-\hat{\Sigma}^{(+)}_P(E_i)}P_\beta\hat \Delta\big|\Psi_0\big\rangle.
\]
Using the matrix elements of $\hat \Delta$, $\langle
\Psi_\beta^{2+}|c_\vec{q}c_\vec{p}\hat \Delta|\Psi_0\rangle =\big\langle
\vec{p}\vec{q}|\hat \Delta|\phi_\beta^{(2)}\rangle$ (cf. Eq.~\eqref{eq:me_dpe}),
and the properties of the two-particle Green's functions (Appendix~\ref{ap:GF}) 
\begin{multline}
G_{\vec{p}\vec{q},\vec{k_1}\vec{k_2},\beta}^{(\mathrm{pp})}(\omega+\varepsilon_\beta^{(2)}\pm i\eta)\\
=\big\langle\Psi_\beta^{2+}\big|c_{\vec{p}}c_{\vec{q}}
\frac{1}{E_i-\hat{H}_P-\hat{\Sigma}^{(+)}_P(E_i)}c^\dagger_\vec{k_2}c^\dagger_\vec{k_1}\big|\Psi_\beta^{2+}\big\rangle\\
=\frac{1}{\omega+\varepsilon_\beta^{(2)}-\varepsilon_{\vec{k_1}}-\varepsilon_{\vec{k_2}}\pm i\eta}\langle
\vec{p}\vec{q}|\psi_{\vec{k_1}\vec{k_2},\beta}^{(-)}\rangle,
\label{eq:j2n}
\end{multline}
we finally obtain for Eq.~\eqref{eq:jresp}
\begin{multline}
J_{\vec{k_1},\vec{k_2}}=2\pi\int_{-\infty}^{\mu^{(2)}}\!d\zeta\,
\delta(\omega+\zeta-\varepsilon_{\vec{k_1}}-\varepsilon_{\vec{k_2}})\\
\times\langle \psi_{\vec{k_1}\vec{k_2},\beta}^{(-)}|\hat \Delta
A^{(2)}(\zeta) \hat \Delta^\dagger|\psi_{\vec{k_1}\vec{k_2},\beta}^{(-)}\rangle,
\label{eq:jresp3}
\end{multline}
where $\mu^{(2)}=\max_\beta(E_0-E_\beta^{2+})$ is the negative of second ionization potential,
$|\psi_{\vec{k_1}\vec{k_2},\beta}^{(-)}\rangle$ is the incoming damped two-electron scattering
state in the optical potential of doubly ionized target and $\hat A^{(2)}(\zeta)$ is the two-particle spectral function, which can be written in
terms of two-hole Dyson orbitals:
\be
 \hat A^{(2)}(\zeta)=\sum_\beta
\delta(\zeta-\varepsilon^{(2)}_\beta)|\phi_\beta^{(2)}\rangle\langle
\phi_\beta^{(2)}|,\label{eq:a2}
\ee
with $\varepsilon^{(2)}_\beta=E_0-E_\beta^{2+}$.

Notice that the current has been obtained using the
approximation~\eqref{eq:approx1}. Exact calculation leads to the appearance of the vertex
functions resulting from $Q_\beta\hat \Delta|\Psi_0\rangle$ and describing a screening of
the optical field by the electrons of the target~\cite{almbladh_theory_1985}.

In the valence shell the DPE mechanism is typically due to ground state electron
correlation, i.e. due to the correlated two-particle spectral function entering
\eqref{eq:jresp3}. In contrast, when core electrons are involved a dominant mechanism for
DPE is due to the final state relaxation (so called shake-off). Multiple stages are then
described by introducing corresponding projection operators for each intermediate
stage. In the following, we focus on the diagrammatic approach because it allows us to
treat all these effects on equal footing.
\section{Diagrammatic approach\label{sec:diag}}
Treatment of the off-diagonal part of the Hamiltonian resolvent is the main difficulty of
the Feshbach projection algebra. It is even more aggravated in the two-particle case.  The
diagrammatic technique provides a natural and practical solution to this problem.
\subsection{Derivation}
\begin{figure}[t!]
\includegraphics[width=0.99\columnwidth]{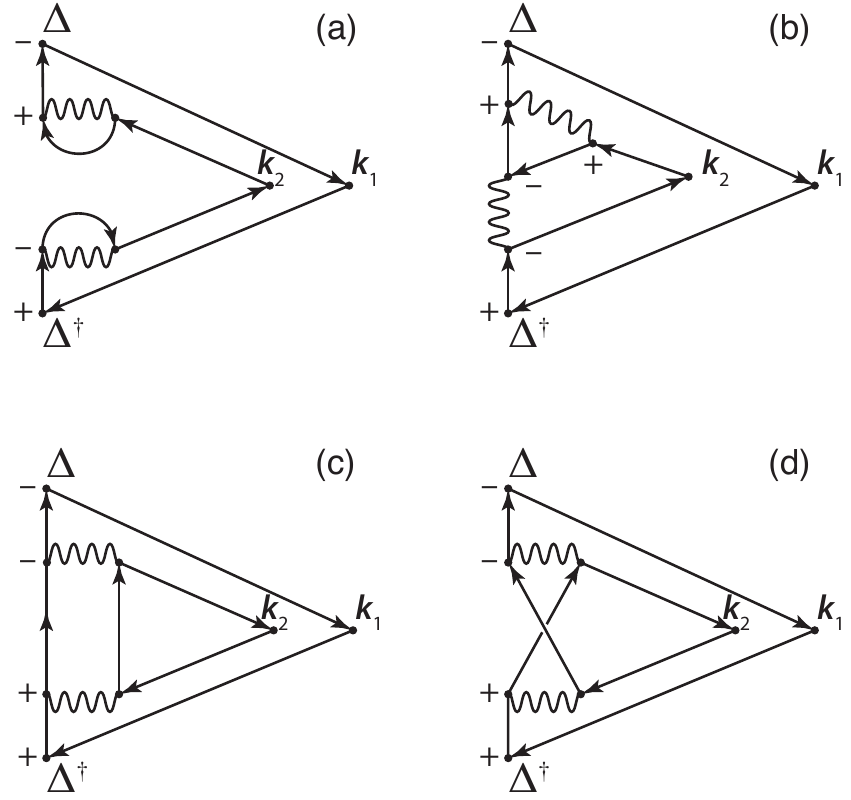}
\caption{Second order diagrams (in bare Coulomb interaction) representing the DPE
  process. The dots labeled $\vec{k_1}$ and $\vec{k_2}$ correspond to the scattering state
  of two electrons observed in a coincidence measurement by the detector. Notice that not
  all combinations of pluses and minuses are possible because Coulomb interaction can only
  connect vertices on the same branch of the Keldysh contour. (a) Diagram vanishes
  according to the assumption~\eqref{eq:restriction} for dressed GFs. (b) Diagram vanishes
  because it contains an isolated island of minuses. (c) and (d) are the lowest order
  nonzero diagrams. The remaining two are obtained by permuting $\vec{k_1}$ and
  $\vec{k_2}$.
\label{fig:diag}}
\end{figure}
Eq.~\eqref{eq:jresp} when transformed to the time domain gives rise to the following
ground state correlator:
\begin{multline}
Z(t,t^\prime)=\langle\Psi_0|c_b^\dagger(t)c_a(t)c_\vec{k_1}^\dagger(0) c_\vec{k_2}^\dagger(0)
c_\vec{k_2}(0)c_\vec{k_1}(0)\\ \times c_c^\dagger(t^\prime)c_d(t^\prime)|\Psi_0\rangle,
\end{multline}
where the field operators are in the Heisenberg representation and
$t,\,t^\prime\in(-\infty,0]$ are \emph{physical} times. For clarity, we omitted the
  indices in the notation of the correlator. It can be evaluated diagrammatically by
  adiabatically switching on the interaction in the remote past,
  i.~e. $\hat{H}_\delta=\hat{H}_0+e^{-\delta|t|} \hat{H}_1$. Now the average is performed
  over the noninteracting ground state $|\Phi_0\rangle$ and the times $t_2^-\prec t_1^+$
  lie on forward, backward branches of Keldysh contour $\gamma$ (Fig.\,\ref{fig:contour}),
  respectively:
\begin{multline}
Z(t,t^\prime)=\big\langle\Phi_0\big|\mathcal{T}
\big\{
e^{-i\int_{\gamma}\!\hat{H}_\delta(t)\, dt}c_b^\dagger(t_+)c_a(t_+)\\
\times c_\vec{k_1}^\dagger(0) c_\vec{k_2}^\dagger(0) c_\vec{k_2}(0)c_\vec{k_1}(0)
c_c^\dagger(t^\prime_-)c_d(t^\prime_-)
\big\}
\big|\Phi_0\big\rangle.
\label{eq:Z}
\end{multline}
$\mathcal{T}$ here is the usual contour ordering
operator~\cite{stefanucci_nonequilibrium_2013} with the order relation
$\prec$. $\hat{H}_\delta$ is such that it is equal to the Hamiltonian of noninteracting
system $H_0$ in the remote past and is identical to $\hat{H}$ at $t=0$. Notice that it is
different from adiabatic switching on of the electromagnetic field in
Eq.~\eqref{eq:pt}. $|\Phi_0\rangle$ is the ground state of $\hat{H}_0$. Using Wick's
theorem we can contract the product of field operators in order to express the correlator
in terms of products of single-particle Green's functions. Zeroth order obviously yields
four fermionic lines. However, if we use the same assumption as in Sec.~\ref{sec:dpeI} any
zeroth order diagram vanishes. This is easy to understand by comparing with SPE
case. There, no-zero contributions are coming from the following contraction:
\[
\big\langle
\contraction{c_b^\dagger(t_+)}{c_a(t_+)}{}{c^\dagger_{\vec{p}}(0)}
\contraction{c_b^\dagger(t_+)c_a(t_+)c^\dagger_{\vec{p}}(0)}{c_{\vec{p}}(0)}{}{c_c^\dagger(t^\prime_-)}
\contraction[2ex]{}{c_b^\dagger(t_+)}{c_a(t_+)c^\dagger_{\vec{p}}(0)c_{\vec{p}}(0)c_c^\dagger(t^\prime_-)}{c_d(t^\prime_-)}
c_b^\dagger(t_+)c_a(t_+)c^\dagger_{\vec{p}}(0)c_{\vec{p}}(0)c_c^\dagger(t^\prime_-)c_d(t^\prime_-)
\big\rangle.
\]
This is the only combination that results in greater GFs when one of the arguments is a
scattering state (and is compatible with~\eqref{eq:restriction}). In particular, the above
contraction equals to
\[
g_{a\vec{p}}^>(t)g_{db}^<(t'-t)g_{\vec{p}c}^>(-t').
\]
In DPE two creation operators with continuum state indices need to be contracted with two
annihilation operators on the positive track. However, there is only one such
operator. Hence, 0th order in interaction is zero. The argument that excludes the first
order diagram is slightly different and is based on the fact that bare interaction is
instantaneous, i.e. corresponding time-arguments necessarily lie on the same, positive or
negative, track.

Second order nonvanishing contributions contain products of two Coulomb interaction
operators (e.g. at contour times $\bar{t}_+$ and $\bar{\bar{t}}_-$) and already a familiar
product of six operators as in Eq.~\eqref{eq:Z}. From all possible contractions (they
yield eight fermionic lines) we have to exclude many terms. Some of them immediately
vanish because of the assumption~\eqref{eq:restriction} for noninteracting GF. Others,
represent the Hartree-Fock renormalization of two fermionic lines and likewise vanish
because of the same assumption for the full fermionic propagators,
Fig.\,\ref{fig:diag}\,(a). Then, there are diagrams (Fig.\,\ref{fig:diag}\,(b)) containing
isolated islands of pluses and minuses which also vanish because otherwise the
two-particle current cannot be written in the Fermi Golden rule
form~\cite{stefanucci_diagrammatic_2014,uimonen_diagrammatic_2015}. Finally, there are
only four (times two for exchange) nonzero diagrams. Two of them are depicted at
Fig.\,\ref{fig:diag}\,(c,d).

\begin{figure}[]
\centering
       \includegraphics[width=0.99\columnwidth]{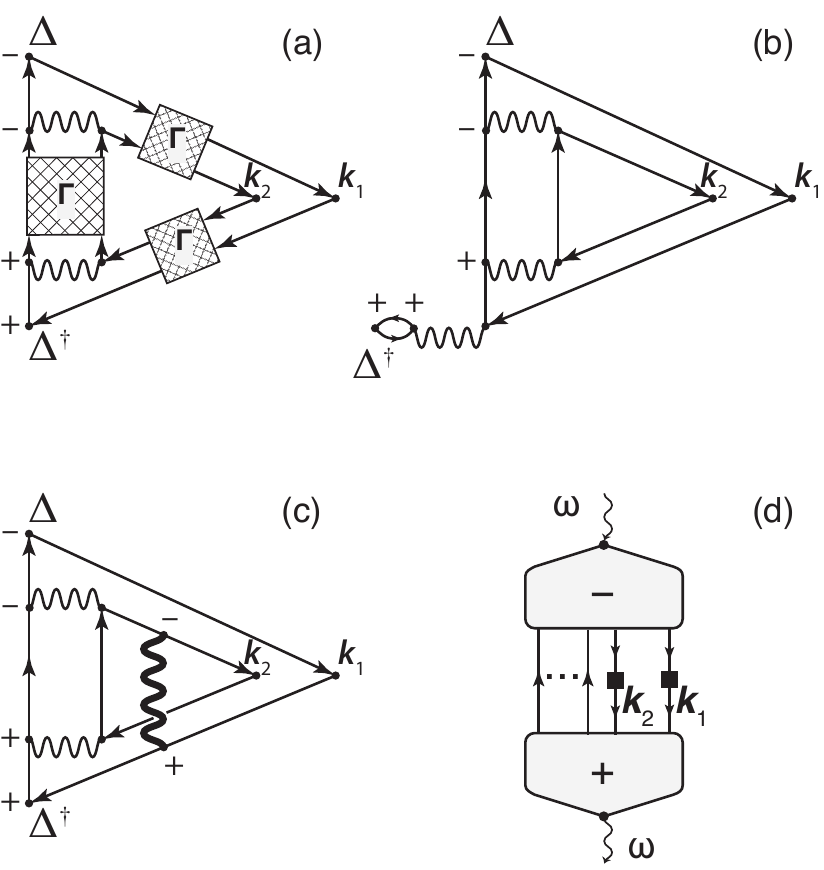}
       \caption[]{(a) Diagram for the two-particle current involving dressed two-particle
         propagators.  (b) Simplest diagram where the optical field is screened. (c) Example
         of a diagram describing external losses. Thick wavy line denotes the screened
         Coulomb interaction. (d) Generic diagram for the two-particle current.
      \label{fig:J2}}
\end{figure}
It is clear now how more general diagrams for the two-electron current can be constructed:
i) One replaces all bare fermionic propagators and interaction lines with the dressed
ones; ii) Each pair of parallel fermionic lines are replaced by the corresponding
two-particle propagator, Fig.\,\ref{fig:J2}\,(a). In doing so one obtains, in principle,
diagrams given by Fig.\,(1b) of Fominykh \emph{et al}.~\cite{fominykh_theory_2000} with a
small correction that zeroth and the first-order two-particle GF should be excluded from
the vertical track; iii) Next class of the diagrams are those that describe the screening
of the optical field, Fig.\,\ref{fig:J2}\,(b); iv) Processes involving intrinsic or
extrinsic losses are given by the diagrams with interaction lines connecting points on
different tracks, i.~e. ``$+\,-$'', ``$+\,0$'', ``$-\,0$''.  They cannot be obtained by
the renormalization of fermionic or bosonic propagators, one such example shown at
Fig.\,\ref{fig:J2}\,(c) reveals a process with extrinsic losses.

Finally, we give a description of a general diagram for a photoemission process. Examining
SPE and DPE diagrams we see that all of them are constructed from the common ancestor: the
density-density response function $\chi^<\equiv\chi^{-+}$ having a form of two islands
with time arguments belonging to either forward or backward tracks of the Keldysh
contour. Now we introduce detectors (shown as black squares at Fig.\,\ref{fig:J2}\,(d)
measuring $J_{\vec{k_1},\vec{k_2}}$. As explained before i) the lesser GF with one of the
indices being a continuum state vanishes because of the
assumptions~(\ref{eqckzero},\ref{eqckzero2}); and ii) observation is made at the rightmost
point of the contour (i.~e. at $t_-=t_+=0$ in our notations), thus, each detector
measuring particle numbers $N_{\vec{k}_i}$ is connected to two greater GF. In view of
this, the detectors ``lie'' on the fermionic lines flowing from the ``$-$'' (forward
track) to ``$+$'' (backward track) islands. Each response function constructed in this way
has an important property that it can be represented in the Fermi Golden rule form, such
construction obviously generalizes to an arbitrary number ($n$) of emitted
particles. Simple counting shows that these processes are of at least $2(n-1)$ order in
the Coulomb interaction.

The diagram in Fig.\,\ref{fig:J2}\,(d) is a generic one describing all the DPE processes
including the ones with losses such as shown at Fig.\,\ref{fig:J2}\,(c). One can go a step
further and give a prescription for classes of lossless diagrams. A detailed analysis of
this particular situation is possible and will be done elsewhere.  Here, we mention
without a derivation that such diagrams can be split into the scattering part (the
two-particle propagators can be written in terms of the scattering states
$|\psi_{\vec{k_1}\vec{k_2},\beta}^{(-)}\rangle$, cf. Eq.~\eqref{eq:j2n}) and the spectral
part (containing the two-particle spectral function, Eq.~\eqref{eq:a2}).
\subsection{Example of plasmon assisted DPE}
\begin{figure}[t!]
\centering
       \includegraphics[width=0.99\columnwidth]{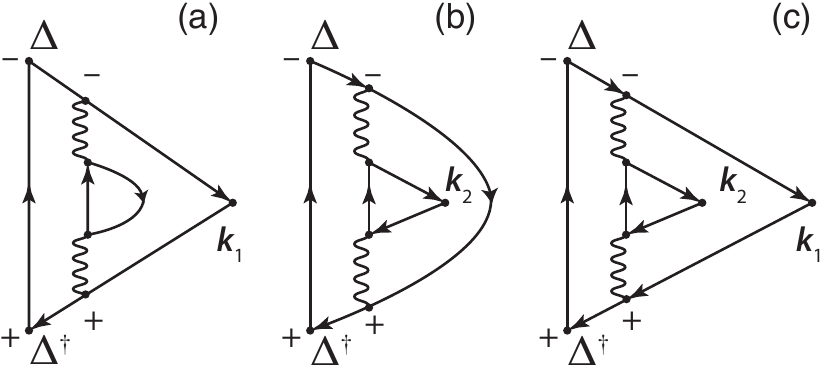}
       \caption[]{Diagrams for the plasmon assisted photoemission. SPE setup: only the
         primary (a), secondary electron (b) is observed, the fate of another electron is
         not specified. (c) DPE setup: both, primary and secondary electrons are observed
         in coincidence. \label{fig:J1J2}}
\end{figure}

As an example we consider the processes depicted in Fig.\,\ref{fig:J1J2}. The diagrams show
a very common situation where a primary electron excited by the laser pulse is loosing its
energy on the way to the detector by exciting a secondary electron. There could be either
bare or screened Coulomb interaction between the two electrons.  In the latter case some
resonant phenomena related to the excitation of e.~g. plasmon are expected. The SPE case
(Figs.~\ref{fig:J1J2}\,(a,b)) is identical to the process of secondary electron excitation
considered by Caroli \emph{et al.}~\cite{caroli_inelastic_1973}. All DPE processes covered
by the diagram at Fig.\,\ref{fig:J1J2}\,(c) form a subset of the SPE process.  The only
difference between the two scenarios is whether primary, secondary or both electrons are
observed in the detector. It is obvious that one reduces the DPE diagram to the SPE ones
by integration over the energy and momentum of the secondary, or primary electrons,
respectively.

Since we do not take into account the interaction between the two emitted electrons (as
given, for, instance by two $\Gamma$-blocks at Fig.\,\ref{fig:J2}\,(a) one can express the
final result for the current as a matrix element over the direct product of two
single-particle scattering states. This is typically a good approximation for the case
when two electrons have different energies (momenta), or for approximately equal $\vec
k_1$ and $\vec k_2$ in the case of larger energies~\cite{berakdar_concepts_2003}.

To work this out consider a part of the DPE diagram that contains a product of two GFs
involving the external momentum $\vec k$. Introducing the Fourier representations for each
of the GFs
$G_{a\vec{k}}^>(\tau)=\int_{-\infty}^{\infty}\frac{d\nu}{2\pi}e^{-i\nu\tau}G_{a\vec{k}}^>(\nu)$,
$G_{\vec{k}b}^>(-\tau')=\int_{-\infty}^{\infty}\frac{d\nu'}{2\pi}e^{i\nu'\tau'}G_{\vec{k}b}^>(\nu')$,
expressing the interacting GF as a product of the M{\o}ller operator and the free-particle
Green's function (see Appendix~\ref{ap:GF}) we obtain expressions similar to
Eqs.~\eqref{eq:GFM}. Thus, in the time domain the product of two interacting
single-particle GFs reduces to a simple propagator computed on the scattering states with
incoming boundary conditions:
 \begin{multline}
G_{a\vec{k}}^>(\tau)G_{\vec{k}b}^>(-\tau')=
\langle\chi^{(-)}_{\vec k}|b\rangle e^{-i \varepsilon_{\vec{k}}(\tau-\tau')}
\langle a|\chi^{(-)}_{\vec{k}}\rangle\\
\times\theta(-\tau)\theta(-\tau')e^{\delta(\tau+\tau')}.
\label{eq:GG}
\end{multline}

\begin{figure}[t!]
\centering
       \includegraphics[width=0.99\columnwidth]{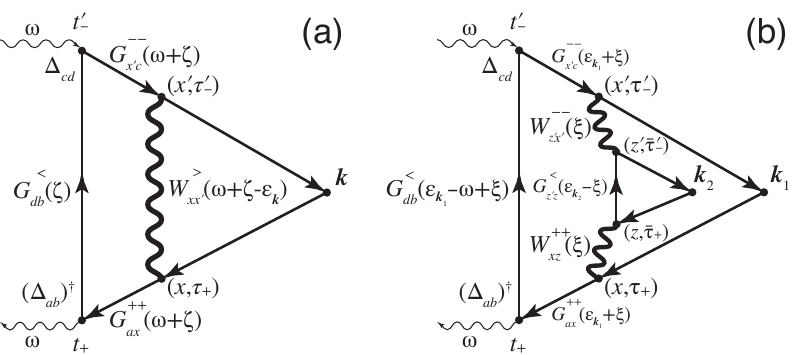}
       \caption[]{Energy flows in (a) SPE diagram with external plasmonic losses, (b) DPE
         diagram describing a related plasmon assisted process. Analytical expressions
         corresponding to these diagrams are first written in the time domain, then the
         integrations are performed by Fourier transforming all the propagators, and lastly
         the limits $\eta\rightarrow 0$ and $\delta\rightarrow 0$ are
         taken.\label{fig:J1J2m}}
\end{figure}

As an exercise let us evaluate the diagram at Fig.\,\ref{fig:J1J2m}\,(a) describing the
SPE process with extrinsic plasmon losses. The current is given by the following
expression in the time domain:
\bea
J_{\vec{k}}&=&\lim_{\eta\rightarrow0}2\eta\lim_{\delta\rightarrow0}\sum_{abcd}\int\!\!d(xx^\prime)
\int_{-\infty}^0\! \!d(tt^\prime)e^{\eta(t+t^\prime)}
\int_{-\infty}^0\! \!d(\tau\tau^\prime)\nn\\
&&\times e^{i\omega(t-t^\prime)} \Delta_{cd}
G_{db}^<(t',t)G_{x'c}^{--}(\tau',t')W_{xx'}^>(\tau,\tau')\nn\\
&&\times G_{\vec{k} x'}^>(0,\tau')G_{x\vec{k}}^>(\tau,0)G_{ax}^{++}(t,\tau)(\Delta_{ab})^\dagger.
\eea
Representing the lesser Green's function on the vertical track in terms of the electron
spectral function (normalized as $\sum_{b}\int_{-\infty}^{\mu}
\frac{d\zeta}{2\pi}A_{bb}(\zeta)=N$, $N$ is the number of electrons in the system)
\be
G_{db}^<(t',t)=i\int_{-\infty}^{\mu} \frac{d\zeta}{2\pi}A_{db}(\zeta)e^{-i\zeta(t'-t)},
\label{eq:G<}
\ee
and the greater component of the screened interaction in terms of the plasmon spectral function
\be
W_{xx'}^>(\tau,\tau')=-i\int_{0}^{\infty} \frac{d\xi}{2\pi}B_{xx'}(\xi)e^{-i\xi(\tau-\tau')},
\label{eq:W>}
\ee
representing time-ordered $G_{x'c}^{--}(\tau',t')$ and anti-time-ordered
$G_{ax}^{++}(t,\tau)$ as Fourier integrals and using expression~\eqref{eq:GG} we obtain:
\bea
J_{\vec{k}}&=&\lim_{\eta\rightarrow0}\lim_{\delta\rightarrow0}\sum_{abcd}\int\!\!d(xx^\prime)
 \int_{-\infty}^{\mu}\!\!\frac{d\zeta}{2\pi}
\int_{0}^{\infty}\!\!\frac{d\xi}{2\pi}B_{xx'}(\xi)\nonumber\\
&\times&\int\!\!d(\omega_1\omega_2)2\eta
\frac{1}{\omega+\zeta-\omega_1-i\eta}\frac{1}{\omega+\zeta-\omega_2+i\eta}\nonumber\\
&\times&\frac{1}{\omega_1-\xi-\varepsilon_{\vec{k}}-i\delta}\frac{1}{\omega_2-\xi-\varepsilon_{\vec{k}}+i\delta}
G_{x'c}^{--}(\omega_2)G_{ax}^{++}(\omega_1)\nonumber\\
&\times&\langle\chi^{(-)}_{\vec k}|x'\rangle
\Delta_{cd}A_{db}(\zeta)(\Delta_{ab})^\dagger
\langle x|\chi^{(-)}_{\vec{k}}\rangle.
\eea
Now the limits can be taken making use of an identity discovered by
C.~O.~Almbladh~\cite{almbladh_theory_1985} (see Appendix~\ref{ap:Sokh-Plem}). It
transforms the product of four fractions in the equation above into the product of three
$\delta$-functions
$(2\pi)^3\delta(\omega_1-\omega-\zeta)\delta(\omega_2-\omega-\zeta)\delta(\xi+\epsilon_{\vec{k}}-\omega+\zeta)$,
and after the frequency integration we obtain
\bea
J_{\vec{k}}&=&2\pi\int_{-\infty}^{\mu}\!\!\frac{d\zeta}{2\pi}
\int_{0}^{\infty}\!\!\frac{d\xi}{2\pi}\,\,\delta(\xi+\varepsilon_{\vec{k}}-\omega-\zeta)\nonumber\\
&\times&\int\!\!d(xx^\prime)\langle\chi^{(-)}_{\vec k}|x'\rangle
B_{xx'}(\xi)\langle x|\chi^{(-)}_{\vec{k}}\rangle\nonumber\\
&\times&\Big[\hat G^{--}(\omega+\zeta)\hat \Delta \hat A (\zeta)\hat \Delta^\dagger \hat G^{++}(\omega+\zeta)\Big]_{x'x}.
\label{eq:Jk_Pl}
\eea
The two-particle current is obtained along the same lines using the energy flow as shown
on Fig.\,\ref{fig:J1J2m}\,(b).
\bea
J_{\vec{k_1}\vec{k_2}}&=&2\pi\int_{-\infty}^{\mu}\!\frac{d\zeta}{2\pi}\int_{-\infty}^{\mu}\!\frac{d\bar{\zeta}}{2\pi}
\int_{0}^{\infty}\!\frac{d\xi}{2\pi}\,\,\delta(\xi+\varepsilon_{\vec{k_1}}-\omega-\zeta)\nonumber\\
&\times&\int\!\!d(xx'zz')\langle\chi^{(-)}_{\vec k_1}|x'\rangle W^{--}_{z'x'}(\xi)
W^{++}_{xz}(\xi)\langle x|\chi^{(-)}_{\vec{k_1}}\rangle\nonumber\\
&\times&\langle\chi^{(-)}_{\vec k_2}|z'\rangle A_{z'z}(\bar{\zeta})\langle z|\chi^{(-)}_{\vec{k_2}}\rangle
\,\delta(\varepsilon_{\vec{k_2}}-\xi-\bar{\zeta})\nonumber\\
&\times&\Big[\hat G^{--}(\omega+\zeta)\hat \Delta \hat A (\zeta)\hat \Delta^\dagger \hat G^{++}(\omega+\zeta)\Big]_{x'x}.
\label{eq:Jkk_Pl}
\eea
Similarly to the previous case, the limits $\eta\rightarrow 0$, $\delta\rightarrow 0$
yield a product (of five) $\delta$-function which were subsequently used to perform three
frequency integrations here (see Appendix~\ref{ap:Sokh-Plem}). All the quantities in
Eqs.~(\ref{eq:Jk_Pl},\ref{eq:Jkk_Pl}) can be expressed in terms of the spectral
functions. We can, for instance, start with a general expresion for the time-ordered
function in terms of functions on the Keldysh contour:
\begin{subequations}
\bea
\hat f^{--}(\tau)&=&\hat f^\delta\delta(\tau)+\theta(\tau)\hat f^>(\tau)+\theta(-\tau)\hat f^<(\tau),\\
\hat f^{++}(\tau)&=&-\hat f^\delta\delta(\tau)+\theta(-\tau)\hat f^>(\tau)+\theta(\tau)\hat f^<(\tau),
\eea
\end{subequations}
where in the first equation $\tau\equiv t_--t_-'$ is equal to the time-difference on the
forward branch of the contour, and $\tau\equiv t_+-t_+'$ is equal to the time-difference
on the backward branch of the contour in the second equation. After the Fourier transform
$\hat f(\omega)=\int_{-\infty}^{\infty}\!d\tau\, e^{i\omega \tau} \hat f(\tau)$, we have
\be
\hat f^{--}(\omega)=\hat f^\delta+\int\limits_{-\infty}^{\infty}\!\frac{d\omega'}{2\pi}
\left[\frac{i \hat f^>(\omega')}{\omega-\omega'+i\delta}
-\frac{i \hat f^<(\omega')}{\omega-\omega'-i\delta}\right].
\label{eq:fmm}
\ee
The fluctuation-dissipation theorem at zero temperature allows to express the lesser and
greater propagators in terms of the corresponding spectral functions
(Kubo-Martin-Schwinger (KMS) conditions~\cite{kadanoff_quantum_1962}):
\[
\hat G^{<}(\omega)=i\theta(\mu-\omega)\hat A(\omega),\quad
\hat G^{>}(\omega)=-i\theta(\omega-\mu)\hat A(\omega).
\]
The screened interaction obeys KMS conditions for bosonic propagators:
\[
\hat W^{<}(\omega)=i\theta(-\omega)\hat B(\omega),\quad
\hat W^{>}(\omega)=-i\theta(\omega)\hat B(\omega),
\]
with the symmetry property for the spectral function $\hat B(-\omega)=-\hat B(\omega)$
(follows e.~g. from the fact that $\hat W^R(t,t')$ is a real function or, more precisely,
a Hermitian matrix). We have already used these equations
(cf. Eqs.~(\ref{eq:G<},\ref{eq:W>})) to express SPE current in terms of spectral
functions. Using Eq.~\eqref{eq:fmm} we can write the spectral representation of the
fermionic propagator
\[
\hat G^{--}(\omega)=\int_{-\infty}^{\infty}\frac{d\omega'}{2\pi}\hat A(\omega')
\left[\frac{\theta(\mu-\omega')}{\omega-\omega'-i\delta}
+\frac{\theta(\omega'-\mu)}{\omega-\omega'+i\delta}\right],
\]
where $\mu$ is the Fermi energy. The anti-time-ordered GF is obtained similarly
$G^{++}(\omega)=-[G^{--}(\omega)]^\dagger$.  The screened interaction is expressed as an
integral over the positive frequencies:
\[
\hat W^{--}(\omega)=v+\int_0^\infty\frac{d\omega'}{2\pi}\hat B(\omega')\frac{2\omega'}{\omega^2-(\omega'-i\delta)^2},
\]
while $\hat W^{++}(\omega)=-[W^{--}(\omega)]^\dagger$.

Let us consider plasmon-mediated DPE. This process is of relevance for metallic and large
molecular systems. Since plasmon is a long wavelength or small momentum electronic
excitation it is useful to go from the abstract basis to momentum representation and write
$W^{--}(k,\omega)$ in a short form as
\be
W(k,\omega)=v_k\left[1+\frac{\omega_\mathrm{p}^2}{\omega^2-\omega_\mathrm{p}^2(k)}\right],
\ee
where $\omega_\mathrm{p}(k)$ is the plasmon dispersion,
$\omega_\mathrm{p}\equiv\omega_\mathrm{p}(0)$ is the classical plasmon frequency, and
$v_k=\frac{4\pi}{k^2}$ is the matrix element of Coulomb interaction. It is clear that in
this form the plasmon peak completely exhausts the $f$-sum rule. Such plasmon pole
approximation for the screened interaction is broadly used in the electronic structure
calculation when full-fledged calculations are not feasible. Similarly, it can be used to
simplify Eq.~\eqref{eq:Jkk_Pl}.

\subsection{Numerical results}
Let us make some simplifications. Usually it is a good approximation to start with the
mean-field Green's functions
\be
G_{xy}^{--}(\omega)=\sum_{a\in \mathrm{occ}}\frac{\langle x|a\rangle n_\a \langle a|y\rangle}{\omega-\varepsilon_a-i\delta}
+\sum_{a\in \mathrm{unocc}}\frac{\langle x|a\rangle \overline{n}_a\langle a|y\rangle}{\omega-\varepsilon_a+i\delta},
\label{eq:mf}
\ee
where $n_a$ is the occupation number of the state $a$ and $\overline{n}_a\equiv 1-n_a$.
After straightforward, but tedious calculation the frequency integrations in
Eq.~\eqref{eq:Jkk_Pl} can be performed (for technical reasons it is better to start from
the time rather then frequency expression, and it can be obtained by directly transcribing
the diagram at Fig.\,\ref{fig:J1J2m}\,(b) using standard rules) yielding the following
expression for the two-particle current:
\bea
\lefteqn{J_{\vec{k_1}\vec{k_2}}=4\pi\!\sum_{abcd}
\frac{n_b n_d\Delta_{cb}\Delta_{ba}\delta(\omega+\varepsilon_b+\varepsilon_d-\varepsilon_{\vec{k_1}}-\varepsilon_{\vec{k_2}})}
{(\varepsilon_c+\varepsilon_{d}-\varepsilon_{\vec{k_1}}-\varepsilon_{\vec{k_2}})
(\varepsilon_{\vec{k_1}}+\varepsilon_{\vec{k_2}}-\varepsilon_a-\varepsilon_{d})}}&&\nonumber\\
&\times&\sum_{\vec{q_1}\vec{q_2}}\!\!
\Bigg[\frac{f_{\vec{k_1}c}^{\vec{q_1}}\left(f_{\vec{k_2}d}^\vec{q_1}\right)^\ast\! v_{\vec{q_1}}\omega_\mathrm{p}^2}
{(\varepsilon_d-\varepsilon_{\vec{k_2}})^2-\omega_\mathrm{p}^2(q_1)}\Bigg]\!\!
\Bigg[\frac{f_{a\vec{k_1}}^{\vec{q_2}}\left(f_{d\vec{k_2}}^\vec{q_2}\right)^*\!v_{\vec{q_2}}\omega_\mathrm{p}^2}
{(\varepsilon_d-\varepsilon_{\vec{k_2}})^2-\omega_\mathrm{p}^2(q_2)}\Bigg],
\label{eq:Jkk_fin}
\eea
with the following matrix elements
\be
f_{a\vec{k}}^{\vec{q}}=\int \!d^3r\, \langle a|r\rangle e^{-i\vec{q}\cdot\vec{r}}\langle r|\chi^{(-)}_{\vec{k}}\rangle.
\label{eq:me}
\ee
Notice that it is not necessary to separately treat the bare Coulomb interaction, it can
be recovered as $\omega_\mathrm{p}\rightarrow\infty$ limit as explained
in~\cite{pavlyukh_time_2013}.

Let us compare Eq.~\eqref{eq:Jkk_fin} with the general result obtained using the Feshbach
projection formalism~\eqref{eq:jresp3}. For the mean-field approximation~\eqref{eq:mf} the
two-particle spectral function is diagonal and is given by the convolution of two
single-particle spectral densities:
\begin{multline}
A^{(2)}_{bd}(\zeta)=\int\!d\overline{\zeta}\, A_{bb}(\zeta-\overline{\zeta})A_{dd}(\overline{\zeta})\\
=\int\!d\overline{\zeta}\, n_bn_d\delta(\zeta-\overline{\zeta}-\varepsilon_b)\delta(\overline{\zeta}-\varepsilon_d)\\
=n_bn_d\delta(\varepsilon_b+\varepsilon_d-\zeta).
\end{multline}
The energy conservation for the whole process, which is given by the $\delta$-function in
the numerator of \eqref{eq:Jkk_fin}, is expressed in terms of the two-particle spectral
function $A^{(2)}(\varepsilon_{\vec{k_1}}+\varepsilon_{\vec{k_2}}-\omega)$,
(cf. Eq.~\eqref{eq:a2}). The denominator of the first line reflects the resonant character
of the considered two-step process. From the resonance conditions (zeroes of the
denominator) we see that the double photoemission is enhanced when $a$ and $c$ are
continuum states and therefore we denote them as $\vec{k}_a$ and $\vec{k}_c$. We replace
the scattering states $|\chi_{\vec{k_1}}^{(-)}\rangle$ and $|\chi_{\vec{k_2}}^{(-)}\rangle$ entering the
matrix elements~\eqref{eq:me} by the plane-waves and perform the integration yielding
$f_{\vec{k}_a\vec{k}}^{\vec{q}}=\delta(\vec{k}-\vec{k}_a-\vec{q})$. Combining all together
we obtain the following concise expression for the plasmon-assisted DPE process:
\bea
J_{\vec{k_1}\vec{k_2}}&=&4\pi\!\sum_{\vec{k}_a\vec{k}_c}\!\!\sum_{bd}
\Delta_{\vec{k}_cb}\Delta_{b\vec{k}_a} A^{(2)}_{bd}(\varepsilon_{\vec{k_1}}+\varepsilon_{\vec{k_2}}-\omega)\nonumber\\
&\times&\frac{\langle \vec{k_1}+ \vec{k_2}- \vec{k}_a|d\rangle\langle d| \vec{k_1}+ \vec{k_2}- \vec{k}_c\rangle}
{(\varepsilon_{\vec{k}_c}+\varepsilon_{d}-\varepsilon_{\vec{k_1}}-\varepsilon_{\vec{k_2}})
(\varepsilon_{\vec{k_1}}+\varepsilon_{\vec{k_2}}-\varepsilon_{\vec{k}_a}-\varepsilon_{d})}\nonumber\\
&\times&W(\vec{k_1}-\vec{k}_c,\varepsilon_d-\varepsilon_{\vec{k_2}})
W(\vec{k_1}-\vec{k}_a,\varepsilon_d-\varepsilon_{\vec{k_2}}).
\label{eq:Jkk_fin2}
\eea
We have seen that the plane-wave approximation for the scattering states (i.~e. the
M{\o}ller operator is given by the identity operator) results in a great simplification
for the two-particle current: it is given by a sum over two bound states (they correspond
to two lesser propagators in the diagrammatic representation of this process) and by the
two momentum integrals corresponding to the propagators of the secondary electron. In
contrast, in the full-fledged calculations based on Eq.~\eqref{eq:Jkk_fin} the momenta of
the secondary electron and the emitted electrons are not rigidly related. Therefore, in
general two additional momentum integrations are required. This will be the subject of a
forthcoming publication where this formalism is applied to a large molecular system.

The DPE process described by Eq.~\eqref{eq:Jkk_fin2} is suited to probe the plasmon
dispersion and damping. First, let us look at the classical plasmon that carries vanishing
momentum and otherwise is strongly damped. This leads us to consider the case
$\vec{k}_a\approx\vec{k}_c\approx\vec{k}_1$, and
$\varepsilon_d-\varepsilon_{\vec{k_2}}=\omega_\mathrm{p}$ is the condition for the plasmon
resonance. In this case the second line reduces to
$|\langle\vec{k_2}|d\rangle|^2/\omega_{\mathrm{p}}^2$, and is clearly off-resonance. The
situation greatly changes if we allow for the plasmon to carry \emph{finite momentum}
$q_\mathrm{c}$ and consider a large momentum of the secondary electron
$\vec{k}_a\approx\vec{k}_c\approx\vec{k}_1>\sqrt{\omega_\mathrm{p}}$. For simplicity take
a symmetric situation when both screened interaction lines carry approximately the same
energy and momentum and denote
$\vec{K}\approx\frac12(\vec{k}_a+\vec{k}_1)\approx\frac12(\vec{k}_c+\vec{k}_1)$ and
$\vec{q}\approx\vec{k_a}-\vec{k}_1\approx\vec{k_c}-\vec{k}_1$. In this case one achieves
the resonant enhancement when
\[
\varepsilon_{\vec{k}_a}-\varepsilon_{\vec{k}_1}\approx
\varepsilon_{\vec{k}_c}-\varepsilon_{\vec{k}_1}=2 (\vec{q}\cdot\vec{K})=\omega_\mathrm{p}.
\]
Thus for colinear $\vec{k}_a$, $\vec{k}_c$ and $\vec{k}_1$ the probability for the
plasmon-assisted emission of the secondary electron is enhanced when $K$ reaches the value
of $\omega_\mathrm{p}/q_\mathrm{c}$.

In order to illustrate the features arising due to the plasmon-assisted process in an
experiment, we computed the current for a simple model system. To be concrete, we consider
the basic jellium model for the C$_{60}$ molecule (treated as spherically symmetric)
\cite{madjet_photoionization_2008, pavlyukh_kohn-sham_2010}, which is known for its
pronounced (dipolar) plasmon resonance at $\omega_p \sim 22$~eV. Inserting a smoothed
box-like potential as approximation to the Kohn-Sham potential, we solved the
Schr\"{o}dinger equation for the 120 orbitals required (240 electrons in total). This
procedure yields the single-particle energies $\varepsilon_d$ associated to the orbitals
$\phi_d(\vec r)$, from which we can compute all quantities in
Eq.~\eqref{eq:Jkk_fin2}. Because of the spherical symmetry, we can separate the radial and
the angular dependence, that is $\phi_d(\vec r) = \frac{u_d(r)}{r} Y_{\ell_d m_d}(\hat r)$
($Y_{\ell m}(\hat r)$ are the spherical harmonics) and only solve the radial
Schr\"{o}dinger equation. For the optical matrix elements, we choose the length gauge and
assume a linear polarization along the $z$ axis ($\hat \Delta = z$). Since we are not
interested in the absolute scale a prefactor proportional to the field strength will not
be included. The matrix elements $\Delta_{\vec k b}$ attain the form
\begin{align*}
\Delta_{\vec k b} &= 4\pi \sum_{\ell m} C_{\ell m \ell_b m_b}
s_{b\ell}(k) Y_{\ell m}(\hat{k}), \\ s_{b\ell}(k) &= \int^\infty_0 d r
\, r^2 u_b(r) j_\ell(k r) ,
\end{align*}
where $j_\ell$ denotes the spherical Bessel function. The coefficients $C_{\ell m \ell_b
  m_b}$ are obtained from the standard Clebsch-Gordan
algebra~\cite{edmonds_angular_1996,varshalovich_quantum_1988}. Similarly, the
Fourier-transformed orbitals $\langle \vec k | d \rangle = \widetilde{\phi}_d(\vec k)$ can
be expressed in terms of the Bessel transformation: $\widetilde{\phi}_d(\vec k) = 4\pi
\widetilde{u}_d(k) Y_{\ell_d m_d}(\hat k)$ with $\widetilde{u}_d(k) =\int^\infty_0 d r \,
r u_d(r) j_{\ell_d}(k r)$.

Next we transform the summation over $\vec{k}_a$ and $\vec{k}_c$ into integrations and
substitute them by the integration over the momentum transfer vectors $\vec q_{a,c} =
\vec{k}_1 - \vec{k}_{a,c}$. At this stage, no further simplification can be made, such
that the six-dimensional integral has to be evaluated. However, it is reasonable to
consider $\vec q_{a,c} $ as small, since the plasmon branch enters the particle-hole
continuum for growing momentum, where it is strongly damped. Hence, we introduce the
momentum cutoff $q_\mathrm{max}$ and assume $k_1,k_2 \gg q_\mathrm{max}$.  Thus, we
approximate $\Delta_{\vec {k}_{a,c} b} = \Delta_{\vec {k}_1 - \vec {q}_{a,c} b} \approx
\Delta_{\vec {k}_1 b}$ and $\widetilde{\phi}_d(\vec{k}_1 + \vec{k}_2 - \vec{k}_{a,c})
=\widetilde{\phi}_d(\vec{k}_2 + \vec{q}_{a,c}) \approx \widetilde{\phi}_d(\vec{k}_2)
$. Furthermore, we integrate over the spherical angles of $\vec{k}_1$ and $\vec{k}_2$,
keeping only the dependence on their magnitude. Thus, the two-electron current can be
written as
\begin{equation}
\label{eq:modeljk1k2}
\begin{split}
J_{k_1,k_2} &\propto \sum_{bd} \sum_{\ell m} \left| C_{\ell m \ell_b m_b}
  s_{b\ell}(k_1) \right|^2 \left| \widetilde{u}_d(k_2) \right|^2 \\
&\times \left(
  1+ \Re\frac{\omega^2_p}{(\varepsilon_d-\varepsilon_{k_2} - i \Gamma)^2
    - \omega^2_p} \right)^2 F_d(k_1,k_2) ,
\end{split}
\end{equation}
where
\[
F_d(k_1,k_2) = \left(\int^{q_\mathrm{max}}_0 d q \frac{1}{q^2+2k_1 q -k^2_2 + 2
\varepsilon_d}\right)^2 .
\]
Note that we inserted the imaginary shift $i \Gamma$ in the energy argument accounting for
a finite width (lifetime in the time domain) of the plasmon resonance (which is assumed
dispersionless for simplicity).

\begin{figure}[t]
\centering
\includegraphics[width=\columnwidth]{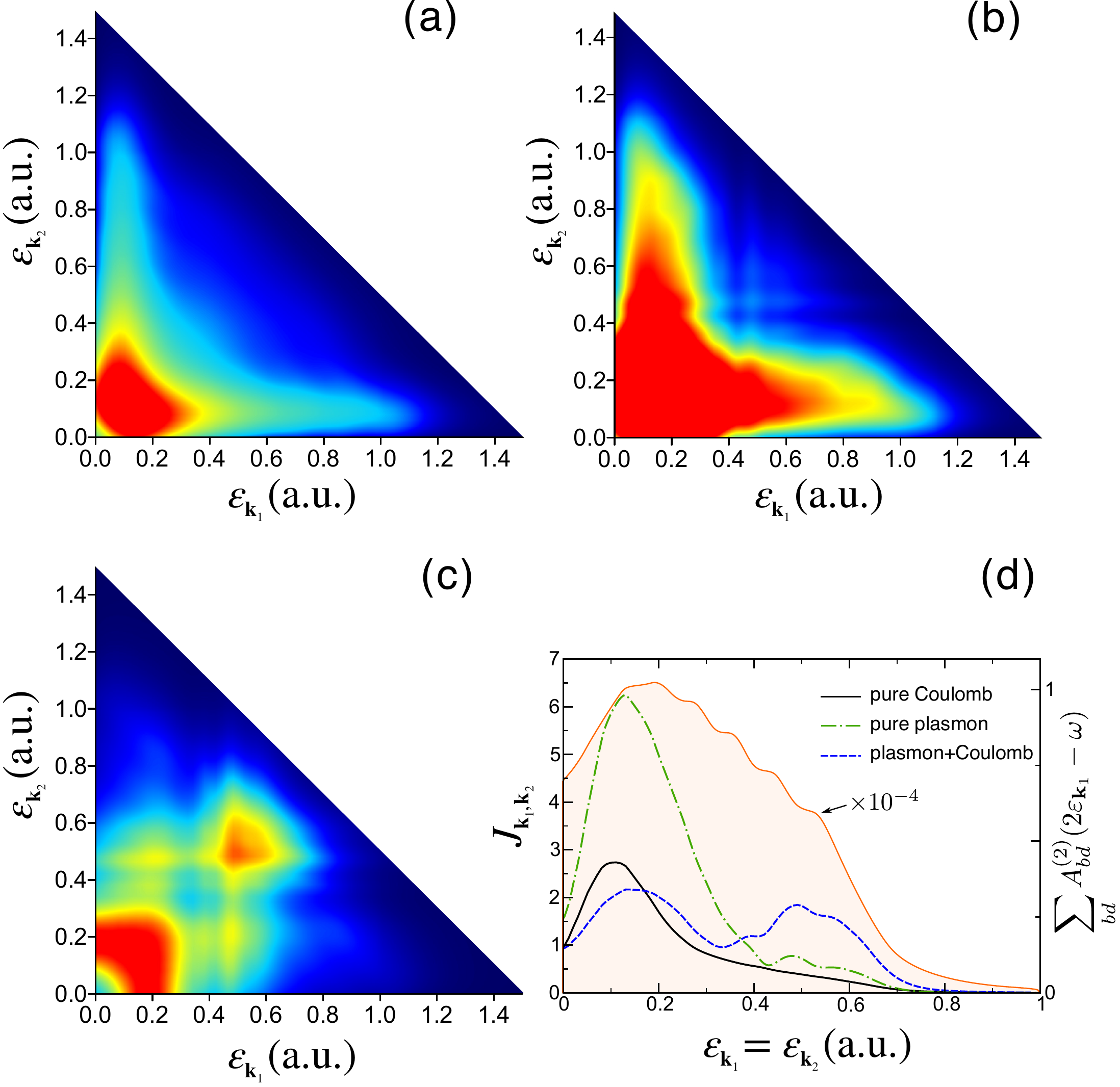}
\caption{(Color online) The symmetrized two-electron current as a function of the
  photoelectron energies (energy-sharing diagram) for typical parameters: $\omega = 2.0$
  and $\omega_p=0.8$. The color scale is the same for all three panels and runs from dark
  blue to red, indicating increasing values. (a) The process is mediated by the pure
  Coulomb interaction. (b) Pure plasmonic contribution. (c) Total (bare Coulomb and
  plasmonic contributions) signal including the interference terms. (d) Equal energy
  sharing ($\varepsilon_{\vec{k}_1} = \varepsilon_{\vec{k}_2}$) for the current and trace
  of the two-particle spectral density (shaded curve). \label{fig:je1e2maps}}
\end{figure}

In an experiment, the distinction between primary ($\vec{k}_1$) and secondary electron
($\vec{k}_2$) is, of course, not possible. For this reason, the photo-current needs to be
symmetrized (let us denote it by $J^\mathrm{sym}$). Representing the $J^\mathrm{sym}$ as a
function of $\varepsilon_{\vec{k}_1}$ and $\varepsilon_{\vec{k}_2}$ yields the typical
energy-sharing diagrams (Fig.\,\ref{fig:je1e2maps}). Spectral properties of the system
(dominated by $A^{(2)}(\varepsilon)$) display themselves along the main diagonal, as only
the sum $\varepsilon_{\vec{k}_1}+\varepsilon_{\vec{k}_2}$ enters. Dominant scattering
events mediated by the (screened) interaction on the other hand are visible along lines
$\varepsilon_{\vec{k}_1} = \mathrm{const.}$ (or $\varepsilon_{\vec{k}_2} =
\mathrm{const.}$). As Eq.~\eqref{eq:modeljk1k2}, indicates the two-particle current
contains contributions from (i) the bare Coulomb (two interacting lines in
Fig.\,\ref{fig:J1J2m}\,(b) are not screened), (ii) plasmonic scattering (both lines are
screened), and the interference terms. (i), (ii) and the total contribution is shown at
panels (a), (b) and (c) of Fig.\,\ref{fig:je1e2maps}, respectively. For vanishing $\Gamma$
the current is dominated by sharp plasmonic resonances. For finite damping parameter such
as used for the present simulations ($\Gamma=0.1$, we use a realistic value as in
Ref.~\cite{moskalenko_attosecond_2012}) the interference terms are important: we still
have a large plasmonic contribution (viz. Fig.\,\ref{fig:J1J2m}\,(b)), however, the bare
Coulomb contributes with the opposite sign. Therefore, in total current the large peak at
$\varepsilon_{\vec{k}_2} \approx 0.15$ becomes less pronounced and additional peaks at
higher energies (e.g. at $\varepsilon_{\vec{k}_2} \approx 0.5$) appear. The whole spectral
width of the signal is limited by the two-particle spectral function shown at
Fig.\,\ref{fig:J1J2m}\,(d) as a shade curve.

\section{Conclusions}
There is a large number of theoretical works devoted to the interaction of light and
matter which involves the emission of one or more electrons. This contribution is meant to
expose parallels between the single and the double electron photoemission in a formal
way. We started by defining corresponding observables and deriving expressions for one-
and two- particle currents based on the first-order time-dependent perturbation
theory. These expressions are suitable if exact formulas in terms of many-body states are
required. In order to obtain computationally useful expressions many-body effects should
also be accounted for in a perturbative fashion. Thus, in the first part of the manuscript
we applied the projection operator formalism. Starting from the explicit form of the
projection operators dividing the whole Hilbert space of the system into that of the
emitted electron(s) and the target we derived the effective one- and two-particle
Hamiltonian, discussed integral equations for the Green's functions describing emitted
particles and demonstrated a close connection of this formalism to the nonequilibrium
Green's function theory.  For the latter, one can easily derive the diagrammatic
expansions for one- and two-particle currents starting from the time-dependent
perturbation theory and using the adiabatic switching of the electron-electron
interaction. Hence, we have electromagnetic field switched on at the remote past (as
$e^{\eta t}$) and independently adiabatically switched on the interaction such that the
total Hamiltonian takes a form $\hat{H}_\delta=\hat{H}_0+e^{-\delta |t|}\hat{H}_1$. We
analyzed in details the diagrammatic structure of one- and two-particle currents. It is
surprisingly simple: one starts with the density-density response function $\chi^{<}$
which necessarily contains two blocks associated with the forward (``$-$'') and backward
(``$+$'') parts of the Keldysh contour. Requesting that one or two lines flowing from
``$-$'' to ``$+$'' blocks are associated with scattering states (with momenta $\vec{k}_i$)
one obtains exactly the diagrams for SPE and DPE currents showing the close connection
between these types of light-matter interaction. It is not difficult to generalize this
approach to an arbitrary number of particles. Finally, we presented a detailed analysis of
the plasmon-assisted DPE and showed that if one of the emitted particles is unobserved,
its diagrammatic representation reduces to the one describing external losses in the SPE
process considered by Caroli \emph{et al.}~\cite{caroli_inelastic_1973}. Plasmon pole
approximation was employed to derive computationally manageable expressions. We
illustrated the distinct features to be expected in an experiment by analyzing the simple
and yet realistic jellium model for the C$_{60}$ molecule. This will extended used in the
forthcoming paper devoted to the \emph{ab-initio} treatment of this large molecular
system.

\section*{Acknowledgments}
This work is supported by the DFG grants No. SFB\,762 and No. PA 1698/1-1.
\appendix
\section{Particle-impact ionization\label{ap:EI}}
Under some circumstances the formalism developed in the main text can be extended to other
mechanisms of ionization, e.g., particle-impact ionization. The basic requirement we
impose is the \emph{distinguishability} of the projectile from the target electrons. This applies also for
a  projectile electron  if the  impact energy is high  and the small momentum transfer is small (optical limit).

The target we describe by the Hamiltonian
Eq.~\eqref{eqhamiltonian2ndq2}. The Coulomb-interaction  between the projectile (with charge $Z$)
and the sample reads
\begin{equation}
  \label{eqvprojsys}
  \hat{V} = \frac{Z}{2} \sum_{ab} \sum_{\nu\mu}
  v_{a \nu b \mu} c^\dagger_a d^\dagger_{\nu} d_{\mu} c_b  \ .
\end{equation}
 $d_{\nu}$ ($d^\dagger_{\nu}$) is the annihilation (creation) operator of the
projectile states $|\nu\rangle$. These states can be chosen as the eigenstates of the
projectile Hamiltonian $\hat{h}_\mathrm{p}$ with energy $\varepsilon_\nu$.

Assuming that the projectile initially possesses the momentum $\vec{k}_i$, we can construct the
asymptotic state prior to the interaction (that is, at $t=-\infty$) as the product state
\begin{align*}
  |\Psi_{0,\vec{k}_i} \rangle = |\vec{k}_i\rangle\otimes |\Psi_0\rangle \ .
\end{align*}
  $\hat{V}$ is switched on reaching its  full
strength at $t=0$.  Assuming that this strength is still way smaller than the internal interaction within the sample,
 we can apply the first order perturbation theory (i.e., the first Born
approximation in the projectile-target interaction~\cite{joachain_quantum_1975}). Denoting  the full
Hamiltonian  by $\hat{H} + \hat{h}_\mathrm{p}$ one may write
\begin{equation}
  \label{eq:ptproj}
  |\tilde{\Psi}^{(+)}\rangle=|\Psi_{0,\vec{k}_i}\rangle+\lim_{\eta\rightarrow 0}\frac{1}
 {E_0+\varepsilon_{\vec{k}_i}-\hat{H}-\hat{h}_\mathrm{p}+i\eta}\hat{V}|\Psi_{0,\vec{k}_i} \rangle \ .
\end{equation}
The projectile has a well defined final momentum $\vec{k}_f$.  In analogy to
Sec.~\ref{sec:spe} we introduce the particle number operator
\begin{align*}
  \hat{N}_{\vec k} \rightarrow P_f \hat{N}_{\vec k} P_f
\end{align*}
with $P_f = |\vec{k}_f \rangle \langle \vec{k}_f |$ projecting only onto the projectile
space.  $\hat{N}_{\vec k}$ acts on the system's states only (including the ejected
electrons upon particle impact). Evaluating then the current as in Sec.~\ref{sec:basic}
and approximating the projectile states by plane waves $\langle \vec r| \vec k \rangle =
e^{i\vec{k}\cdot \vec{r}}$ yields
\begin{equation}
  \begin{split}
  J_{\vec k} &= \lim_{\eta\rightarrow0}2\eta
  \Big\langle \Psi_{0,\vec{k}_i}\Big|\hat{V}^\dagger\frac{1} {E_0 +\varepsilon_{\vec{k}_i}
    -\hat{H}-\hat{h}_\mathrm{p}-i\eta} P_f c^\dagger_{\vec k} c_{\vec k} P_f \\
  &\quad \frac{1} {E_0 +\varepsilon_{\vec{k}_i}
    -\hat{H}-\hat{h}_\mathrm{p}+i\eta} \hat{V} \Big|\Psi_{0,\vec{k}_i}\Big\rangle \\
  &= \lim_{\eta\rightarrow0}2\eta  \Big\langle \Psi_{0}\Big|\hat{V}^\mathrm{eff}(\vec q)^\dagger
  \frac{1} {E_0 +\varepsilon_{\vec{k}_i} - \varepsilon_{\vec{k}_f}
    -\hat{H}-i\eta} c^\dagger_{\vec k} c_{\vec k} \\ & \quad
  \frac{1} {E_0 +\varepsilon_{\vec{k}_i} - \varepsilon_{\vec{k}_f}
    -\hat{H}+i\eta} \hat{V}^{\mathrm{eff}}(\vec q)\Big|\Psi_0\Big\rangle \ ,
  \end{split}
\end{equation}
where $\vec q = \vec{k}_i - \vec{k}_f$ is the \emph{momentum transfer}, and
$\hat{V}^{\mathrm{eff}}(\vec q)$ is the effective single-particle operator acting on the
target, explicitly
\begin{equation}
  \label{eqveffkikf}
  \hat{V}^{\mathrm{eff}}(\vec{k}_i-\vec{k_f}) = \langle \vec{k}_i | \hat{V} | \vec{k}_f
  \rangle = \frac{Z}{2} \sum_{ab} v_{a \vec{k}_f b\vec{k}_i} c^\dagger_a c_b \ .
\end{equation}
In this optical limit 
\be
\hat{V}^{\mathrm{eff}}(\vec q)=\frac{4\pi Z}{q^2}e^{i \vec{q}\cdot\vec{r}}.
\ee
acts similar to the light-matter interaction $\hat{\Delta}$; the transferred energy (or
\emph{energy loss}) $\varepsilon_{\vec{k}_i} - \varepsilon_{\vec{k}_f}$ resembles the
photon energy.
\section{Green's functions\label{ap:GF}}
Let us recast the following many-body correlators from  Sec.~\ref{sec:spe}
\[
G^{(\mathrm{p})}_{\vec{p}\vec{q},\alpha}(z)=\big\langle
\Psi_\alpha^+\big|c_\vec{p}\frac{1}{z-\hat{H}_p-\hat{\Sigma}_p(z)}c_{\vec q}^\dagger\big|\Psi_\alpha^+\big\rangle,
\]
in the form of one-particle averages. We define the particle propagator of one-particle
system in the presence of optical potential $\hat{W}(z)$:
\[
\mathcal{G}_{\vec{p}\vec{q}}(z)=\big\langle
\vec{p}\big|\frac{1}{z-\hat{H}_f-\hat{W}(z)}\big|\vec{q}\big\rangle.
\]
Consider $G^{(\mathrm{p})}_{\vec{p}\vec{q},\alpha}(\omega+E_0\pm i\eta)$. The matrix element of the
effective Hamiltonian operator in its definition can be simplified to
\[
\langle \Psi_\alpha^+|c_\vec{p}(\hat{H}_P+\hat{\Sigma}_{P}(z))c_\vec{q}^\dagger|\Psi_\alpha^+\rangle
=E_\alpha^+ + \langle \vec{p} |\hat{H}_{f}+\hat{W}_\alpha(z)|\vec{q}\rangle,
\]
where we decompose the total $N$-particle Hamiltonian $H$ as a sum of three terms:
\[
\hat H=\hat{H}_f+\hat{H}^++ \hat{V}.
\]
Here $\hat{H}_f$ is the free-particle Hamiltonian, $\hat{H}^+$ is the Hamiltonian of ionized system:
\[
\hat{H}^+|\Psi_\alpha^+\rangle=E_\alpha^+|\Psi_\alpha^+\rangle,
\]
and $\hat{V}$ is the frequency independent part of the self-energy.  If the optical
potential is identified with the self-energy then one we can relate two propagators
\[
G^{(\mathrm{p})}_{\vec{p}\vec{q},\alpha}(\omega+E_0\pm i\eta)
=\mathcal{G}^{(\pm)}_{\vec{p}\vec{q},\alpha}(\omega+\varepsilon_\alpha),
\]
where we introduced the Green's functions
$\mathcal{G}^{(\pm)}_{\vec{p}\vec{q}}(\omega)=\mathcal{G}^{(\pm)}_{\vec{p}\vec{q}}(\omega\pm
i\eta)$ and $\varepsilon_\alpha=E_0-E_\alpha^+$.  From the formal scattering theory (see
Sec.~20 of Joachain~\cite{joachain_quantum_1975}) and independent of the concrete choice
of the representation we can express them in terms of the M{\o}ller operator and the
free-particle Green's function
\begin{equation}
\mathcal{G}^{\pm}(\omega)=\hat{\Omega}^{(\pm)}\mathcal{G}^{(\pm)}_0(\omega).
\label{eq:GO}
\end{equation}
\paragraph{two-particle case:}
For DPE the two-particle Green's function over the excited state $\Psi_\beta^{2+}$ is
required:
\[
G_{\vec{p}\vec{q},\vec{k_1}\vec{k_2},\beta}^{(\mathrm{pp})}(z)=\big\langle
\Psi_\beta^{2+}\big|c_\vec{p}c_\vec{q}\frac{1}{z-\hat{H}_p-\hat{\Sigma}_P(z)}
c_{\vec{k_1}}^\dagger c_{\vec{k_2}}^\dagger\big|\Psi_\beta^{2+}\big\rangle,
\]
where the projection operator is defined by Eq.~\eqref{eqprojDPE}. This propagator can be
related to the scattering Green's function of the two-particle system in the presence of
the optical potential of doubly-ionized target:
\[
G^{(\mathrm{pp})}_{\vec{p}\vec{q},\vec{k_1}\vec{k_2},\beta}(\omega+E_0\pm i\eta)
=\mathcal{G}^{(\pm)}_{\vec{p}\vec{q},\vec{k_1}\vec{k_2},\beta}(\omega+\varepsilon^{(2)}_\beta),
\]
with
$\varepsilon^{(2)}_\beta=E_0-E_\beta^{2+}$. $\mathcal{G}^{(\pm)}_{\vec{p}\vec{q},\vec{k_1}\vec{k_2},\beta}$
can be likewise expressed in the form~\eqref{eq:GO}.
\section{Matrix identities\label{sec:matrix_id}}
The formalism of presented here works in finite as well as in infinite-dimensional Hilbert
spaces. For illustration we formulate it in the matrix form. Given $\mathcal M$ is square
block matrix:
\begin{equation}
\mathcal M=\left[
\begin{array}{cc}
\mathcal{A}&\mathcal{B}\\
\mathcal{C}&\mathcal{D}
\end{array}
\right],
\label{eq:M}
\end{equation}
where $\mathcal D$ is square invertible matrix the Schur
complement~\cite{zhang_schur_2005} (also known in physics as the Feshbach
map~\cite{feshbach_unified_1962,capuzzi_projection_1996,escher_one-body_2002}) is defined
as:
\[
\tilde{\mathcal{A}}=\mathcal{A}-\mathcal{B}\mathcal{D}^{-1}\mathcal{C}.
\]
We might think of $\mathcal M$ as a Hamiltonian operator acting in some larger Hilbert
space, whereas $\mathcal A$ is the same operator, but acting in a physically relevant
subspace. Be $P$ the projection operator onto this subspace ($P\mathcal{M}P=\mathcal{A}$)
and $Q=I-P$ is its complement ($Q\mathcal{M}Q=\mathcal{D}$). For definiteness we may take
$\mathcal M$ to be a compact self-adjoint operator on the Hilbert space describing an
$N$-fermion system $\mathcal{H}^{(N)}$ and $\mathcal{A}$ its projection upon the Hilbert
space of two particles $\mathcal{H}^{(2)}$. Because of the couplings between subspaces
(for physical Hamiltonians obviously holds $\mathcal B=\mathcal C^\dagger$) $\mathcal{M}$
and $\mathcal{A}$ have different spectral properties. Nonetheless, one \emph{can} show the
following equivalence
\begin{equation}
\mathcal{M}V=0\quad \Longleftrightarrow \quad \tilde{\mathcal{A}}PV=0,
\label{eq:Fmap}
\end{equation}
for a vector $V\in \mathcal{H}^{(N)}$. If $\mathcal{M}\equiv H-EI$ the first part implies that $V$
is an eigenvector of $H$ with the energy $E$. The second part implies that $PV$ is a
corresponding eigenvector of $\tilde{\mathcal{A}}(E)$ with the same energy:
\begin{equation}
(H_P+\Sigma_P(E)-EI_P)PV=0.
\label{eq:Heff}
\end{equation}
Expression for the self-energy~\eqref{eq:Sgm} is derived for instance in Sec.~20.2.3 of
Joachain~\cite{joachain_quantum_1975}.  A mathematically rigorous proof of the
theorem~\eqref{eq:Fmap} as well as other properties of the Feshbach-Schur map can be found
in Chap.~11 of Gustafson and Sigal~\cite{gustafson_mathematical_2011}.  It is further
possible to write the inverse of the matrix $\mathcal M$ explicitly~\footnote{According to
  Zhang~\cite{zhang_schur_2005} it was a Polish astronomer Banachiewicz who obtained this
  formula for the first time. However, it was reinvented many times (see a short
  historical review at the top of p. 699 of Ref.~\cite{jensen_fermi_2006} where the
  authors suggest to use the name Schur-Livsic-Feshbach-Grushin for the equation)}:
\begin{equation}
\mathcal{M}^{-1}=\left[
\begin{array}{cc}
\tilde{\mathcal{A}}^{-1}&-\tilde{\mathcal{A}}^{-1}\mathcal{B}\mathcal{D}^{-1}\\
-\mathcal{D}^{-1}\mathcal{C}\tilde{\mathcal{A}}^{-1}&
\mathcal{D}^{-1}+\mathcal{D}^{-1}\mathcal{C}\tilde{\mathcal{A}}^{-1}\mathcal{B}\mathcal{D}^{-1}
\end{array}
\right].
\label{eq:Ban}
\end{equation}
This identity is natural to apply to compute resolvents. For instance, Eq.~\eqref{eq:res}
is given the first line of Eq.~\eqref{eq:Ban}. This formula can also be found in Almbladh
as Eq.~(19)~\cite{almbladh_theory_1985}.

\section{Properties of projection operators\label{sec:idem}}
Our basic assumptions for operators with continuum indices
$c_\vec{p}|\Psi_\alpha^{+}\rangle=0$ and $c_\vec{p}|\Psi_\beta^{2+}\rangle=0$ imply that
final states of the target are the vacuum states for these operators. Thus, standard
Wick's theorem can be used for the calculation of various correlators. It follows
%
\bea c_\vec{p}c_\vec{q}^\dagger
|\Psi_\alpha^{+}\rangle&=&\delta_{\vec{p}\vec{q}}|\Psi_\alpha^{+}\rangle,\\ c_\vec{k_2}
c_\vec{k_1}c_\vec{p}^\dagger c_\vec{q}^\dagger|\Psi_\beta^{2+}\rangle&=& (\delta_{\vec
  k_1\vec p}\delta_{\vec k_2\vec q}-\delta_{\vec k_1\vec q}\delta_{\vec k_2\vec p})
|\Psi_\beta^{2+}\rangle.
\eea
%
These equations lead to the indempontency relations $P_\alpha P_\alpha=P_\alpha$ and
$P_\beta P_\beta=P_\beta$ and to the properties
%
\bea
c_\vec{k}^\dagger |\Psi_\alpha^{+}\rangle
\langle \Psi_\alpha^{+}|c_\vec{k}
&=&P_\alpha c_\vec{k}^\dagger c_\vec{k}P_\alpha,\\
c_\vec{k_1}^\dagger c_\vec{k_2}^\dagger |\Psi_\beta^{2+}\rangle
\langle \Psi_\beta^{2+}|c_\vec{k_2} c_\vec{k_1}
&=&P_\beta c_\vec{k_1}^\dagger c_\vec{k_2}^\dagger c_\vec{k_2} c_\vec{k_1}P_\beta.
\eea
%
The matrix element of a one-particle operator $\hat{O}=\hat{O}(x_1)+\hat{O}(x_2)$ over the
determinant two-particle states $\langle\x_1x_2|ab\rangle
=\frac1{\sqrt2}(\phi_a(x_1)\phi_b(x_2)-\phi_b(x_1)\phi_a(x_2))$
can be verified by direct evaluation:
%
\bea
\langle ab|\hat O|cd\rangle&=&\bra a|\hat O |c\ket\delta_{bd}+\bra b|\hat O |d\ket\delta_{ac}\nonumber\\
&-&\bra a|\hat O |d\ket\delta_{bc}-\bra b|\hat O |c\ket\delta_{ad}.
\eea
%
If one of the states  is a two-hole Dyson orbital the matrix element is computed similarly:
%
\bea
\lefteqn{\langle ab|\hat O|\varphi_\beta^{(2)}\rangle
=\frac12 \sum_{cd}\langle ab|\hat O|cd \rangle\big\langle\Psi^{2+}_\beta|
c_cc_d|\Psi_0\big\rangle}\nonumber\\
&=&\sum_{cd}\big(\bra a|\hat O |c\ket\delta_{bd}-\bra b|\hat O |c\ket\delta_{ad}\big)\big\langle\Psi^{2+}_\beta|
c_cc_d|\Psi_0\big\rangle.
\eea
%
Using this result and the vacuum assumption for the initial states we can compute a matrix
element entering the Fermi golden rule formula for SPE:
\begin{multline}
\big\langle \Psi^{+}_\alpha|c_\vec{k}\hat \Delta|\Psi_0\big\rangle=
\sum_{ab}\Delta_{ab}\big\langle \Psi^{+}_\alpha|c_\vec{k}c^\dagger_ac_b|\Psi_0\big\rangle\\
\approx\sum_{b}\langle \vec{k}|\hat \Delta|b\rangle
\big\langle \Psi^{+}_\alpha|c_b|\Psi_0\big\rangle
=\big\langle \vec{k}|\hat \Delta|\phi_\alpha\big\rangle,\label{eq:me_spe}
\end{multline}
and DPE:
\begin{multline}
\big\langle \Psi^{2+}_\beta|c_\vec{k_1}c_\vec{k_2}\hat \Delta|\Psi_0\big\rangle=
\sum_{ab}\Delta_{ab}\big\langle \Psi^{2+}_\beta|c_\vec{k_1}c_\vec{k_2}c^\dagger_ac_b|\Psi_0\big\rangle\\
\approx\sum_{bc}\big[\langle \vec{k_1}|\hat \Delta|b\rangle\delta_{\vec{k_2}c}-
\langle \vec{k_2}|\hat \Delta|b\rangle\delta_{\vec{k_1}c}\big]
\big\langle \Psi^{2+}_\beta|c_bc_c|\Psi_0\big\rangle\\
=\big\langle \vec{k_1}\vec{k_2}|\hat \Delta|\phi_\beta^{(2)}\big\rangle.
\label{eq:me_dpe}
\end{multline}
We used an assumption $c_\vec{k}|\Psi_0\rangle \approx 0$ to derive \eqref{eq:me_spe}
and $c_\vec{k_1}c_\vec{k_2}|\Psi_0\rangle \approx 0$ to derive \eqref{eq:me_dpe}.
\section{Sokhotski-Plemelj-type identities\label{ap:Sokh-Plem}}
Following identities were used to perform frequency integrations leading to
Eqs.~(\ref{eq:Jk_Pl},\ref{eq:Jkk_Pl}).
\bea
\lim_{\eta\rightarrow0}\lim_{\delta\rightarrow0}2\eta\frac{1}{\omega_1-z_1-i\eta}\frac{1}{\omega_2-z_2+i\eta}
\frac{1}{z_3-\omega_3-i\delta}\nonumber\\\times\frac{1}{z_3+z_2-z_1-\omega_3+i\delta}
=\prod_{i=1}^3 2\pi\delta(z_i-\omega_i),
\eea
for  $\omega_1=\omega_2$, and
\bea
\lim_{\eta\rightarrow0}\lim_{\delta\rightarrow0}2\eta\frac{1}{\omega_1-z_1-i\eta}\frac{1}{\omega_2-z_2+i\eta}
\frac{1}{z_3-z_2+\omega_2-\omega_3-i\delta}\nonumber\\\times\frac{1}{z_4-z_1+\omega_1-\omega_4+i\delta}
\frac{1}{\omega_4+\omega_5-z_4-z_5+i\delta}\nonumber\quad\\
\times\frac{1}{\omega_3+\omega_5-z_3-z_5-i\delta}
=\prod_{i=1}^5 2\pi\delta(z_i-\omega_i),\quad\quad\quad
\eea
for $\omega_1=\omega_2$, $\omega_3=\omega_4$. The first equation appears in
\cite{almbladh_theory_1985}. To the best of our knowledge the second equation has not been
addressed in the literature. These identities can be verified by the Fourier
transformation with respect to $z_i$ variables.  

\begin{thebibliography}{73}%
\makeatletter
\providecommand \@ifxundefined [1]{%
 \@ifx{#1\undefined}
}%
\providecommand \@ifnum [1]{%
 \ifnum #1\expandafter \@firstoftwo
 \else \expandafter \@secondoftwo
 \fi
}%
\providecommand \@ifx [1]{%
 \ifx #1\expandafter \@firstoftwo
 \else \expandafter \@secondoftwo
 \fi
}%
\providecommand \natexlab [1]{#1}%
\providecommand \enquote  [1]{``#1''}%
\providecommand \bibnamefont  [1]{#1}%
\providecommand \bibfnamefont [1]{#1}%
\providecommand \citenamefont [1]{#1}%
\providecommand \href@noop [0]{\@secondoftwo}%
\providecommand \href [0]{\begingroup \@sanitize@url \@href}%
\providecommand \@href[1]{\@@startlink{#1}\@@href}%
\providecommand \@@href[1]{\endgroup#1\@@endlink}%
\providecommand \@sanitize@url [0]{\catcode `\\12\catcode `\$12\catcode
  `\&12\catcode `\#12\catcode `\^12\catcode `\_12\catcode `\%12\relax}%
\providecommand \@@startlink[1]{}%
\providecommand \@@endlink[0]{}%
\providecommand \url  [0]{\begingroup\@sanitize@url \@url }%
\providecommand \@url [1]{\endgroup\@href {#1}{\urlprefix }}%
\providecommand \urlprefix  [0]{URL }%
\providecommand \Eprint [0]{\href }%
\providecommand \doibase [0]{http://dx.doi.org/}%
\providecommand \selectlanguage [0]{\@gobble}%
\providecommand \bibinfo  [0]{\@secondoftwo}%
\providecommand \bibfield  [0]{\@secondoftwo}%
\providecommand \translation [1]{[#1]}%
\providecommand \BibitemOpen [0]{}%
\providecommand \bibitemStop [0]{}%
\providecommand \bibitemNoStop [0]{.\EOS\space}%
\providecommand \EOS [0]{\spacefactor3000\relax}%
\providecommand \BibitemShut  [1]{\csname bibitem#1\endcsname}%
\let\auto@bib@innerbib\@empty
\bibitem [{\citenamefont {Cardona}\ and\ \citenamefont
  {Ley}(1978)}]{cardona_photoemission_1978}%
  \BibitemOpen
  \bibinfo {editor} {\bibfnamefont {M.}~\bibnamefont {Cardona}}\ and\ \bibinfo
  {editor} {\bibfnamefont {L.}~\bibnamefont {Ley}},\ eds.,\ \href@noop {}
  {\emph {\bibinfo {title} {Photoemission in {Solids} {I} {General}
  {Principles}}}}\ (\bibinfo  {publisher} {Springer},\ \bibinfo {address}
  {Berlin},\ \bibinfo {year} {1978})\BibitemShut {NoStop}%
\bibitem [{\citenamefont {H\"{u}fner}(2003)}]{hufner_photoelectron_2003}%
  \BibitemOpen
  \bibfield  {author} {\bibinfo {author} {\bibfnamefont {S.}~\bibnamefont
  {H\"{u}fner}},\ }\href@noop {} {\emph {\bibinfo {title} {Photoelectron
  spectroscopy: principles and applications}}},\ \bibinfo {edition} {3rd}\
  ed.,\ Advanced texts in physics\ (\bibinfo  {publisher} {Springer},\ \bibinfo
  {address} {Berlin; New York},\ \bibinfo {year} {2003})\BibitemShut {NoStop}%
\bibitem [{\citenamefont {Schattke}\ and\ \citenamefont
  {Van~Hove}(2002)}]{schattke_solid-state_2002}%
  \BibitemOpen
  \bibinfo {editor} {\bibfnamefont {W.}~\bibnamefont {Schattke}}\ and\ \bibinfo
  {editor} {\bibfnamefont {M.~A.}\ \bibnamefont {Van~Hove}},\ eds.,\ \href@noop
  {} {\emph {\bibinfo {title} {Solid-state photoemission and related methods:
  theory and experiment}}}\ (\bibinfo  {publisher} {Wiley-VCH},\ \bibinfo
  {address} {Weinheim},\ \bibinfo {year} {2002})\BibitemShut {NoStop}%
\bibitem [{\citenamefont {Schmidt}(1997)}]{schmidt_electron_1997}%
  \BibitemOpen
  \bibfield  {author} {\bibinfo {author} {\bibfnamefont {V.}~\bibnamefont
  {Schmidt}},\ }\href@noop {} {\emph {\bibinfo {title} {Electron spectrometry
  of atoms using synchrotron radiation}}},\ \bibinfo {series} {Cambridge
  monographs on atomic, molecular, and chemical physics}\ No.~\bibinfo {number}
  {6}\ (\bibinfo  {publisher} {Cambridge University Press},\ \bibinfo {address}
  {Cambridge; New York},\ \bibinfo {year} {1997})\BibitemShut {NoStop}%
\bibitem [{\citenamefont {Weigold}(1999)}]{weigold_electron_1999}%
  \BibitemOpen
  \bibfield  {author} {\bibinfo {author} {\bibfnamefont {E.}~\bibnamefont
  {Weigold}},\ }\href@noop {} {\emph {\bibinfo {title} {Electron momentum
  spectroscopy}}},\ Physics of atoms and molecules\ (\bibinfo  {publisher}
  {Kluwer Academic/Plenum Publishers},\ \bibinfo {address} {New York},\
  \bibinfo {year} {1999})\BibitemShut {NoStop}%
\bibitem [{\citenamefont {Berakdar}\ \emph {et~al.}(2003)\citenamefont
  {Berakdar}, \citenamefont {Lahmam-Bennani},\ and\ \citenamefont
  {Dal~Cappello}}]{berakdar_electron-impact_2003}%
  \BibitemOpen
  \bibfield  {author} {\bibinfo {author} {\bibfnamefont {J.}~\bibnamefont
  {Berakdar}}, \bibinfo {author} {\bibfnamefont {A.}~\bibnamefont
  {Lahmam-Bennani}}, \ and\ \bibinfo {author} {\bibfnamefont {C.}~\bibnamefont
  {Dal~Cappello}},\ }\href@noop {} {\bibfield  {journal} {\bibinfo  {journal}
  {Phys. Rep.}\ }\textbf {\bibinfo {volume} {374}},\ \bibinfo {pages} {91}
  (\bibinfo {year} {2003})}\BibitemShut {NoStop}%
\bibitem [{\citenamefont {Manson}\ and\ \citenamefont
  {Starace}(1982)}]{manson_photoelectron_1982}%
  \BibitemOpen
  \bibfield  {author} {\bibinfo {author} {\bibfnamefont {S.~T.}\ \bibnamefont
  {Manson}}\ and\ \bibinfo {author} {\bibfnamefont {A.~F.}\ \bibnamefont
  {Starace}},\ }\href@noop {} {\bibfield  {journal} {\bibinfo  {journal} {Rev.
  Mod. Phys.}\ }\textbf {\bibinfo {volume} {54}},\ \bibinfo {pages} {389}
  (\bibinfo {year} {1982})}\BibitemShut {NoStop}%
\bibitem [{\citenamefont {Amusia}(1990)}]{amusia_atomic_1990}%
  \BibitemOpen
  \bibfield  {author} {\bibinfo {author} {\bibfnamefont {M.~Y.}\ \bibnamefont
  {Amusia}},\ }\href@noop {} {\emph {\bibinfo {title} {Atomic photoeffect}}},\
  Physics of atoms and molecules\ (\bibinfo  {publisher} {Plenum Press},\
  \bibinfo {address} {New York},\ \bibinfo {year} {1990})\BibitemShut {NoStop}%
\bibitem [{\citenamefont {Eland}(2009)}]{eland_dynamics_2009}%
  \BibitemOpen
  \bibfield  {author} {\bibinfo {author} {\bibfnamefont {J.~H.~D.}\
  \bibnamefont {Eland}},\ }in\ \href@noop {} {\emph {\bibinfo {booktitle}
  {Advances in {Chemical} {Physics}}}},\ \bibinfo {editor} {edited by\ \bibinfo
  {editor} {\bibfnamefont {S.~A.}\ \bibnamefont {Rice}}}\ (\bibinfo
  {publisher} {John Wiley \& Sons, Inc.},\ \bibinfo {year} {2009})\ pp.\
  \bibinfo {pages} {103--151}\BibitemShut {NoStop}%
\bibitem [{\citenamefont {Kadanoff}\ and\ \citenamefont
  {Baym}(1962)}]{kadanoff_quantum_1962}%
  \BibitemOpen
  \bibfield  {author} {\bibinfo {author} {\bibfnamefont {L.}~\bibnamefont
  {Kadanoff}}\ and\ \bibinfo {author} {\bibfnamefont {G.}~\bibnamefont
  {Baym}},\ }\href@noop {} {\emph {\bibinfo {title} {Quantum statistical
  mechanics {Green}'s function methods in equilibrium and nonequilibrium
  problems}}}\ (\bibinfo  {publisher} {W.A. Benjamin},\ \bibinfo {address} {New
  York},\ \bibinfo {year} {1962})\BibitemShut {NoStop}%
\bibitem [{\citenamefont {Langreth}(1976)}]{devreese_linear_1976}%
  \BibitemOpen
  \bibfield  {author} {\bibinfo {author} {\bibfnamefont {D.~C.}\ \bibnamefont
  {Langreth}},\ }in\ \href@noop {} {\emph {\bibinfo {booktitle} {Linear and
  {Nonlinear} {Electron} {Transport} in {Solids}}}},\ \bibinfo {series} {{NATO}
  {Advanced} {Study} {Institutes} {Series}}, Vol.~\bibinfo {volume} {17},\
  \bibinfo {editor} {edited by\ \bibinfo {editor} {\bibfnamefont
  {J.}~\bibnamefont {Devreese}}\ and\ \bibinfo {editor} {\bibfnamefont
  {V.}~\bibnamefont {Doren}}}\ (\bibinfo  {publisher} {Springer US},\ \bibinfo
  {year} {1976})\ pp.\ \bibinfo {pages} {3--32}\BibitemShut {NoStop}%
\bibitem [{\citenamefont {Danielewicz}(1984)}]{danielewicz_quantum_1984}%
  \BibitemOpen
  \bibfield  {author} {\bibinfo {author} {\bibfnamefont {P.}~\bibnamefont
  {Danielewicz}},\ }\href@noop {} {\bibfield  {journal} {\bibinfo  {journal}
  {Ann. Phys.}\ }\textbf {\bibinfo {volume} {152}},\ \bibinfo {pages} {239}
  (\bibinfo {year} {1984})}\BibitemShut {NoStop}%
\bibitem [{\citenamefont {Stefanucci}\ and\ \citenamefont {van
  Leeuwen}(2013)}]{stefanucci_nonequilibrium_2013}%
  \BibitemOpen
  \bibfield  {author} {\bibinfo {author} {\bibfnamefont {G.}~\bibnamefont
  {Stefanucci}}\ and\ \bibinfo {author} {\bibfnamefont {R.}~\bibnamefont {van
  Leeuwen}},\ }\href@noop {} {\emph {\bibinfo {title} {Nonequilibrium
  {Many}-{Body} {Theory} of {Quantum} {Systems}: {A} {Modern}
  {Introduction}}}}\ (\bibinfo  {publisher} {Cambridge University Press},\
  \bibinfo {address} {Cambridge},\ \bibinfo {year} {2013})\BibitemShut
  {NoStop}%
\bibitem [{\citenamefont {Wohlfarth}\ and\ \citenamefont
  {Cederbaum}(2002)}]{wohlfarth_ionization_2002}%
  \BibitemOpen
  \bibfield  {author} {\bibinfo {author} {\bibfnamefont {M.~N.~R.}\
  \bibnamefont {Wohlfarth}}\ and\ \bibinfo {author} {\bibfnamefont {L.~S.}\
  \bibnamefont {Cederbaum}},\ }\href@noop {} {\bibfield  {journal} {\bibinfo
  {journal} {Phys. Rev. A}\ }\textbf {\bibinfo {volume} {65}},\ \bibinfo
  {pages} {052703} (\bibinfo {year} {2002})}\BibitemShut {NoStop}%
\bibitem [{\citenamefont {Inglesfield}(1983)}]{inglesfield_plasmon_1983}%
  \BibitemOpen
  \bibfield  {author} {\bibinfo {author} {\bibfnamefont {J.~E.}\ \bibnamefont
  {Inglesfield}},\ }\href@noop {} {\bibfield  {journal} {\bibinfo  {journal}
  {J. Phys. C}\ }\textbf {\bibinfo {volume} {16}},\ \bibinfo {pages} {403}
  (\bibinfo {year} {1983})}\BibitemShut {NoStop}%
\bibitem [{\citenamefont {Almbladh}(1985)}]{almbladh_theory_1985}%
  \BibitemOpen
  \bibfield  {author} {\bibinfo {author} {\bibfnamefont {C.-O.}\ \bibnamefont
  {Almbladh}},\ }\href@noop {} {\bibfield  {journal} {\bibinfo  {journal}
  {Phys. Scr.}\ }\textbf {\bibinfo {volume} {32}},\ \bibinfo {pages} {341}
  (\bibinfo {year} {1985})}\BibitemShut {NoStop}%
\bibitem [{\citenamefont {Berglund}\ and\ \citenamefont
  {Spicer}(1964)}]{berglund_photoemission_1964}%
  \BibitemOpen
  \bibfield  {author} {\bibinfo {author} {\bibfnamefont {C.~N.}\ \bibnamefont
  {Berglund}}\ and\ \bibinfo {author} {\bibfnamefont {W.~E.}\ \bibnamefont
  {Spicer}},\ }\href@noop {} {\bibfield  {journal} {\bibinfo  {journal} {Phys.
  Rev.}\ }\textbf {\bibinfo {volume} {136}},\ \bibinfo {pages} {A1030}
  (\bibinfo {year} {1964})}\BibitemShut {NoStop}%
\bibitem [{\citenamefont {Mahan}(1970)}]{mahan_theory_1970}%
  \BibitemOpen
  \bibfield  {author} {\bibinfo {author} {\bibfnamefont {G.~D.}\ \bibnamefont
  {Mahan}},\ }\href@noop {} {\bibfield  {journal} {\bibinfo  {journal} {Phys.
  Rev. B}\ }\textbf {\bibinfo {volume} {2}},\ \bibinfo {pages} {4334} (\bibinfo
  {year} {1970})}\BibitemShut {NoStop}%
\bibitem [{\citenamefont {Schaich}\ and\ \citenamefont
  {Ashcroft}(1971)}]{schaich_model_1971}%
  \BibitemOpen
  \bibfield  {author} {\bibinfo {author} {\bibfnamefont {W.~L.}\ \bibnamefont
  {Schaich}}\ and\ \bibinfo {author} {\bibfnamefont {N.~W.}\ \bibnamefont
  {Ashcroft}},\ }\href@noop {} {\bibfield  {journal} {\bibinfo  {journal}
  {Phys. Rev. B}\ }\textbf {\bibinfo {volume} {3}},\ \bibinfo {pages} {2452}
  (\bibinfo {year} {1971})}\BibitemShut {NoStop}%
\bibitem [{\citenamefont {Langreth}(1971)}]{langreth_scattering_1971}%
  \BibitemOpen
  \bibfield  {author} {\bibinfo {author} {\bibfnamefont {D.~C.}\ \bibnamefont
  {Langreth}},\ }\href@noop {} {\bibfield  {journal} {\bibinfo  {journal}
  {Phys. Rev. B}\ }\textbf {\bibinfo {volume} {3}},\ \bibinfo {pages} {3120}
  (\bibinfo {year} {1971})}\BibitemShut {NoStop}%
\bibitem [{\citenamefont {Caroli}\ \emph {et~al.}(1973)\citenamefont {Caroli},
  \citenamefont {Lederer-Rozenblatt}, \citenamefont {Roulet},\ and\
  \citenamefont {Saint-James}}]{caroli_inelastic_1973}%
  \BibitemOpen
  \bibfield  {author} {\bibinfo {author} {\bibfnamefont {C.}~\bibnamefont
  {Caroli}}, \bibinfo {author} {\bibfnamefont {D.}~\bibnamefont
  {Lederer-Rozenblatt}}, \bibinfo {author} {\bibfnamefont {B.}~\bibnamefont
  {Roulet}}, \ and\ \bibinfo {author} {\bibfnamefont {D.}~\bibnamefont
  {Saint-James}},\ }\href@noop {} {\bibfield  {journal} {\bibinfo  {journal}
  {Phys. Rev. B}\ }\textbf {\bibinfo {volume} {8}},\ \bibinfo {pages} {4552}
  (\bibinfo {year} {1973})}\BibitemShut {NoStop}%
\bibitem [{\citenamefont {Almbladh}(1986)}]{almbladh_importance_1986}%
  \BibitemOpen
  \bibfield  {author} {\bibinfo {author} {\bibfnamefont {C.-O.}\ \bibnamefont
  {Almbladh}},\ }\href@noop {} {\bibfield  {journal} {\bibinfo  {journal}
  {Phys. Rev. B}\ }\textbf {\bibinfo {volume} {34}},\ \bibinfo {pages} {3798}
  (\bibinfo {year} {1986})}\BibitemShut {NoStop}%
\bibitem [{\citenamefont {Campbell}\ \emph {et~al.}(2002)\citenamefont
  {Campbell}, \citenamefont {Hedin}, \citenamefont {Rehr},\ and\ \citenamefont
  {Bardyszewski}}]{campbell_interference_2002}%
  \BibitemOpen
  \bibfield  {author} {\bibinfo {author} {\bibfnamefont {L.}~\bibnamefont
  {Campbell}}, \bibinfo {author} {\bibfnamefont {L.}~\bibnamefont {Hedin}},
  \bibinfo {author} {\bibfnamefont {J.~J.}\ \bibnamefont {Rehr}}, \ and\
  \bibinfo {author} {\bibfnamefont {W.}~\bibnamefont {Bardyszewski}},\
  }\href@noop {} {\bibfield  {journal} {\bibinfo  {journal} {Phys. Rev. B}\
  }\textbf {\bibinfo {volume} {65}},\ \bibinfo {pages} {064107} (\bibinfo
  {year} {2002})}\BibitemShut {NoStop}%
\bibitem [{\citenamefont {Guzzo}\ \emph {et~al.}(2011)\citenamefont {Guzzo},
  \citenamefont {Lani}, \citenamefont {Sottile}, \citenamefont {Romaniello},
  \citenamefont {Gatti}, \citenamefont {Kas}, \citenamefont {Rehr},
  \citenamefont {Silly}, \citenamefont {Sirotti},\ and\ \citenamefont
  {Reining}}]{guzzo_valence_2011}%
  \BibitemOpen
  \bibfield  {author} {\bibinfo {author} {\bibfnamefont {M.}~\bibnamefont
  {Guzzo}}, \bibinfo {author} {\bibfnamefont {G.}~\bibnamefont {Lani}},
  \bibinfo {author} {\bibfnamefont {F.}~\bibnamefont {Sottile}}, \bibinfo
  {author} {\bibfnamefont {P.}~\bibnamefont {Romaniello}}, \bibinfo {author}
  {\bibfnamefont {M.}~\bibnamefont {Gatti}}, \bibinfo {author} {\bibfnamefont
  {J.~J.}\ \bibnamefont {Kas}}, \bibinfo {author} {\bibfnamefont {J.~J.}\
  \bibnamefont {Rehr}}, \bibinfo {author} {\bibfnamefont {M.~G.}\ \bibnamefont
  {Silly}}, \bibinfo {author} {\bibfnamefont {F.}~\bibnamefont {Sirotti}}, \
  and\ \bibinfo {author} {\bibfnamefont {L.}~\bibnamefont {Reining}},\
  }\href@noop {} {\bibfield  {journal} {\bibinfo  {journal} {Phys. Rev. Lett.}\
  }\textbf {\bibinfo {volume} {107}},\ \bibinfo {pages} {166401} (\bibinfo
  {year} {2011})}\BibitemShut {NoStop}%
\bibitem [{\citenamefont {Cini}(1976)}]{cini_density_1976}%
  \BibitemOpen
  \bibfield  {author} {\bibinfo {author} {\bibfnamefont {M.}~\bibnamefont
  {Cini}},\ }\href@noop {} {\bibfield  {journal} {\bibinfo  {journal} {Solid
  State Commun.}\ }\textbf {\bibinfo {volume} {20}},\ \bibinfo {pages} {605}
  (\bibinfo {year} {1976})}\BibitemShut {NoStop}%
\bibitem [{\citenamefont {Sawatzky}(1977)}]{sawatzky_quasiatomic_1977}%
  \BibitemOpen
  \bibfield  {author} {\bibinfo {author} {\bibfnamefont {G.~A.}\ \bibnamefont
  {Sawatzky}},\ }\href@noop {} {\bibfield  {journal} {\bibinfo  {journal}
  {Phys. Rev. Lett.}\ }\textbf {\bibinfo {volume} {39}},\ \bibinfo {pages}
  {504} (\bibinfo {year} {1977})}\BibitemShut {NoStop}%
\bibitem [{\citenamefont {Tarantelli}\ \emph {et~al.}(1991)\citenamefont
  {Tarantelli}, \citenamefont {Sgamellotti},\ and\ \citenamefont
  {Cederbaum}}]{tarantelli_many_1991}%
  \BibitemOpen
  \bibfield  {author} {\bibinfo {author} {\bibfnamefont {F.}~\bibnamefont
  {Tarantelli}}, \bibinfo {author} {\bibfnamefont {A.}~\bibnamefont
  {Sgamellotti}}, \ and\ \bibinfo {author} {\bibfnamefont {L.~S.}\ \bibnamefont
  {Cederbaum}},\ }\href@noop {} {\bibfield  {journal} {\bibinfo  {journal} {J.
  Chem. Phys.}\ }\textbf {\bibinfo {volume} {94}},\ \bibinfo {pages} {523}
  (\bibinfo {year} {1991})}\BibitemShut {NoStop}%
\bibitem [{\citenamefont {Tarantelli}\ \emph {et~al.}(1994)\citenamefont
  {Tarantelli}, \citenamefont {Sgamellotti},\ and\ \citenamefont
  {Cederbaum}}]{tarantelli_aggregation_1994}%
  \BibitemOpen
  \bibfield  {author} {\bibinfo {author} {\bibfnamefont {F.}~\bibnamefont
  {Tarantelli}}, \bibinfo {author} {\bibfnamefont {A.}~\bibnamefont
  {Sgamellotti}}, \ and\ \bibinfo {author} {\bibfnamefont {L.~S.}\ \bibnamefont
  {Cederbaum}},\ }\href@noop {} {\bibfield  {journal} {\bibinfo  {journal}
  {Phys. Rev. Lett.}\ }\textbf {\bibinfo {volume} {72}},\ \bibinfo {pages}
  {428} (\bibinfo {year} {1994})}\BibitemShut {NoStop}%
\bibitem [{\citenamefont {Verdozzi}\ \emph {et~al.}(2001)\citenamefont
  {Verdozzi}, \citenamefont {Cini},\ and\ \citenamefont
  {Marini}}]{verdozzi_auger_2001}%
  \BibitemOpen
  \bibfield  {author} {\bibinfo {author} {\bibfnamefont {C.}~\bibnamefont
  {Verdozzi}}, \bibinfo {author} {\bibfnamefont {M.}~\bibnamefont {Cini}}, \
  and\ \bibinfo {author} {\bibfnamefont {A.}~\bibnamefont {Marini}},\
  }\href@noop {} {\bibfield  {journal} {\bibinfo  {journal} {Journal of
  Electron Spectroscopy and Related Phenomena}\ }\textbf {\bibinfo {volume}
  {117-118}},\ \bibinfo {pages} {41} (\bibinfo {year} {2001})}\BibitemShut
  {NoStop}%
\bibitem [{\citenamefont {Cini}(1978)}]{cini_theory_1978}%
  \BibitemOpen
  \bibfield  {author} {\bibinfo {author} {\bibfnamefont {M.}~\bibnamefont
  {Cini}},\ }\href@noop {} {\bibfield  {journal} {\bibinfo  {journal} {Phys.
  Rev. B}\ }\textbf {\bibinfo {volume} {17}},\ \bibinfo {pages} {2486}
  (\bibinfo {year} {1978})}\BibitemShut {NoStop}%
\bibitem [{\citenamefont {Tarantelli}\ \emph {et~al.}(1985)\citenamefont
  {Tarantelli}, \citenamefont {Tarantelli}, \citenamefont {Sgamellotti},
  \citenamefont {Schirmer},\ and\ \citenamefont
  {Cederbaum}}]{tarantelli_greens_1985}%
  \BibitemOpen
  \bibfield  {author} {\bibinfo {author} {\bibfnamefont {F.}~\bibnamefont
  {Tarantelli}}, \bibinfo {author} {\bibfnamefont {A.}~\bibnamefont
  {Tarantelli}}, \bibinfo {author} {\bibfnamefont {A.}~\bibnamefont
  {Sgamellotti}}, \bibinfo {author} {\bibfnamefont {J.}~\bibnamefont
  {Schirmer}}, \ and\ \bibinfo {author} {\bibfnamefont {L.~S.}\ \bibnamefont
  {Cederbaum}},\ }\href@noop {} {\bibfield  {journal} {\bibinfo  {journal} {J.
  Chem. Phys.}\ }\textbf {\bibinfo {volume} {83}},\ \bibinfo {pages} {4683}
  (\bibinfo {year} {1985})}\BibitemShut {NoStop}%
\bibitem [{\citenamefont {Herrmann}\ \emph {et~al.}(1998)\citenamefont
  {Herrmann}, \citenamefont {Samarin}, \citenamefont {Schwabe},\ and\
  \citenamefont {Kirschner}}]{herrmann_two_1998}%
  \BibitemOpen
  \bibfield  {author} {\bibinfo {author} {\bibfnamefont {R.}~\bibnamefont
  {Herrmann}}, \bibinfo {author} {\bibfnamefont {S.}~\bibnamefont {Samarin}},
  \bibinfo {author} {\bibfnamefont {H.}~\bibnamefont {Schwabe}}, \ and\
  \bibinfo {author} {\bibfnamefont {J.}~\bibnamefont {Kirschner}},\ }\href@noop
  {} {\bibfield  {journal} {\bibinfo  {journal} {Phys. Rev. Lett.}\ }\textbf
  {\bibinfo {volume} {81}},\ \bibinfo {pages} {2148} (\bibinfo {year}
  {1998})}\BibitemShut {NoStop}%
\bibitem [{\citenamefont {Bardyszewski}\ and\ \citenamefont
  {Hedin}(1985)}]{bardyszewski_new_1985}%
  \BibitemOpen
  \bibfield  {author} {\bibinfo {author} {\bibfnamefont {W.}~\bibnamefont
  {Bardyszewski}}\ and\ \bibinfo {author} {\bibfnamefont {L.}~\bibnamefont
  {Hedin}},\ }\href@noop {} {\bibfield  {journal} {\bibinfo  {journal} {Phys.
  Scr.}\ }\textbf {\bibinfo {volume} {32}},\ \bibinfo {pages} {439} (\bibinfo
  {year} {1985})}\BibitemShut {NoStop}%
\bibitem [{\citenamefont {Fujikawa}\ and\ \citenamefont
  {Hedin}(1989)}]{fujikawa_theory_1989}%
  \BibitemOpen
  \bibfield  {author} {\bibinfo {author} {\bibfnamefont {T.}~\bibnamefont
  {Fujikawa}}\ and\ \bibinfo {author} {\bibfnamefont {L.}~\bibnamefont
  {Hedin}},\ }\href@noop {} {\bibfield  {journal} {\bibinfo  {journal} {Phys.
  Rev. B}\ }\textbf {\bibinfo {volume} {40}},\ \bibinfo {pages} {11507}
  (\bibinfo {year} {1989})}\BibitemShut {NoStop}%
\bibitem [{\citenamefont {Hedin}\ \emph {et~al.}(1998)\citenamefont {Hedin},
  \citenamefont {Michiels},\ and\ \citenamefont
  {Inglesfield}}]{hedin_transition_1998}%
  \BibitemOpen
  \bibfield  {author} {\bibinfo {author} {\bibfnamefont {L.}~\bibnamefont
  {Hedin}}, \bibinfo {author} {\bibfnamefont {J.}~\bibnamefont {Michiels}}, \
  and\ \bibinfo {author} {\bibfnamefont {J.}~\bibnamefont {Inglesfield}},\
  }\href@noop {} {\bibfield  {journal} {\bibinfo  {journal} {Phys. Rev. B}\
  }\textbf {\bibinfo {volume} {58}},\ \bibinfo {pages} {15565} (\bibinfo {year}
  {1998})}\BibitemShut {NoStop}%
\bibitem [{\citenamefont {Brand}\ and\ \citenamefont
  {Cederbaum}(1996)}]{brand_extended_1996}%
  \BibitemOpen
  \bibfield  {author} {\bibinfo {author} {\bibfnamefont {J.}~\bibnamefont
  {Brand}}\ and\ \bibinfo {author} {\bibfnamefont {L.~S.}\ \bibnamefont
  {Cederbaum}},\ }\href@noop {} {\bibfield  {journal} {\bibinfo  {journal}
  {Ann. Phys.}\ }\textbf {\bibinfo {volume} {252}},\ \bibinfo {pages} {276}
  (\bibinfo {year} {1996})}\BibitemShut {NoStop}%
\bibitem [{\citenamefont {Bell}\ and\ \citenamefont
  {Squires}(1959)}]{bell_formal_1959}%
  \BibitemOpen
  \bibfield  {author} {\bibinfo {author} {\bibfnamefont {J.~S.}\ \bibnamefont
  {Bell}}\ and\ \bibinfo {author} {\bibfnamefont {E.~J.}\ \bibnamefont
  {Squires}},\ }\href@noop {} {\bibfield  {journal} {\bibinfo  {journal} {Phys.
  Rev. Lett.}\ }\textbf {\bibinfo {volume} {3}},\ \bibinfo {pages} {96}
  (\bibinfo {year} {1959})}\BibitemShut {NoStop}%
\bibitem [{\citenamefont {Cederbaum}(2000)}]{cederbaum_optical_2000}%
  \BibitemOpen
  \bibfield  {author} {\bibinfo {author} {\bibfnamefont {L.~S.}\ \bibnamefont
  {Cederbaum}},\ }\href@noop {} {\bibfield  {journal} {\bibinfo  {journal}
  {Phys. Rev. Lett.}\ }\textbf {\bibinfo {volume} {85}},\ \bibinfo {pages}
  {3072} (\bibinfo {year} {2000})}\BibitemShut {NoStop}%
\bibitem [{\citenamefont {Cederbaum}(2001)}]{cederbaum_optical_2001}%
  \BibitemOpen
  \bibfield  {author} {\bibinfo {author} {\bibfnamefont {L.~S.}\ \bibnamefont
  {Cederbaum}},\ }\href@noop {} {\bibfield  {journal} {\bibinfo  {journal}
  {Ann. Phys.}\ }\textbf {\bibinfo {volume} {291}},\ \bibinfo {pages} {169}
  (\bibinfo {year} {2001})}\BibitemShut {NoStop}%
\bibitem [{\citenamefont {Berakdar}(1998)}]{berakdar_emission_1998}%
  \BibitemOpen
  \bibfield  {author} {\bibinfo {author} {\bibfnamefont {J.}~\bibnamefont
  {Berakdar}},\ }\href@noop {} {\bibfield  {journal} {\bibinfo  {journal}
  {Phys. Rev. B}\ }\textbf {\bibinfo {volume} {58}},\ \bibinfo {pages} {9808}
  (\bibinfo {year} {1998})}\BibitemShut {NoStop}%
\bibitem [{\citenamefont {Schwarzkopf}\ \emph {et~al.}(1993)\citenamefont
  {Schwarzkopf}, \citenamefont {Kr\"{a}ssig}, \citenamefont {Elmiger},\ and\
  \citenamefont {Schmidt}}]{schwarzkopf_energy-_1993}%
  \BibitemOpen
  \bibfield  {author} {\bibinfo {author} {\bibfnamefont {O.}~\bibnamefont
  {Schwarzkopf}}, \bibinfo {author} {\bibfnamefont {B.}~\bibnamefont
  {Kr\"{a}ssig}}, \bibinfo {author} {\bibfnamefont {J.}~\bibnamefont
  {Elmiger}}, \ and\ \bibinfo {author} {\bibfnamefont {V.}~\bibnamefont
  {Schmidt}},\ }\href@noop {} {\bibfield  {journal} {\bibinfo  {journal} {Phys.
  Rev. Lett.}\ }\textbf {\bibinfo {volume} {70}},\ \bibinfo {pages} {3008}
  (\bibinfo {year} {1993})}\BibitemShut {NoStop}%
\bibitem [{\citenamefont {Briggs}\ and\ \citenamefont
  {Schmidt}(2000)}]{briggs_differential_2000}%
  \BibitemOpen
  \bibfield  {author} {\bibinfo {author} {\bibfnamefont {J.~S.}\ \bibnamefont
  {Briggs}}\ and\ \bibinfo {author} {\bibfnamefont {V.}~\bibnamefont
  {Schmidt}},\ }\href@noop {} {\bibfield  {journal} {\bibinfo  {journal} {J.
  Phys. B}\ }\textbf {\bibinfo {volume} {33}},\ \bibinfo {pages} {R1} (\bibinfo
  {year} {2000})}\BibitemShut {NoStop}%
\bibitem [{\citenamefont {Esposito}\ and\ \citenamefont
  {Galperin}(2009)}]{esposito_transport_2009}%
  \BibitemOpen
  \bibfield  {author} {\bibinfo {author} {\bibfnamefont {M.}~\bibnamefont
  {Esposito}}\ and\ \bibinfo {author} {\bibfnamefont {M.}~\bibnamefont
  {Galperin}},\ }\href@noop {} {\bibfield  {journal} {\bibinfo  {journal}
  {Phys. Rev. B}\ }\textbf {\bibinfo {volume} {79}},\ \bibinfo {pages} {205303}
  (\bibinfo {year} {2009})}\BibitemShut {NoStop}%
\bibitem [{\citenamefont {Demuth}(2005)}]{demuth_determining_2005}%
  \BibitemOpen
  \bibfield  {author} {\bibinfo {author} {\bibfnamefont {M.}~\bibnamefont
  {Demuth}},\ }\href@noop {} {\emph {\bibinfo {title} {Determining spectra in
  quantum theory}}},\ \bibinfo {series} {Progress in mathematical physics}\
  No.\ \bibinfo {number} {v. 44}\ (\bibinfo  {publisher} {Birkh\"{a}user},\
  \bibinfo {address} {Boston},\ \bibinfo {year} {2005})\BibitemShut {NoStop}%
\bibitem [{\citenamefont {Marzari}\ \emph {et~al.}(2012)\citenamefont
  {Marzari}, \citenamefont {Mostofi}, \citenamefont {Yates}, \citenamefont
  {Souza},\ and\ \citenamefont {Vanderbilt}}]{marzari_maximally_2012}%
  \BibitemOpen
  \bibfield  {author} {\bibinfo {author} {\bibfnamefont {N.}~\bibnamefont
  {Marzari}}, \bibinfo {author} {\bibfnamefont {A.~A.}\ \bibnamefont
  {Mostofi}}, \bibinfo {author} {\bibfnamefont {J.~R.}\ \bibnamefont {Yates}},
  \bibinfo {author} {\bibfnamefont {I.}~\bibnamefont {Souza}}, \ and\ \bibinfo
  {author} {\bibfnamefont {D.}~\bibnamefont {Vanderbilt}},\ }\href@noop {}
  {\bibfield  {journal} {\bibinfo  {journal} {Rev. Mod. Phys.}\ }\textbf
  {\bibinfo {volume} {84}},\ \bibinfo {pages} {1419} (\bibinfo {year}
  {2012})}\BibitemShut {NoStop}%
\bibitem [{\citenamefont {Brouder}\ \emph {et~al.}(2007)\citenamefont
  {Brouder}, \citenamefont {Panati}, \citenamefont {Calandra}, \citenamefont
  {Mourougane},\ and\ \citenamefont {Marzari}}]{brouder_exponential_2007}%
  \BibitemOpen
  \bibfield  {author} {\bibinfo {author} {\bibfnamefont {C.}~\bibnamefont
  {Brouder}}, \bibinfo {author} {\bibfnamefont {G.}~\bibnamefont {Panati}},
  \bibinfo {author} {\bibfnamefont {M.}~\bibnamefont {Calandra}}, \bibinfo
  {author} {\bibfnamefont {C.}~\bibnamefont {Mourougane}}, \ and\ \bibinfo
  {author} {\bibfnamefont {N.}~\bibnamefont {Marzari}},\ }\href@noop {}
  {\bibfield  {journal} {\bibinfo  {journal} {Phys. Rev. Lett.}\ }\textbf
  {\bibinfo {volume} {98}},\ \bibinfo {pages} {046402} (\bibinfo {year}
  {2007})}\BibitemShut {NoStop}%
\bibitem [{\citenamefont {Joachain}(1975)}]{joachain_quantum_1975}%
  \BibitemOpen
  \bibfield  {author} {\bibinfo {author} {\bibfnamefont {C.~J.}\ \bibnamefont
  {Joachain}},\ }\href@noop {} {\emph {\bibinfo {title} {Quantum collision
  theory}}}\ (\bibinfo  {publisher} {North-Holland Pub. Co.; American Elsevier
  Pub. Co.},\ \bibinfo {address} {Amsterdam; New York},\ \bibinfo {year}
  {1975})\BibitemShut {NoStop}%
\bibitem [{\citenamefont {Berakdar}(2003)}]{berakdar_concepts_2003}%
  \BibitemOpen
  \bibfield  {author} {\bibinfo {author} {\bibfnamefont {J.}~\bibnamefont
  {Berakdar}},\ }\href@noop {} {\emph {\bibinfo {title} {Concepts of highly
  excited electronic systems}}},\ \bibinfo {edition} {1st}\ ed.\ (\bibinfo
  {publisher} {Wiley-VCH},\ \bibinfo {address} {Weinheim},\ \bibinfo {year}
  {2003})\BibitemShut {NoStop}%
\bibitem [{\citenamefont {Napitu}\ and\ \citenamefont
  {Berakdar}(2010)}]{napitu_two-particle_2010}%
  \BibitemOpen
  \bibfield  {author} {\bibinfo {author} {\bibfnamefont {B.~D.}\ \bibnamefont
  {Napitu}}\ and\ \bibinfo {author} {\bibfnamefont {J.}~\bibnamefont
  {Berakdar}},\ }\href@noop {} {\bibfield  {journal} {\bibinfo  {journal}
  {Phys. Rev. B}\ }\textbf {\bibinfo {volume} {81}},\ \bibinfo {pages} {195108}
  (\bibinfo {year} {2010})}\BibitemShut {NoStop}%
\bibitem [{\citenamefont {Pavlyukh}\ and\ \citenamefont
  {Berakdar}(2011)}]{pavlyukh_communication:_2011}%
  \BibitemOpen
  \bibfield  {author} {\bibinfo {author} {\bibfnamefont {Y.}~\bibnamefont
  {Pavlyukh}}\ and\ \bibinfo {author} {\bibfnamefont {J.}~\bibnamefont
  {Berakdar}},\ }\href@noop {} {\bibfield  {journal} {\bibinfo  {journal} {J.
  Chem. Phys.}\ }\textbf {\bibinfo {volume} {135}},\ \bibinfo {pages} {201103}
  (\bibinfo {year} {2011})}\BibitemShut {NoStop}%
\bibitem [{\citenamefont {Bang}\ \emph {et~al.}(1985)\citenamefont {Bang},
  \citenamefont {Gareev}, \citenamefont {Pinkston},\ and\ \citenamefont
  {Vaagen}}]{bang_one-_1985}%
  \BibitemOpen
  \bibfield  {author} {\bibinfo {author} {\bibfnamefont {J.~M.}\ \bibnamefont
  {Bang}}, \bibinfo {author} {\bibfnamefont {F.~G.}\ \bibnamefont {Gareev}},
  \bibinfo {author} {\bibfnamefont {W.~T.}\ \bibnamefont {Pinkston}}, \ and\
  \bibinfo {author} {\bibfnamefont {J.~S.}\ \bibnamefont {Vaagen}},\
  }\href@noop {} {\bibfield  {journal} {\bibinfo  {journal} {Phys. Rep.}\
  }\textbf {\bibinfo {volume} {125}},\ \bibinfo {pages} {253} (\bibinfo {year}
  {1985})}\BibitemShut {NoStop}%
\bibitem [{\citenamefont {Cederbaum}\ \emph {et~al.}(1986)\citenamefont
  {Cederbaum}, \citenamefont {Domcke}, \citenamefont {Schirmer},\ and\
  \citenamefont {Von~Niessen}}]{cederbaum_correlation_1986}%
  \BibitemOpen
  \bibfield  {author} {\bibinfo {author} {\bibfnamefont {L.~S.}\ \bibnamefont
  {Cederbaum}}, \bibinfo {author} {\bibfnamefont {W.}~\bibnamefont {Domcke}},
  \bibinfo {author} {\bibfnamefont {J.}~\bibnamefont {Schirmer}}, \ and\
  \bibinfo {author} {\bibfnamefont {W.}~\bibnamefont {Von~Niessen}},\
  }\href@noop {} {\bibfield  {journal} {\bibinfo  {journal} {Adv. Chem. Phys}\
  }\textbf {\bibinfo {volume} {65}},\ \bibinfo {pages} {115} (\bibinfo {year}
  {1986})}\BibitemShut {NoStop}%
\bibitem [{\citenamefont {Deleuze}\ and\ \citenamefont
  {Cederbaum}(1999)}]{deleuze_new_1999}%
  \BibitemOpen
  \bibfield  {author} {\bibinfo {author} {\bibfnamefont {M.}~\bibnamefont
  {Deleuze}}\ and\ \bibinfo {author} {\bibfnamefont {L.}~\bibnamefont
  {Cederbaum}},\ }\href@noop {} {\bibfield  {journal} {\bibinfo  {journal}
  {Adv. Quantum Chem.}\ }\textbf {\bibinfo {volume} {35}},\ \bibinfo {pages}
  {77} (\bibinfo {year} {1999})}\BibitemShut {NoStop}%
\bibitem [{\citenamefont {Fominykh}\ \emph {et~al.}(2000)\citenamefont
  {Fominykh}, \citenamefont {Henk}, \citenamefont {Berakdar}, \citenamefont
  {Bruno}, \citenamefont {Gollisch},\ and\ \citenamefont
  {Feder}}]{fominykh_theory_2000}%
  \BibitemOpen
  \bibfield  {author} {\bibinfo {author} {\bibfnamefont {N.}~\bibnamefont
  {Fominykh}}, \bibinfo {author} {\bibfnamefont {J.}~\bibnamefont {Henk}},
  \bibinfo {author} {\bibfnamefont {J.}~\bibnamefont {Berakdar}}, \bibinfo
  {author} {\bibfnamefont {P.}~\bibnamefont {Bruno}}, \bibinfo {author}
  {\bibfnamefont {H.}~\bibnamefont {Gollisch}}, \ and\ \bibinfo {author}
  {\bibfnamefont {R.}~\bibnamefont {Feder}},\ }\href@noop {} {\bibfield
  {journal} {\bibinfo  {journal} {Solid State Commun.}\ }\textbf {\bibinfo
  {volume} {113}},\ \bibinfo {pages} {665} (\bibinfo {year}
  {2000})}\BibitemShut {NoStop}%
\bibitem [{\citenamefont {Fominykh}\ \emph {et~al.}(2002)\citenamefont
  {Fominykh}, \citenamefont {Berakdar}, \citenamefont {Henk},\ and\
  \citenamefont {Bruno}}]{fominykh_spectroscopy_2002}%
  \BibitemOpen
  \bibfield  {author} {\bibinfo {author} {\bibfnamefont {N.}~\bibnamefont
  {Fominykh}}, \bibinfo {author} {\bibfnamefont {J.}~\bibnamefont {Berakdar}},
  \bibinfo {author} {\bibfnamefont {J.}~\bibnamefont {Henk}}, \ and\ \bibinfo
  {author} {\bibfnamefont {P.}~\bibnamefont {Bruno}},\ }\href@noop {}
  {\bibfield  {journal} {\bibinfo  {journal} {Phys. Rev. Lett.}\ }\textbf
  {\bibinfo {volume} {89}},\ \bibinfo {pages} {086402} (\bibinfo {year}
  {2002})}\BibitemShut {NoStop}%
\bibitem [{\citenamefont {Domcke}(1991)}]{domcke_theory_1991}%
  \BibitemOpen
  \bibfield  {author} {\bibinfo {author} {\bibfnamefont {W.}~\bibnamefont
  {Domcke}},\ }\href@noop {} {\bibfield  {journal} {\bibinfo  {journal} {Phys.
  Rep.}\ }\textbf {\bibinfo {volume} {208}},\ \bibinfo {pages} {97} (\bibinfo
  {year} {1991})}\BibitemShut {NoStop}%
\bibitem [{\citenamefont {Capuzzi}\ and\ \citenamefont
  {Mahaux}(1996)}]{capuzzi_projection_1996}%
  \BibitemOpen
  \bibfield  {author} {\bibinfo {author} {\bibfnamefont {F.}~\bibnamefont
  {Capuzzi}}\ and\ \bibinfo {author} {\bibfnamefont {C.}~\bibnamefont
  {Mahaux}},\ }\href@noop {} {\bibfield  {journal} {\bibinfo  {journal} {Ann.
  Phys.}\ }\textbf {\bibinfo {volume} {245}},\ \bibinfo {pages} {147} (\bibinfo
  {year} {1996})}\BibitemShut {NoStop}%
\bibitem [{\citenamefont {Roulet}\ \emph {et~al.}(1969)\citenamefont {Roulet},
  \citenamefont {Gavoret},\ and\ \citenamefont
  {Nozi\`{e}res}}]{roulet_singularities_1969}%
  \BibitemOpen
  \bibfield  {author} {\bibinfo {author} {\bibfnamefont {B.}~\bibnamefont
  {Roulet}}, \bibinfo {author} {\bibfnamefont {J.}~\bibnamefont {Gavoret}}, \
  and\ \bibinfo {author} {\bibfnamefont {P.}~\bibnamefont {Nozi\`{e}res}},\
  }\href@noop {} {\bibfield  {journal} {\bibinfo  {journal} {Phys. Rev.}\
  }\textbf {\bibinfo {volume} {178}},\ \bibinfo {pages} {1072} (\bibinfo {year}
  {1969})}\BibitemShut {NoStop}%
\bibitem [{\citenamefont {Feuerbacher}\ and\ \citenamefont
  {Cederbaum}(2005)}]{feuerbacher_direct_2005}%
  \BibitemOpen
  \bibfield  {author} {\bibinfo {author} {\bibfnamefont {B.}~\bibnamefont
  {Feuerbacher}}\ and\ \bibinfo {author} {\bibfnamefont {L.~S.}\ \bibnamefont
  {Cederbaum}},\ }\href@noop {} {\bibfield  {journal} {\bibinfo  {journal}
  {Phys. Rev. A}\ }\textbf {\bibinfo {volume} {72}},\ \bibinfo {pages} {022731}
  (\bibinfo {year} {2005})}\BibitemShut {NoStop}%
\bibitem [{\citenamefont {Stefanucci}\ \emph {et~al.}(2014)\citenamefont
  {Stefanucci}, \citenamefont {Pavlyukh}, \citenamefont {Uimonen},\ and\
  \citenamefont {van Leeuwen}}]{stefanucci_diagrammatic_2014}%
  \BibitemOpen
  \bibfield  {author} {\bibinfo {author} {\bibfnamefont {G.}~\bibnamefont
  {Stefanucci}}, \bibinfo {author} {\bibfnamefont {Y.}~\bibnamefont
  {Pavlyukh}}, \bibinfo {author} {\bibfnamefont {A.-M.}\ \bibnamefont
  {Uimonen}}, \ and\ \bibinfo {author} {\bibfnamefont {R.}~\bibnamefont {van
  Leeuwen}},\ }\href@noop {} {\bibfield  {journal} {\bibinfo  {journal} {Phys.
  Rev. B}\ }\textbf {\bibinfo {volume} {90}},\ \bibinfo {pages} {115134}
  (\bibinfo {year} {2014})}\BibitemShut {NoStop}%
\bibitem [{\citenamefont {Uimonen}\ \emph {et~al.}(2015)\citenamefont
  {Uimonen}, \citenamefont {Stefanucci}, \citenamefont {Pavlyukh},\ and\
  \citenamefont {van Leeuwen}}]{uimonen_diagrammatic_2015}%
  \BibitemOpen
  \bibfield  {author} {\bibinfo {author} {\bibfnamefont {A.-M.}\ \bibnamefont
  {Uimonen}}, \bibinfo {author} {\bibfnamefont {G.}~\bibnamefont {Stefanucci}},
  \bibinfo {author} {\bibfnamefont {Y.}~\bibnamefont {Pavlyukh}}, \ and\
  \bibinfo {author} {\bibfnamefont {R.}~\bibnamefont {van Leeuwen}},\
  }\href@noop {} {\bibfield  {journal} {\bibinfo  {journal} {Phys. Rev. B}\
  }\textbf {\bibinfo {volume} {91}},\ \bibinfo {pages} {115104} (\bibinfo
  {year} {2015})}\BibitemShut {NoStop}%
\bibitem [{\citenamefont {Pavlyukh}\ \emph {et~al.}(2013)\citenamefont
  {Pavlyukh}, \citenamefont {Rubio},\ and\ \citenamefont
  {Berakdar}}]{pavlyukh_time_2013}%
  \BibitemOpen
  \bibfield  {author} {\bibinfo {author} {\bibfnamefont {Y.}~\bibnamefont
  {Pavlyukh}}, \bibinfo {author} {\bibfnamefont {A.}~\bibnamefont {Rubio}}, \
  and\ \bibinfo {author} {\bibfnamefont {J.}~\bibnamefont {Berakdar}},\
  }\href@noop {} {\bibfield  {journal} {\bibinfo  {journal} {Phys. Rev. B}\
  }\textbf {\bibinfo {volume} {87}},\ \bibinfo {pages} {205124} (\bibinfo
  {year} {2013})}\BibitemShut {NoStop}%
\bibitem [{\citenamefont {Madjet}\ \emph {et~al.}(2008)\citenamefont {Madjet},
  \citenamefont {Chakraborty}, \citenamefont {Rost},\ and\ \citenamefont
  {Manson}}]{madjet_photoionization_2008}%
  \BibitemOpen
  \bibfield  {author} {\bibinfo {author} {\bibfnamefont {M.~E.}\ \bibnamefont
  {Madjet}}, \bibinfo {author} {\bibfnamefont {H.~S.}\ \bibnamefont
  {Chakraborty}}, \bibinfo {author} {\bibfnamefont {J.~M.}\ \bibnamefont
  {Rost}}, \ and\ \bibinfo {author} {\bibfnamefont {S.~T.}\ \bibnamefont
  {Manson}},\ }\href@noop {} {\bibfield  {journal} {\bibinfo  {journal} {J.
  Phys. B}\ }\textbf {\bibinfo {volume} {41}},\ \bibinfo {pages} {105101}
  (\bibinfo {year} {2008})}\BibitemShut {NoStop}%
\bibitem [{\citenamefont {Pavlyukh}\ and\ \citenamefont
  {Berakdar}(2010)}]{pavlyukh_kohn-sham_2010}%
  \BibitemOpen
  \bibfield  {author} {\bibinfo {author} {\bibfnamefont {Y.}~\bibnamefont
  {Pavlyukh}}\ and\ \bibinfo {author} {\bibfnamefont {J.}~\bibnamefont
  {Berakdar}},\ }\href@noop {} {\bibfield  {journal} {\bibinfo  {journal}
  {Phys. Rev. A}\ }\textbf {\bibinfo {volume} {81}},\ \bibinfo {pages} {042515}
  (\bibinfo {year} {2010})}\BibitemShut {NoStop}%
\bibitem [{\citenamefont {Edmonds}(1996)}]{edmonds_angular_1996}%
  \BibitemOpen
  \bibfield  {author} {\bibinfo {author} {\bibfnamefont {A.~R.}\ \bibnamefont
  {Edmonds}},\ }\href@noop {} {\emph {\bibinfo {title} {Angular momentum in
  quantum mechanics}}}\ (\bibinfo  {publisher} {Princeton University Press},\
  \bibinfo {year} {1996})\BibitemShut {NoStop}%
\bibitem [{\citenamefont {Varshalovich}\ and\ \citenamefont
  {Moskalev}(1988)}]{varshalovich_quantum_1988}%
  \BibitemOpen
  \bibfield  {author} {\bibinfo {author} {\bibfnamefont {D.~A.}\ \bibnamefont
  {Varshalovich}}\ and\ \bibinfo {author} {\bibfnamefont {A.~N.}\ \bibnamefont
  {Moskalev}},\ }\href@noop {} {\emph {\bibinfo {title} {Quantum {Theory} of
  {Angular} {Momentum}}}}\ (\bibinfo  {publisher} {World Scientific Pub Co
  Inc},\ \bibinfo {year} {1988})\BibitemShut {NoStop}%
\bibitem [{\citenamefont {Moskalenko}\ \emph {et~al.}(2012)\citenamefont
  {Moskalenko}, \citenamefont {Pavlyukh},\ and\ \citenamefont
  {Berakdar}}]{moskalenko_attosecond_2012}%
  \BibitemOpen
  \bibfield  {author} {\bibinfo {author} {\bibfnamefont {A.~S.}\ \bibnamefont
  {Moskalenko}}, \bibinfo {author} {\bibfnamefont {Y.}~\bibnamefont
  {Pavlyukh}}, \ and\ \bibinfo {author} {\bibfnamefont {J.}~\bibnamefont
  {Berakdar}},\ }\href@noop {} {\bibfield  {journal} {\bibinfo  {journal}
  {Phys. Rev. A}\ }\textbf {\bibinfo {volume} {86}},\ \bibinfo {pages} {013202}
  (\bibinfo {year} {2012})}\BibitemShut {NoStop}%
\bibitem [{\citenamefont {Zhang}(2005)}]{zhang_schur_2005}%
  \BibitemOpen
  \bibfield  {author} {\bibinfo {author} {\bibfnamefont {F.}~\bibnamefont
  {Zhang}},\ }\href@noop {} {\emph {\bibinfo {title} {The {Schur} complement
  and its applications}}},\ \bibinfo {series} {Numerical methods and
  algorithms}\ No.~\bibinfo {number} {4}\ (\bibinfo  {publisher} {Springer},\
  \bibinfo {address} {New York},\ \bibinfo {year} {2005})\BibitemShut {NoStop}%
\bibitem [{\citenamefont {Feshbach}(1962)}]{feshbach_unified_1962}%
  \BibitemOpen
  \bibfield  {author} {\bibinfo {author} {\bibfnamefont {H.}~\bibnamefont
  {Feshbach}},\ }\href@noop {} {\bibfield  {journal} {\bibinfo  {journal} {Ann.
  Phys.}\ }\textbf {\bibinfo {volume} {19}},\ \bibinfo {pages} {287} (\bibinfo
  {year} {1962})}\BibitemShut {NoStop}%
\bibitem [{\citenamefont {Escher}\ and\ \citenamefont
  {Jennings}(2002)}]{escher_one-body_2002}%
  \BibitemOpen
  \bibfield  {author} {\bibinfo {author} {\bibfnamefont {J.}~\bibnamefont
  {Escher}}\ and\ \bibinfo {author} {\bibfnamefont {B.~K.}\ \bibnamefont
  {Jennings}},\ }\href@noop {} {\bibfield  {journal} {\bibinfo  {journal}
  {Phys. Rev. C}\ }\textbf {\bibinfo {volume} {66}},\ \bibinfo {pages} {034313}
  (\bibinfo {year} {2002})}\BibitemShut {NoStop}%
\bibitem [{\citenamefont {Gustafson}(2011)}]{gustafson_mathematical_2011}%
  \BibitemOpen
  \bibfield  {author} {\bibinfo {author} {\bibfnamefont {S.~J.}\ \bibnamefont
  {Gustafson}},\ }\href@noop {} {\emph {\bibinfo {title} {Mathematical concepts
  of quantum mechanics}}},\ \bibinfo {edition} {2nd}\ ed.,\ Universitext\
  (\bibinfo  {publisher} {Springer},\ \bibinfo {address} {Heidelberg},\
  \bibinfo {year} {2011})\BibitemShut {NoStop}%
\bibitem [{Note1()}]{Note1}%
  \BibitemOpen
  \bibinfo {note} {According to Zhang~\cite {zhang_schur_2005} it was a Polish
  astronomer Banachiewicz who obtained this formula for the first time.
  However, it was reinvented many times (see a short historical review at the
  top of p. 699 of Ref.~\cite {jensen_fermi_2006} where the authors suggest to
  use the name Schur-Livsic-Feshbach-Grushin for the equation)}\BibitemShut
  {NoStop}%
\bibitem [{\citenamefont {Jensen}\ and\ \citenamefont
  {Nenciu}(2006)}]{jensen_fermi_2006}%
  \BibitemOpen
  \bibfield  {author} {\bibinfo {author} {\bibfnamefont {A.}~\bibnamefont
  {Jensen}}\ and\ \bibinfo {author} {\bibfnamefont {G.}~\bibnamefont
  {Nenciu}},\ }\href@noop {} {\bibfield  {journal} {\bibinfo  {journal}
  {Commun. Math. Phys.}\ }\textbf {\bibinfo {volume} {261}},\ \bibinfo {pages}
  {693} (\bibinfo {year} {2006})}\BibitemShut {NoStop}%
\end{thebibliography}
\end{document}